\documentclass[vecphys]{svmult}

\usepackage{makeidx}         % allows index generation
\usepackage{graphicx}        % standard LaTeX graphics tool
\usepackage{multicol}        % used for the two-column index
\usepackage{cite}            % adjusts the "syntax" of the refs in the
\usepackage[bottom]{footmisc}% places footnotes at page bottom
\makeindex             % used for the subject index
\begin{document}

\title{Nonlinear interactions and non-classical light}
\author{Dmitry V. Strekalov \and Gerd Leuchs}
\institute{Max Planck Institute for the Science of Light, Staudstra\ss e 2, 90158 Erlangen, Germany
\texttt{strekalov@yahoo.com}
%\and Name and Address of your Institute \texttt{name@email.address}
}
% Use the package "url.sty" to avoid problems with special characters
% used in your e-mail or web address.
% Addresses should be removed from contribution and entered into
% blist.tex" (by the compiler).

\maketitle

\begin{abstract}

Non-classical concerns light whose properties cannot be
explained by classical electrodynamics and which requires invoking quantum
principles to be understood. Its existence is a direct consequence of field quantization; its study is a source of our understanding of
many quantum phenomena. Non-classical light also has properties that may be
of technological significance. We start this chapter by discussing the
definition of non-classical light and basic examples. Then some of the most
prominent applications of non-classical light are reviewed. After that, as
the principal part of our discourse, we review the most common sources of
non-classical light. We will find them surprisingly diverse, including
physical systems of various sizes and complexity, ranging from single
atoms to optical crystals and to semiconductor lasers. Putting all these
dissimilar optical devices in the common perspective we attempt to
establish a trend in the field and to foresee the new cross-disciplinary
approaches and techniques of generating non-classical light.

\end{abstract}

\noindent

\section{Introduction}\label{sec:Intro}

\subsection{Classical and non-classical light}\label{sec:nonclas}

In historical perspective, light doubtlessly is among the most classical phenomena of physics. The oldest known treatise on this subject, ``Optics" by Euclid, dates back to approximately 300 B.C. Yet in contemporary physics, light is one of the strongest manifestations of quantum. Electromagnetic field quanta, the \textit{photons}, are certainly real: they can be emitted and detected one by one, delivering discrete portions of energy and momentum, and in this sense may be viewed as particles of light. At the same time, wave properties of light are most readily observed in diffraction and interference experiments. 

Before the quantum mechanical principle of \textit{duality} was understood, this twofold nature of light has lead to curious oscillations in its understanding by scientists and philosophers starting from antiquity. Pythagoras believed light to
consist of moving particles, but Aristotle later compared it to ocean waves. Much later Sir Isaac Newton has revived the concept of corpuscles, which again yielded the ground to the wave theory when interference and diffraction were discovered. Then the sum of evidence for each, the wave and the corpuscular nature of light, became undeniable and quantum optics emerged.

So what is classical and what is non-classical light? 
Before going on we note that nature is as it is, and any distinction between quantum and classical is somewhat artificial and is merely a result of our desire to describe nature by models with which we can make quantitative predictions. There is a whole class of models labeled classical because they have or could have been formulated before the invention of quantum physics. Nevertheless it can help our perception to see how far a particular model, be it in the classical or in the quantum class, can be stretched before it fails making correct predictions. No matter which type of light one studies, it seems obvious that one can in principle find some departure between the model and the observation. %The harder it is to find a discrepancy, the wider the applicability and 

It appears logical to call ``classical" the phenomena that can be  quantitatively described without invoking quantum mechanics, e.g. in terms of Maxwell's equations. Interference and diffraction are obvious examples from classical optics. Somewhat less obvious examples are ``photon bunching", Hanbury Brown and Twiss type interference of thermal light, and a few other phenomena that occasionally raise the quantum-or-classical debate in conference halls and in the literature. 

Conversely, non-classical (quantum) are those phenomena that can \textit{only} be described in quantum mechanics. It should be noted that in many cases it is \textit{convenient} to describe classical light in terms of quantum optics, which, however, does not make it non-classical in the sense mentioned above. This is done in order to use the same language for classical and quantum phenomena analysis. As stated above, nature does not make this distinction. It is our choice if we employ as much as possible a classical model with a limited applicability.

One of the most useful quantum vs. classical distinction criteria is based on the various correlation functions of optical fields. Such correlation functions are computed by averaging the observables over their joint probability distribution. For a simple example let us consider a normalized auto-correlation function of light intensity $I(t)=E^*(t)E(t)$:
\begin{equation}
g^{(2)}(\tau)=\frac{\langle E^*(t)E^*(t+\tau)E(t)E(t+\tau)\rangle}{\langle E^*(t)E(t)\rangle^2}.\label{g2}
\end{equation}
Using the Cauchy-Schwarz inequality, it is easy to see \cite{LoudonBook,Klyshko98concepts} that $g^{(2)}(0)\geq 1$. Smaller values for this observable are impossible in classical optics, but they do occur in nature, e.g. for photon number states and for  amplitude-squeezed light. Therefore \textit{antibunching}  \cite{Walls79review,Paul82rev,Leuchs1986Ch26,Klyshko96rev_nonclas,Kimble77antibunch} $g^{(2)}(0)< 1$ can be taken as a sufficient but not necessary criterion for non-classical light\footnote{In practice, it is often convenient  to measure autocorrelation function (\ref{g2}) using a beam splitter and a pair of detectors. The same or similar set up can be used for measuring a \textit{cross-correlation} function of two optical modes. Note that this measurement yields a different observable whose classical range is $g_{12}^{(2)}(0) > 0.5$ \cite{Klyshko94aspects}.}. 
A similar argument can be made for the intensity correlation (\ref{g2}) as a function of spatial coordinates instead of time. In this case inequalities similar to Cauchy-Schwarz lead to such non-classicality criteria as Bell's inequalities violation \cite{Bell64,chsh1969,Clauser74bell,Clauser78bell} and negative conditional Von Neumann entropy \cite{Cerf97entropy,nielsen2010quantum}. 

Another criterion, likewise sufficient but not necessary, is the negativity of the phase space distribution function \cite{SchleichBk}. In quantum mechanics such distribution functions can be introduced with limiting cases being the Wigner function, or the Glauber-Sudarshan P-function. Negative, complex or irregular values of these functions can also be used as indications of non-classical light \cite{Hillery84rev,Hillery89noise,Lee90nonclas,Klyshko96nonclas}. We note that one of the limiting cases, the Q-function (also referred to as the Husimi function), is always regular. 

Both criteria are sufficient but not necessary as can be seen by the following examples: Photon number states are non-classical according to both criteria; amplitude squeezed states are non-classical according to the first but not the second criterion; superpositions of coherent states, so-called cat states \cite{Yurke86cat,Vlastakis13cat}\footnote{The cat states, named after Schr\"{o}dinger's cat, are  superpositions of two out-of-phase macroscopic ($|\alpha|\gg1$) coherent states, e.g. $|\Psi\rangle_{\rm cat}\propto|\alpha\rangle+|-\alpha\rangle$%$[2(1+e^{-2|\alpha|^2})]^{-1/2}$
.}, are non-classical according to the second but not the first criterion.

The qualifier ``not necessary" in the above criteria is essential. Currently we know of no simple general criterion, which is sufficient \textit{and} necessary. But we can make the following statement: classical states can involve either no fluctuations at all, or only statistical fluctuations. In quantum physics such states either do not exist, or they are described as mixed states. Pure quantum states exhibit a so-called quantum uncertainty, which results in the projection noise when measured. A classical stochastic model can
describe some aspects of a quantum uncertainty. But such models are always limited.
Allowing for all possible experimental scenarios, a pure quantum state can never be
described by one and the same classical stochastic model. In this sense coherent states, being pure quantum states in their own right, have to be classified as non-classical. This statement calls for a more detailed justification, which is provided in the Appendix in the end of this chapter. 

\subsection{Types of non-classical light}\label{sec:types}

A key concept in the following discussion will be an optical mode. This concept is fundamental to electromagnetic field quantization: the spatio-temporal character of a mode is described by a real-valued mode function, which is a function of position and time and is normalized. In classical physics this amplitude function is multiplied by a complex number describing amplitude and phase, or alternatively two orthogonal field quadratures. In quantum optics these amplitudes are described by operators; superpositions of the photon creation and annihilation operators, to be precise.

\begin{figure}[b]
\centering
\includegraphics*[width=\textwidth]{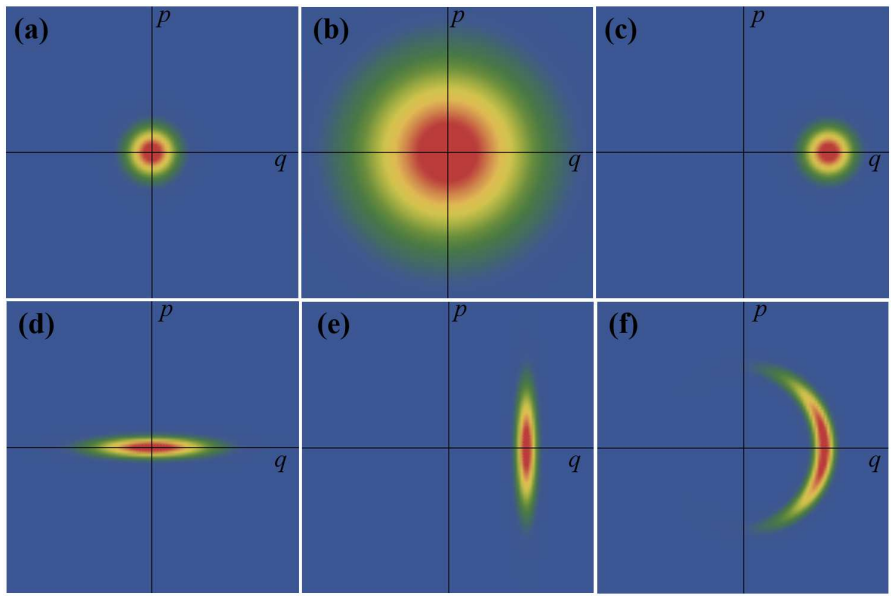}
\caption[]{Phase space diagrams for symmetric ordering (Wigner functions) for some classical and non-classical states: vacuum (a),  thermal (b), coherent (c), squeezed vacuum (d), quadrature squeezed light (e), and photon-number squeezed light (f). Mean photon number for states (b), (c), (e) and (f) is $\langle\hat{N}\rangle=6$, for (d) $\langle\hat{N}\rangle=1$. Squeezing parameter for states (d)-(f) is 7.66 dB.}
\label{fig:pq}       
\end{figure}

Quantum-statistical properties of single-mode light are conveniently illustrated by phase space diagrams where the optical state Wigner function $W(p,q)$ is plotted against the canonical harmonic oscillator coordinates $p$ and $q$. We recall \cite{LoudonBook} that for an optical mode with central frequency $\omega$ the corresponding quadrature  operators are related to the photon creation and annihilation operators $a^\dagger$ and $a$ as
\begin{equation}
\hat{p}=-i\sqrt{\hbar\omega/2}\,(a-a^\dagger),\quad
\hat{q}=-i\sqrt{\hbar/(2\omega)}\,(a+a^\dagger).\label{pq}
\end{equation}
In Fig.~\ref{fig:pq} we show some examples of classical %(a-c), 
and non-classical %(d-f) 
light phase diagrams. Here (a) is the vacuum state, and (b), (c) are the thermal and coherent states with the mean photon number $\langle\hat{N}\rangle=\langle a^\dagger a\rangle=6$, respectively (the plots are scaled for $\hbar\omega=1$). The diagram (d) represents squeezed vacuum with $\langle\hat{N}\rangle=1$. For the squeezed vacuum states the mean photon number is uniquely related to squeezing; our example  $\langle\hat{N}\rangle=1$ requires $20\lg(e){\rm arcsinh}(\langle\hat{N}\rangle^{1/2})\approx 7.66$ dB of squeezing. The diagrams (e) and (f) show the quadrature and amplitude (or photon-number) squeezed states, respectively, with the same squeezing factor as in (d) and the same mean photon number as in (b) and (c). 

The thermal state represented in Fig.~\ref{fig:pq}(b) is clearly classical. The vacuum state (a) and the coherent state (c), which is  also called displaced vacuum state, are said to be at the quantum-classical boundary. They do not violate either of the two criteria formulated above, but the boundary value $g^{(2)}(0)=1$ implies that the optical field does not fluctuate, see Eq.(\ref{g2}). Then the Poissonian statistics of photocounts, observed with a coherent field, must be attributed to stochastic character of the detection process. But if this were the case, it would not be possible to observe sub shot noise correlation of two independent detectors' signals (such as measured in two-mode squeezing experiments) even with 100\%-efficient photodetectors. Therefore one is lead to a conclusion that detection of coherent light cannot be fully described in the semiclassical approximation, and in this sense such light is non-classical. Despite these notational difficulties coherent states as any other pure quantum state can be used as resource for optical quantum engineering, such as in quantum key distribution.

\begin{figure}[htb]
\centering
\includegraphics*[width=\textwidth]{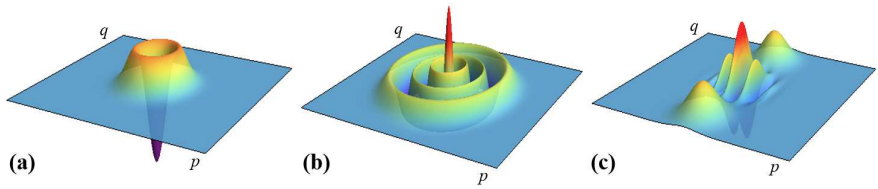}
\caption[]{Phase space diagrams for non-Gaussian states: single-photon (a) and six-photon (b) Fock states, and a cat state corresponding to superposition of two $\langle N\rangle=6$ coherent states such as shown in Fig.~\ref{fig:pq}(c).}
\label{fig:pq1}       
\end{figure}

Wigner functions for the classical and non-classical states shown in Fig.~\ref{fig:pq} are symmetric or distorted Gaussian. Other non-classical states may be non-Gaussian, and clearly displaying Wigner function negativity. Such are the Fock states and the already mentioned cat states, shown in Fig.~\ref{fig:pq1}. Their Wigner functions are respectively
\begin{eqnarray}
W_{\rm Fock}&=&\frac{2}{\pi}(-1)^N{\rm L}_N(2p^2+2q^2)e^{-p^2-q^2},\\
W_{\rm cat}&=&\frac{e^{-p^2-(q-|\alpha|)^2}+e^{-p^2-(q+|\alpha|)^2}+2\cos(2p|\alpha|)e^{-p^2-q^2}}{\pi
\sqrt{2\left(1+e^{-2|\alpha|^2}\right)}
},\nonumber
\end{eqnarray}
where ${\rm L}_N$ is the $N$-th order Laguerre polynomial.
It is interesting to observe that the Wigner function of a photon-number eigenstate indicates a non-zero quasi-probability for the mode to be found in a \textit{different} number-state, and in fact reaches the maximum for the vacuum state $|N\rangle=0$. This reminds us to be cautious with physical interpretations of quasi-probability functions such as the Wigner function. 

The standard way of measuring and gauging the intensity fluctuations is to split the optical beam with a 50/50 beam splitter and either subtract or multiply the photocurrents of two detectors placed at each output. Time dependence may be obtained by introducing a variable optical or electronic delay in one channel. This approach would identify the photon-number squeezed state shown in  Fig.~\ref{fig:pq}(f) as non-classical, but would not reveal the non-classical properties of the squeezed states in diagrams (d) and (e). In fact, an ensemble of measurements on such identically prepared states (which is often equivalent to a time-sequential measurement on one system) would show excessive photon-number fluctuation, above the shot noise limit.
To measure the quadrature squeezing one has to set up a measurement sensitive to the Wigner function projection on the squeezed quadrature, rather than in the radial direction. This can be achieved in a heterodyne measurement, when a coherent local oscillator field is injected into the unused port of the beam splitter. Changing the local oscillator phase one can chose the projection direction. 
The right choice of the phase leads to a sub shot noise measurement revealing the non-classicallity: $g^{(2)}(0)< 1$. The same situation can be described in a different language, by saying that the beam splitter transforms the input (e.g. quadrature-squeezed and coherent) modes to output modes, both of which are photon-number squeezed, or anti-bunched. 

Non-classical phenomena in two or more optical modes are usually associated with the term \textit{entanglement}. A quantum optical system comprising modes labeled $A$ and $B$ is said to be entangled if its wave function does not factorize: $|\Psi\rangle_{AB}\ne|\Psi\rangle_{A}\otimes|\Psi\rangle_{B}$. This concept can be applied to systems of more than two modes, in which case one has multipartite entanglement. Entangled states can also be described in density operators notation, which allows to consider the states that are not quantum-mechanically pure.

Perhaps the most common examples of entangled states in optics are so-called Bell states of a polarization-entangled photon pair\footnote{We use a notation where a vertical or horizontal arrow represents one of the two orthogonal linear polarizations, and subscripts $A$ and $B$ one of the two spatial modes. Hence we work in four-dimensional Hilbert space where single-photon base states can be mapped as follows:  $|\updownarrow\rangle_{A}\longrightarrow|1,0,0,0\rangle$, $|\leftrightarrow\rangle_{A}\longrightarrow|0,1,0,0\rangle$, $|\updownarrow\rangle_{B}\longrightarrow|0,0,1,0\rangle$, $|\leftrightarrow\rangle_{B}\longrightarrow|0,0,0,1\rangle$.}
\begin{eqnarray}
|\Psi^{(-)}\rangle_{AB}&=&\left(|\updownarrow\rangle_{A}|\leftrightarrow\rangle_{B}-|\leftrightarrow\rangle_{A}|\updownarrow\rangle_{B}\right)/\sqrt{2},\nonumber\\
|\Psi^{(+)}\rangle_{AB}&=&\left(|\updownarrow\rangle_{A}|\leftrightarrow\rangle_{B}+|\leftrightarrow\rangle_{A}|\updownarrow\rangle_{B}\right)/\sqrt{2},\label{BellStates}\\
|\Phi^{(\pm)}\rangle_{AB}&=&\left(|\updownarrow\rangle_{A}|\updownarrow\rangle_{B}\pm|\leftrightarrow\rangle_{A}|\leftrightarrow\rangle_{B}\right)/\sqrt{2},\nonumber
\end{eqnarray}
and frequency entangled states such as 
\begin{equation}
|\Psi\rangle_{AB}=\int |\omega_A+\nu\rangle_{A}|\omega_B-\nu\rangle_{B}\,F(\nu)d\nu,\label{FreqEnt}
\end{equation}
where $A$ and $B$ designate spatial modes. It is also possible to have a photon pair simultaneously entangled in \textit{both} polarization and frequency (or equivalently, time) \cite{Kwiat98hyper}.

Not only a single pair of photons may be entangled. It is also possible to create an entangled state with larger certain or uncertain photon numbers. One of the examples is vacuum entangled with a Fock state $|\Psi\rangle_{AB}=\left(|N\rangle_{A}|0\rangle_{B}+|0\rangle_{A}|N\rangle_{B}\right)/\sqrt{2}$, dubbed ``NOON-state" \cite{Dowling2008noon,Kwiat98hyper,Afek10noon}. 
Macroscopical states can be entangled not only in photon numbers, but also in the canonical coordinates (quadratures) $p$ and $q$. To distinguish it from the entanglement in the discrete photon numbers, quadrature entanglement is also called continuous-variable entanglement. It can be generated e.g. by combining two squeezed vacuum states on a beam splitter \cite{Zhang07quad_ent,Yoshino07quad_ent}, and forms a foundation for continuous-variable quantum information processing \cite{Andersen10qi} and quantum state teleportation \cite{Bowen03teleport}. The discrete and continuous variable descriptions correspond to expanding the wave function in two different bases. One and the same state can be represented by either one of them. On the practical side: photon number resolving detectors measure in terms of the discrete Fock state basis and homodyne detection measures in terms of quadrature basis. 

Graphic representation of two- or multi-mode non-classical states on a phase diagram is more complicated than for a single mode. In general, it requires as many diagrams as there are modes, with a color coding indicating the quantum-correlated sub-spaces \textit{within} each diagram \cite{Chekhova15sqz_rev}. This appears to allow for a better photon localization  in phase space than is permitted by Heisenberg uncertainty. However the uncertainty principle in not really violated, because the localization occurs in superpositions of quadratures of different field operators that do commute, e.g. $[x_1-x_2, p_1+p_2]=0$, giving rise to Einstein-Podolski-Rosen correlations. 

The quantum state of any one mode of a system comprising many modes and described by a multi-partite state can be found by taking a trace over the unobserved modes. If initially the entire system was in a pure entangled state, the single mode sub-system will be found in a mixed state, as can e.g. be seen starting from equations (\ref{BellStates}) and (\ref{FreqEnt}). This also can be understood following a von Neumann entropy analysis \cite{Cerf97entropy}. Indeed, if a bipartite system is in a pure state with $S=0$, and the conditional entropy is negative $S_{B|A}<0$ because of entanglement, then the entropy of a sub-system is positive, $S_A=S-S_{B|A}>0$, which means that it is in a mixed state. Remarkably, in some cases this does not preclude this mode from being in a non-classical state. For example, the twin beams of an optical parametric oscillator (OPO) that are well-known to be quantum-correlated (or two-mode squeezed), are predicted \cite{Collett85sqz,Reid88sqz,Fabre89opo} and demonstrated \cite{fuerst11sqz} to be also single-mode squeezed when the OPO is well above the threshold. In this case one finds a mixed squeezed state which occupies a larger area in phase space than required by the uncertainty relation.

It should be noted that two-mode quantum correlation and single spatial mode non-classical photon statistics are often viewed as two sides of the same coin. This affinity, emphasized by the use of the term ``two-mode squeezing" in analogy with the ``two-mode entanglement", arises from the simplicity of conversion between these types of photon statistics. The conversion is performed with a linear beamsplitter, and can be elegantly described \cite{Yurke86interf,Campos89BS} by an SU(2) operator converting two input states to two output states. This operation leads to a conversion of phase fluctuations into amplitude fluctuations, and of two-mode entanglement into single-mode squeezing \cite{Afek10noon,Spasibko14BS,Chekhova15sqz_rev}.

A special case of two-mode squeezing is realized when the modes are associated with orthogonal polarizations of the same optical beam. Just like a spatial mode can be associated with any function from an orthogonal set of Helmholtz equation solutions (e.g. Laguerre-Gauss modes), here we are free to chose any polarization basis to designate polarization modes. It is often convenient to chose a linear basis $(x,y)$. In this case polarization Stokes operators are introduced as 
\begin{eqnarray}
\hat S_1&=&a^\dagger_x a_x-a^\dagger_y a_y,\nonumber\\
\hat S_2&=&a^\dagger_x a_y+a^\dagger_y a_x,\label{StokesOps}\\
\hat S_3&=&i(a^\dagger_y a_x-a^\dagger_x a_y).\nonumber
\end{eqnarray}
Like the canonical coordinate or quadrature operators (\ref{pq}), Stokes operators do not commute. They too span a phase space (three-dimensional instead of two-dimensional, since we now have two independent polarization modes) where a pure state occupies the minimum volume allowed by the uncertainty relations. Its shape, however, can be distorted - squeezed. For example, squeezing in the $S_1$ quadrature can be observed as sub shot noise fluctuations of the difference of the currents generated by two photo detectors set to measure optical powers in the $x$ and $y$ linear polarizations.

With increasing the number of modes, which in optics may be associated with the Hilbert space dimension, the list of possible non-classical states rapidly grows. Some examples are the
entangled states of multiple photons in different modes, such as optical GHZ states \cite{Greenberger90GHZ,Pan2000GHZ}, W states \cite{Dur2000W,Eibl04W,Wen10resolution}, as well as cluster \cite{Briegel01cluster} and graph \cite{Hein04graph} states, Smolin states \cite{Smolin01} and others.

In quantum communications, higher-dimensional entanglement provides a higher
information capacity \cite{BP2000qkd,Walborn06qkd,Dixon12qkd,Wasilewski06sqz}. From a fundamental point of view,
higher-dimensional entanglement leads to stronger violations
of generalized Bell's inequalities \cite{Collins02Bell}. This has been experimentally demonstrated in a 16-dimentional Hilbert space spanned by the optical polarization states \cite{Lo16entangl} and in a 12-dimensional Hilbert space spanned by the optical orbital angular momentum (OAM) states \cite{Dada11Bell}.   

Entanglement in the Hilbert space spanned by OAM states \cite{Mair01OAM} is a relatively novel and very promising approach to generating multi-mode entanglement. Two-photon entanglement in 100$\times$100 - dimensional space was demonstrated following this approach \cite{Krenn14OAM}, as was the \textit{four-photon} entanglement \cite{Hiesmayr16OAM}.

Entanglement of a 100, and with certain allowances of even a 1000 optical modes based on polarization, rather than spacial, variables has also been theoretically discussed \cite{Mitchell14_100part} and shown to be within reach with the existing technology. However applying the entanglement metrics \cite{Horodecki09quant} such as  negativity \cite{Vidal02negativity} or concurrence \cite{Wootters98concur,Hildebrand07concur} shows that such states are very close to classical light.

Let us now review some of the practical applications that make non-classical light such an important topic in optics.

\subsection{Applications of non-classical light}\label{sec:aps}

The fact that light can posses non-classical properties that can only be explained in the framework of quantum mechanics is remarkable and important for our understanding of Nature. Besides that, nonclasical light can have useful technological applications.

\paragraph{Absolute calibration of light detectors and sources.} %%%%%%%%%%%%%%%%%%%%%%%%%%%%%%%%%%%%%%%%%%%%%%%%%%%
Perhaps the oldest application of non-classical light, proposed back in 1969-1970 \cite{Zeldovich69cal,Burnham70cal} and further developed by David N. Klyshko in 1980 \cite{Klyshko80calibration}, is the absolute calibration of the quantum efficiency of photon counting detectors. The concept underlying this method is very simple. Suppose a process generating photon pairs, such as spontaneous parametric down conversion (SPDC), produces $N$ signal and idler photon pairs per second. The photons are sent into photon counting detectors with quantum efficiencies $\eta_1$ for the signal channel and $\eta_2$ for the idler channel. Imperfect detection $\eta_{1,2}<1$ leads to \textit{random} loss of photons in both detectors. Then the mean values for the number of photocounts $N_{1,2}$ and for coincidence counts $N_c$ are found as $N_{1,2}=N\eta_{1,2}$ and $N_{c}=N\eta_{1}\eta_2$. Therefore both quantum efficiencies can be inferred by counting the individual and coincidence detections: $\eta_{1,2}=N_{2,1}/N_{c}$. In practical applications one needs to account for multiple pairs occasionally generated in SPDC during a coincidence window, dark noise and dead time of the detectors, and other factors that make the calibration formula and procedure more complicated \cite{Polyakov07cal,Ware07cal}.

A single-detector implementation of this technique was also discussed in \cite{Klyshko80calibration}. This requires a photon number resolving detector collecting all of SPDC light (both the signal and the idler components) near degeneracy. This technique is based on comparing the single- and double-photon detection probabilities. Like the two-detector method, it also received further development \cite{Czitrovszky2000cal,Lebedev08cal}.

Another possibility that was mention in \cite{Klyshko80calibration} is calibration of photo detectors operating in the photo current (continuous) regime instead of photon counting (Geiger) regime. In this case, a correlation function of two photo currents is used instead of the coincidence counting rate \cite{Brida06cal}. Note that since the discrete character of photo detections is no longer required, this method allows for using the two-mode squeezed light instead of two-photon SPDC light \cite{Brida06cal,Vahlbruch16sqz}. A multimode version of this method was used for calibration of CCD cameras   \cite{Brida10cal}.

Similarly to spontaneous emission by excited atoms, SPDC can be viewed as amplification of vacuum unsertainty of the optical field. This vacuum uncertainty is often referred to as \textit{vacuum fluctuations}. But strictly speaking this is a time independent uncertainty which is stochastically projected on a single value when measured. When repeating the process of state preparation and measurement many times, the uncertainty is transformed into an apparent fluctuation. Note that a measurement does not necessarily involve the action of a human experimenter. Coupling the system under study to some environment which then looses coherence (i.e. which decoheres) has the same effect.

The vacuum uncertainty has a spectral brightness\footnote{We recall that spectral brightness, determining the mean number of photons per mode, in free space is measured in terms of light intensity emitted into a unity solid angle per unity frequency bandwidth.} of $S_{vac}=\hbar c \lambda^{-3}$ \cite{KlyshkoPhotonsBook}. Since parametric amplification of weak signals is linear, it is possible to perform absolute calibration of a light source directly in the units of $S_{vac}$ by seeding its light into a parametric amplifier and comparing the emitted parametric signals with and without seeding \cite{Klyshko87bright}. 

\paragraph{Sub shot noise measurements.} %%%%%%%%%%%%%%%%%%%%%%%%%%%%%%%%%%%%%%%%%%%%%%%%%%%

\begin{figure}[t]
\centering
\includegraphics*[width=\textwidth]{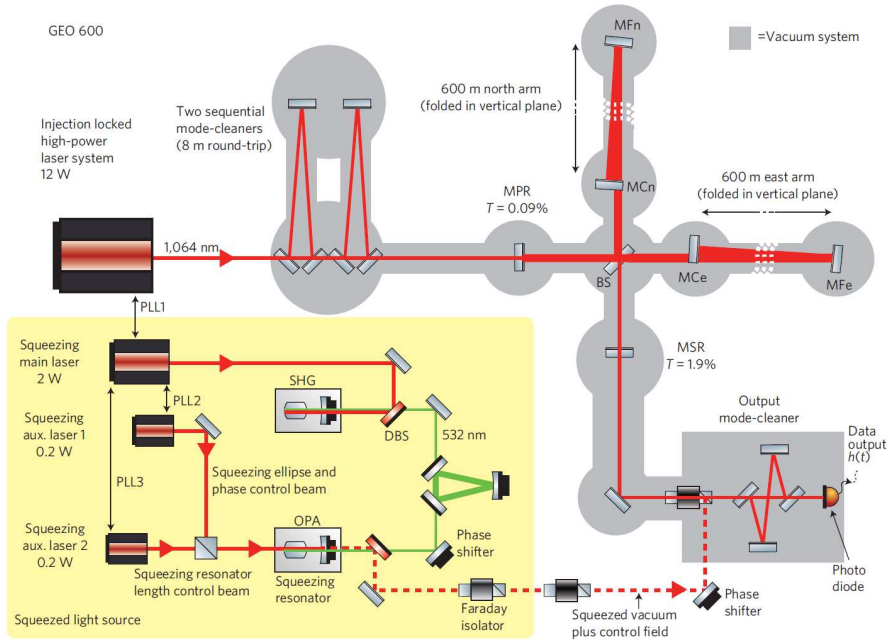}
\caption[]{A complex LIGO interferometer uses a squeezed vacuum input to reduce the measurement noise below the shot-noise limit. Reprinted from  \cite{LIGO11sqz}.}
\label{fig:LIGO}       
\end{figure}

We already noted that the power fluctuations, or noise of non-classical light  may well be reduced below the classical shot noise limit. This effect may be used for low-noise measurements of a variable of interest. First application of squeezed vacuum for sub shot noise interferometric phase measurements has been demonstrated \cite{Xiao87sqz} already in 1987, followed by another publication from a different group\cite{Grangier87sqz}. In these works squeezed vacuum was generated in a degenerate-wavelength OPO pumped below the threshold by the second harmonic of the coherent laser light used in the interferometer. This technique, now commonly used in the field, fixes the frequency and phase relation between the coherent signal and squeezed vacuum. Injecting the squeezed vacuum into a dark port of an interferometer reduces the signal fluctuations below the shot noise by an amount which depends on the degree of squeezing. A reduction figure of 3.5 dB was reached with this approach in the GEO 600 setup of the LIGO project \cite{LIGO11sqz}, see Fig.~\ref{fig:LIGO}. In this case the state-of-the-art 10 dB squeezed vacuum resource was used. However, imperfect transmission of the complex multi-path interferometer $\eta=0.62<1$ increased the observed signal variance (i.e., noise) from the squeezed vacuum source value  $ V^{(0)}_{sqz}=0.1$ to $V_{sqz}=\eta\times V^{(0)}_{sqz}+(1-\eta)\times 1=0.44$. This calculated variance agrees well with the reported 3.5 dB of shot noise suppression.

Besides interferometry, non-classical light can facilitate sub shot noise measurements in spectroscopy \cite{Polzik92sqz,Ribeiro97spectrscopy} and in biological research \cite{Taylor13sqz}. On the other hand, strong intensity fluctuations can enhance the two-photon absorption in atoms and other systems, compared with light of the same average intensity but Poisson or sub-Poisson fluctuations. Theoretical analysis of this phenomenon in two-phoon and squeezed light predicts a linear (rather than quadratic) dependence of the absorption rate on the optical intensity for weak fields, the possibility of a decreasing absorption rate with increasing intensity, and a significant differences between absorption rates for the pase- and amplitude-squeezed beams of the same intensity \cite{Gea-Banacloche89two-phot,Javanainen90two-phot}. Further theoretical analysis including the second harmonic generation is provided in \cite{Dayan07TPA}.

Two-photom absorption of non-classical light has been observed with cesium \cite{Georgiades95twophot} and rubidium \cite{Dayan04TPA} atoms. In both cases atomic two-photon transitions were excited by non-degenerate squeezed light generated in an OPO cavity. Excitation rate scaling as the power 1.3 (instead of 2) of the light intensity was observed in \cite{Georgiades95twophot}. Conversely, it is possible to characterize photon bunching by observing two-photon response in semiconductors \cite{Boitier09two-phot}.

Speaking of spectroscopy, we must mention yet another application of non-classical light, not related to noise reduction but remarkable nonetheless. In this application strongly non-degenerate  SPDC light propagates in a nonlinear interferometer filled with a sample of refractive material \cite{Korystov2001hooks}. As expected, a strong dispersion in e.g. infrared range is indicated by the characteristic distortion patterns of interference fringes in the infrared (idler) port. However it also leads to similar distortions arising in the signal port, which allows for performing infrared spectroscopy using visible light optics and detectors. 

\paragraph{High-resolution imaging.} %%%%%%%%%%%%%%%%%%%%%%%%%%%%%%%%%%%%%%%%%%%
The term ``imaging" may refer to both creating and reading of patterns, as well as to optical detection of small displacements. All these functionalities have been shown to benefit from applications of non-classical light. Creating lithographic images with higher than diffraction-limited resolution has been proposed in year 2000 \cite{Boto00litho}. This proposal is based on using photo-polymers sensitive to $N$-photon absorption in conjunction with already mentioned entangled NOON states. It was theoretically shown that using these states in a Mach-Zehnder interferometer can generate $\lambda/(2N)$-spaced fringes of the $N$th order intensity distribution $\langle I^N\rangle$ that would imprint in the polymer. It should be noted that even with classical light the $N$-photon material response by itself provides a $\sqrt{N}$ reduction of the optical point-spread function. With special modulation techniques this reduction factor can be further pushed to reach the quantum limit of $N$ \cite{Peer04litho}. Therefore the practical benefit of the quantum lithography proposal turned out to be limited. However its originality and intellectual value have stimulated a number of follow-up works. Particularly for $N=2$, it was theoretically proven that not only faint two-photon light, but also stronger two-mode squeezed light can be used for this purpose \cite{Nagasako01opa}. On the experimental side, we would like to acknowledge the success in driving coherent \cite{Dayan05pdc-sfg} and incoherent \cite{Dayan04TPA} two-photon processes with SPDC light.

Discerning the objects' features with resolution exceeding the Rayleigh diffraction limit is possible in setups similar to two-photon \textit{Ghost imaging} setup \cite{Pittman95Qimaging} but relying on multi-photon entangled states such as GHZ or W states \cite{Wen10resolution}. Alternatively, axial resolution  can be enhanced by a factor of two realizing a quantum version of optical coherence tomography measurement with two-photon light \cite{Nasr03QOCT}. In this case one makes use of the signal-idler intensity correlation time being much shorter that their individual coherence times. 

\begin{figure}[htb]
\centering
\includegraphics*[width=0.8\textwidth]{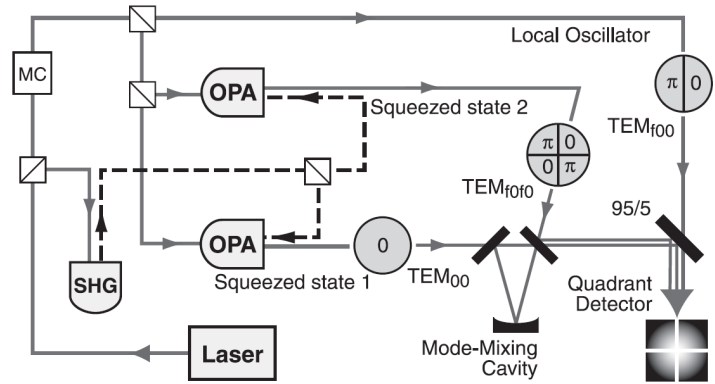}
\caption[]{A superposition of squeezed light and coherent local oscillator with segmented phase shifts enables the beam displacement measurement with precision exceeding the standard quantum limit. SHG is second
harmonic generator, OPA is optical parametric amplifier, MC is mode cleaner, 95/5 is a beamsplitter with 95\% reflectivity. ${\rm TEM_{f0f0}}$ and ${\rm TEM_{f00}}$ designate a formerly ${\rm TEM_{00}}$ mode modified by split phase plates. Dashed
lines show 532 nm light; solid lines show 1064 nm light. Reprinted from  \cite{Treps03pointer}.}
\label{fig:point}       
\end{figure} 

The resolution of small lateral displacement measurements is limited by the shot noise to the value 
\begin{equation}
d_0\approx \sqrt{\frac{\pi}{8N}}\,w_0,\label{d0}
\end{equation}
where $w_0$ is the Gaussian width of a ${\rm TEM_{00}}$ probe beam focused onto a split-field detector, and $N$ is the number of detected photons. It has been shown \cite{Treps03pointer} that by composing the probe beam out of coherent and squeezed optical beams as shown in Fig.~\ref{fig:point}, the shot-noise resolution limit (\ref{d0}) can be improved by approximately a factor of two. 

\paragraph{Quantum information processing.} %%%%%%%%%%%%%%%%%%%%%%%%%%%%%%%%%%%%%%%%%%%
The concept of quantum information processing, or quantum computing, was conceived in 1982 by Richard Feynman \cite{Feynman82qc}. At the heart of this concept is a notion that a quantum superposition principle can be utilized to implement a large number of computations in parallel. To implement such quantum parallelism, logic operations of a quantum computer must be performed by quantum systems. 
It should be noted, however, that in order to access the results of this parallel computation one has to perform a measurement which is equivalent to a projection onto just one result. Therefore, one benefits from this parallelism only if the single measurement already provides an advantage, such as in the Shor algorithm \cite{Shor97} where a quantum interference phenomenon is utilized to find prime factors of a large number faster that it is possible by the classical search. Note that the \textit{classical} optical interference can be used in a similar way \cite{Clauser96factor}.

Instead of encoding information in bits that take on binary values 0 or 1, these systems encode it in \textit{qubits}, allowing any superposition of the binary values.
A qubit may be implemented in various two-level physical systems, such as an atom, ion, spin-1/2 particle, and many others. To distinguish such systems from photons, we will call them \textit{massive}. Polarization of a photon, as well as its localization in two spatial or frequency modes, also can be used as a qubit. 
The advantage of optical qubits over massive ones is slow decoherence of the former: photons hardly interact with ambient electromagnetic or gravity fields. 

This advantage however turns into a disadvantage when it comes to implementation of quantum logical operations that require photon-photon interaction. 
Such interaction can be facilitated using optical nonlinearity at the single photon level. Several approaches to building quantum gates based on nonlinear response of optical media have been theoretically discussed. One of these approaches is the Quantum Zeno Blocade which can be realized based on two-photon absorption \cite{Franson04QZB}, electromagnetically induced transparency \cite{Franson07WGM_QZB,Clader13switch}, or on the second-order polarizability of optical nonlinear crystals enhanced by high-$Q$ cavities \cite{Huang10QZB,Huang10switch_prop,Sun13switch}.
Several experimental demonstrations of these techniques have been performed with multi-photon (typically, weak coherent) states \cite{Hendrickson13switch,Strekalov14switch}, however functional photonic quantum gates so far remain beyond the reach. 

This difficulty has lead to the concept of quantum network \cite{Kimble08internet}, where transmission of information is performed by photonic qubits, while its processing is performed by massive qubits. Various types of massive qubits have been successfully coupled to single photons, including atoms \cite{Aoki09switch,Specht11qmem,Ourjoumtsev11atom,Chen13switch,Baur14switch,Shomroni14switch,Tiecke14switch,Rosenblum15single} and quantum dots \cite{Michler2000qdot1}. Nitrogen vacancy centers in diamonds have been also proposed for this application \cite{Li11nv,Chen12Qgate}.

Building a quantum network requires non-classical light sources whose central wavelength and optical bandwidth are compatible with the massive qubits. In the most straightforward way this can be achieved by using the same atomic transition for the generation of non-classical light (see discussion in section \ref{sec:atoms}), and then for transferring quantum information to atomic qubits \cite{Volz06atom,Beugnon06atom,Maunz07atom,Leong15HOM}. Alternatively, narrow-line parametric light sources discussed in section \ref{sec:spdc} can be used. Note that while generating narrow-band squeezed light or squeezed vacuum is relatively easy by opertating an OPO source above the threshold, generation of equally narrow-band photon pairs below the threshold is more difficult, as it requires tunable resonators with very high $Q$-factor \cite{Bao08pairs,Fekete13src,Schunk15CsRb,Lenhard15,Schunk16CsRb}. For many quantum information applications such sources also need to be strictly single-mode, which has been recently achieved using whispering gallery mode (WGM) \cite{Fortsch15sm} and waveguide \cite{Luo15sm} resonators.
 
Using massive qubits often requires low temperatures, very low pressure vacuum, thorough shielding of ambient fields, and entails other serious technical complications. The concept of \textit{linear} quantum computing \cite{Knill01linearqc} strives to avoid these complications. There are no massive qubits in a linear quantum computer, but there are also no photon-photon interactions. This interaction is replaced by a measurement process followed by feed-forward to or post-selection of the remaining photons. This procedure is certainly nonlinear (and even non-unitary), and can be used to implement quantum logic operations over a sub-space of a larger Hilbert space. 

In higher-dimensional Hilbert spaces photonic \textit{qutrits} \cite{Bogdanov04qutrit,Lanyon08qutrit} and even \textit{ququarts} \cite{Bogdanov06ququart,Luo15ququart} can be introduced as useful notions. As an example, a photon qutrit encoded in polarization has three basis states: $|\updownarrow\updownarrow\rangle$, $(|\leftrightarrow\updownarrow\rangle+|\updownarrow\leftrightarrow\rangle)/\sqrt{2}$, and $|\leftrightarrow\leftrightarrow\rangle$. A ququart basis consists of four states and can be easily envision if we further lift the frequency degeneracy, or couple the photon pair into different spatial modes. Usually these states are discussed in the context of quantum secure communications using alphabets with higher than binary basis. 

Transmission of information by photonic qubits presents sufficient interest by itself, besides being a quantum computer building block. The fundamental property of a qubit is that it cannot be cloned, or duplicated. Such cloning would be incompatible with the linearity of quantum mechanics \cite{Wootters82noclone}. Therefore, the information encoded in qubits can be read only once; in other words, it cannot be covertly intercepted. This property of qubits served as a foundation for the original quantum key distribution (QKD) protocol BB84 \cite{BB84}, and for numerous and diverse QKD protocols that emerged later. QKD is the least demanding application of non-classical light reviewed in this chapter, and the only quantum optics application known to us that has been relatively broadly commercialized to-date. Discrete variables QKD  can be successfully implemented even with weak coherent light, e.g. strongly attenuated laser pulses, which adequately approximate single-photon states. Similarly, non-orthogonal coherent states of light can be successfully used in continuous variables QKD \cite{Braunstein05rev}.

Coherent states are pure quantum states unlike thermal states and thus qualify as non-classical states (see Appendix for discussion). For some quantum protocols coherent states suffice, for others they do not. Furthermore, it is often argued that much of their properties can be described by classical models. For all these reasons we concentrate the discussion on states which are more non-classical than coherent states.

Some proposed quantum information protocols relying on non-classical light fall between the QKD and quantum computing in terms of architecture and complexity. One of such protocols is the \textit{quantum commitment}. It is designed to allow Alice to fix (``commit") an observable value in such a way that Bob cannot learn this value until Alice reveals it. Alice, on the other hand, cannot change her commitment after it has been made. Originally proposed in 1988 \cite{Crepeau88}, this protocol has been experimentally demonstrated \cite{Lunghi13commit} in 2013 with an added benefit of closing a loophole present in the original proposal. Other protocols proposed for implementing quantum secret sharing among multiple parties \cite{Braunstein05rev} and quantum digital signatures \cite{Croal16qsign} may be used in the context of quantum money, quantum voting, and other visionary applications. In the following section we review the sources of non-classical light, which is the main objective of this chapter.

\section{Sources of non-classical light}\label{sec:src}

\subsection{Atoms real and artificial}\label{sec:atoms}

\paragraph{Atoms}%%%%%%%%%%%%%%%%%%%%%%

The early interest in non-classical, and in particular entangled, optical states was stimulated by the quest for experimental violations of Bell's inequalities. The first successful and statistically reliable violation was reported in 1972 by Freedman and Clauser \cite{Freedman72Bell}. They used a cascade two-photon transition in calcium beam producing a polarization-entangled pair of blue and green photons, and performed a polarization-based Bell measurement which has shown a six standard deviations violation. Therefore the conceptually more advanced two-photon entanglement was observed with atomic sources prior to a more straightforward antibunching effect.

Photon antibunching in resonance fluorescence from a coherently driven two-level atom is easy to understand. Once the atom emits a photon, it occupies the
ground state and cannot emit another photon for a period of time of the order of the
excited state lifetime (in the weak excitation regime), until the interaction dynamics drives the atom back to the excited state. Hence the Poissonian statistics of the coherent pump photons is converted to sub-Poissonian statistics of fluorescence photons, leading to a state whose photon-number fluctuation is reduced below the shot noise limit typical for coherent light, such as shown in Fig.~\ref{fig:pq}(f).

Antibunching in resonance atomic fluorescence was predicted back in 1976 by Carmichael and Walls and observed in 1977-78 by two different research groups using beams of sodium atoms, see the review \cite{Walls79review} for details. More recently, four-wave mixing in a rubidium vapor cell was used to produce and characterize heralded Fock-basis qubits $\alpha|0\rangle+\beta|1\rangle$ \cite{Brannan14qbit}.

A sodium atomic beam passing through an optical cavity was also used for the first demonstration of squeezed light in 1985 \cite{Slusher85sqz}. Soon after that the first magneto-optical traps were implemented. They allowed to suppress the thermal motion of atoms and - associated with it - the dephasing, which increased the observed squeezing form 0.3 dB \cite{Slusher85sqz} to 2.2 dB \cite{Lambrecht96sqz}. 
Even stronger was the two-mode squeezing observed by seeding one \cite{McCormick07sqz} or both \cite{Corzo11sqz} of these modes with weak coherent light. In these experiments the squeezing was measured to be 3.5 dB (8.1 dB corrected for losses) and 3 dB  (over 3.5 dB corrected for losses), respectively.  This technique has a potential for tailoring the spatial structure of multimode non-classical light, e.g. generating twin beams carrying orbital angular momentum \cite{Boyer08twin}.   

A single pump laser was used in experiments \cite{Walls79review,Slusher85sqz,Lambrecht96sqz,McCormick07sqz,Corzo11sqz,Boyer08twin}. To suppress the effect of thermal motion, the four-wave mixing process can be driven by two \textit{different}, conterpropagating,  laser beams in a configuration typical for saturation absorption spectroscopy. This technique has allowed for generation of very high flux of photon pairs with controllable waveform, see \cite{Balic05atom} and references therein. Such pairs can be used for heralded preparation of nearly single-photon pulses. Moreover, the gound-state coherence in cold atomic ensembles is sufficiently long-lived to allow the ``read" laser pulse to arrive with a substantial delay after the ``write" pulse, which allows one to control the delay between the emitted heralding and the heralded photons \cite{Chou04heralded,Polyakov04heralded,Eisaman04shaping}. A controlled delay is in fact just a special case of temporal shaping of the biphoton correlation function, which can be achieved with the ``read" pulse profile manipulation \cite{Chen10atom}.

\begin{figure}[t]
\centering
\includegraphics*[width=0.9\textwidth]{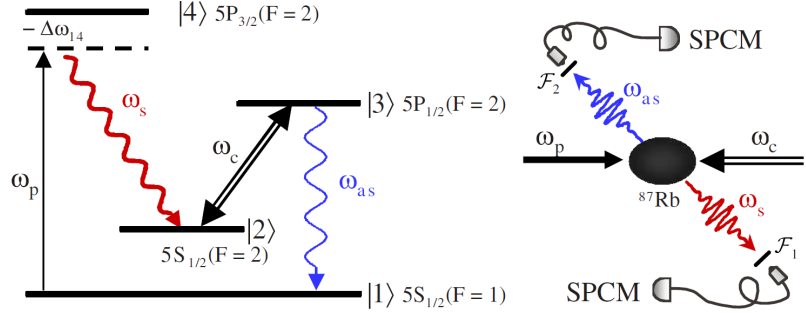}
\caption[]{A double-$\Lambda$ configuration of $^{87}$Rb transitions involved in the four-wave mixing process generating non-classical light and the experimental diagram. 
Reprinted from \cite{Balic05atom}.}
\label{fig:levels}       
\end{figure} 

Quantum optics researchers favored alkali atomic gases because of their strong resonant Kerr response. A typical energy diagram of this process, called a double-$\Lambda$ configuration, is shown in Fig. ~\ref{fig:levels}. This diagram is drawn specifically for $^{87}$Rb D1 and D2 manifolds, but its analogues can be realized in various atomic species. Strong pump and control optical fields have frequencies $\omega_p$ and $\omega_c$, corresponding to D2 and D1 transition wavelengths, respectively. Generated quantum (two mode squeezed) light has the Stokes and anti-Stokes  frequencies $\omega_s$ and $\omega_{as}$, respectively. The energy and momentum conservation requires $\omega_p+\omega_c \approx \omega_s +\omega_{as}$ and $\vec{k}_p+\vec{k}_c \approx \vec{k}_s +\vec{k}_{as}$, where the approximations arise from neglecting the momentum recoil and kinetic energy that maybe carried away by the atom. Note that the momentum conservation allows for a very broad angular spectrum of the emitted light in the case of counter-propagating ($\vec{k}_p+\vec{k}_c \approx 0$) beams.

Another important feature of atomic Kerr media is that its response may be sensitive to light polarization. This can lead to nonlinear phenomena such as polarization self-rotation (see \cite{Matsko02sqz} and references therein), where one polarization is amplified while the orthogonal polarization is deamplified. If the input light is polarized linearly \cite{Matsko02sqz,Barreiro11sqz} or circularly \cite{Ries03sqz}, the vacuum field in the orthogonal polarization becomes squeezed. 

Coupling atomic media with optical cavities opens up the field of cavity quantum electrodynamics (cQED), rich with non-classical phenomena. Even a single atom strongly interacting with an optical mode can generate squeezed light \cite{Ourjoumtsev11atom}. It can also
be used to implement a photonic blockade \cite{Birnbaum05blockade}, leading to a photon turnstile capable of generating single photons on demand \cite{Dayan08turnstile,Aoki09switch}. In terms of quantum systems engineering, this can be considered as a next step after delayed heralded single-photon generation, and two steps after single photons generated at random times. 
A real or artificial atom strongly coupled to a cavity mode is also predicted to be capable of generating the ``N-photon bundles" \cite{Munoz14bundle}, arguably equivalent to flying Fock states \cite{Strekalov14bundle}.
 
Once generated, the non-classical states need to be routed in a decoherence-free manner towards the information-processing nodes or to detectors. The single photon routing controlled by other single-photon states would enable quantum logic operations on photons, and make an optical quantum computer possible. Serious efforts have been made in this direction. An optical transistor was reported \cite{Chen13switch}, in which a single control photon induced a ground-state coherence in a cold Cesium cloud, affecting the transmission of a dealyed probe pulse. In a more recent work \cite{Shomroni14switch}, a
single-photon switch based on a single Rubidium atom interacting with the evanescent field of a fused silica microsphere resonator was demonstrated. This system was shown capable of switching from a high reflection (65\%) to a high transmission (90\%) state triggered by as few as three control photons on average (1.5 photons, if correction for linear losses is made).  

Finally, let us point out that single molecules can be similarly used as sources of single photons, as demonstrated by a significant measured anti-bunching \cite{Basche92antibunch,Brunel99src}. Molecular sources can operate at room temperature in the on-demand mode \cite{Lounis2000single}.

\paragraph{Artificial atoms}%%%%%%%%%%%%%%%%%%%%%% 

Discrete level spectra are available not only in atoms but also in solid-state nanosystems, such as quantum dots, carbon nanotubes, nitrogen vacancy (NV) centers in diamond, or impurities in semiconductors. Because of this property such systems are often referred to as ``artificial atoms".  They too have been actively utilized as sources of non-classical light. The physical mechanism regulating the photons statistics of an artificial atom emission is very similar to that of real atoms. 

While an optical photon absorption by an atom causes an electron transition from the ground to an excited state, in quantum dot it causes generation of an electron-hole pair, called an exciton. The recombination of this exciton is responsible for the resonance fluorescence of a quantum dot. Applications of this process for single-photon sources are reviewed in \cite{Lounis05src,Buckley12qd_rev}. Such sources often require liquid helium cooling, although the first demonstration of non-classical light emitted from a quantum dot was done in year 2000 by Michler \textit{et al.} at room temperature \cite{Michler2000qdot}. In this experiment a single CdSe/ZnS quantum dot was driven by a resonant constant wave (CW) pump laser. Its fluorescence had sub-Poissonian photon-number distribution with $g^{(2)}(0)=0.47\pm 0.02$. More recent quantum dot based sources \cite{Bounouar12qd,Holmes14qd} also can operate at room temperature exhibiting non-classical anti-bunched photon statistics in pulsed regime, although their anti-bunching is significantly stronger at liquid helium temperatures.

In carbon nanotubes, two-photon generation is suppressed due to Auger processes and excitons localization. Antibunching of the light emitted by such systems can be very strong, reaching the value of $g^{(2)}(0)= 0.03$ at 4.2 K \cite{Hogele08nanotube}.  

In contrast with quantum dots and carbon nanotubes, NV centers in diamond provide the most
stable quantum emitters at room temperature. In \cite{Schietinger08nv}, a CW emission from a single NV center in a diamond nanocrystal was coupled to a 4.84 $\mu$m in diameter polystyrene microspherical resonator. The non-classical character of the single quantum emitter was verified by measuring $g^{(2)}(0)\approx 0.3$, while the coupling to the WGMs was evident from a discrete spectrum of the emission.
NV-center based pulsed single-photon sources also operate at room temperature reaching nearly the same anti-bunching  figure \cite{Babinec10diamond}.

The power fluctuation measurements carried out in \cite{Michler2000qdot,Bounouar12qd,Holmes14qd,Babinec10diamond} allowed to probe the Wigner function only in the radial direction (c.f. Fig.~\ref{fig:pq}). A more advanced measurement also providing the access to the orthogonal quadratures was carried out by Schulte \textit{et al.}, who used a local oscillator with variable phase in a heterodyne setup \cite{Schulte15sqz}. They also studied the amount of squeezing as function of the excitation power.

\begin{figure}[b]
\centering
\includegraphics*[width=\textwidth]{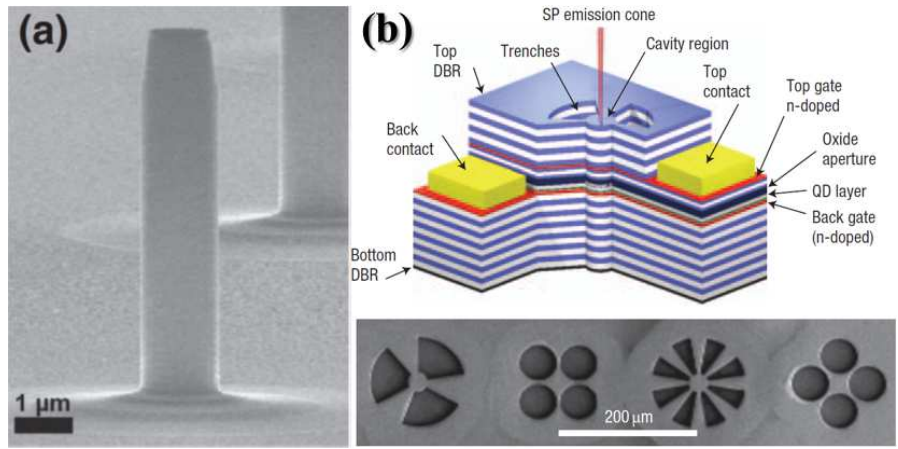}
\caption[]{A micro-pillar optical resonator (a) has Bragg mirrors at the base and on the top, providing strong coupling of quantum dots (embedded near its center) to a vertical mode. In a different design (b) the resonator is formed by cutting trenches of various shapes (shown in the bottom) in a layered structure. Reprinted from \cite{Press07qdot} and \cite{Strauf07single}.}
\label{fig:pillar}       
\end{figure} 

Just as with real atoms, coupling quantum dots to microcavities provides access to the benefits of cQED. One of these benefits is the improved collection efficiency. Because of the high Purcell factor of the microcavities, a quantum dot fluorescence is preferentially radiated into the cavity modes and can be conveniently collected. Press \textit{et al.} \cite{Press07qdot} have been using micro-pillar structures for this purpose. A micro-pillar resonator shown in Fig.~\ref{fig:pillar}(a) measures about a micron in diameter and five micron tall. It is complete with Bragg mirrors at both ends, each consisting of approximately 30 pairs of AlAs/GaAs layers. A layer of InGaAs quantum dots is grown in the central anti-node of the cavity. The structure is cooled to 6-40 K and pumped by a pulsed mode-locked laser. Photons collected from the cavity were antibunched with  $g^{(2)}(0)\approx 0.18$. A different design shown in Fig.~\ref{fig:pillar}(b) uses a layered structure where a pillar cavity is defined by cutting trenches of various shapes \cite{Strauf07single}. This shape allows one to control the polarization dispersion of the resonator and to generate single photons with a desired polarization. Quantum dots have been coupled not only to pillar or planar cavities, but also to WGM resonators. For example, strong coupling regime was achieved with a single GaAs \cite{Peter05qdot} and InAs \cite{Michler2000qdot1,Srinivasan07qdot} quantum dots. 

Instead of a  cavity, a quantum dot can be coupled to a single-mode on-chip waveguide \cite{Makhonin14qdot}. This approach not only allows to generate strongly non-classical ($g^{(2)}(0)< 0.1$) light, but also leverages scalable on-chip photonic technology. Operating these systems in pulsed mode gives them a much desired ``single photon on demand" quality.

Quantum dots can support not only single excitons, but also biexitons. Recombination of a biexciton leads to emission of a photon pair, similarly to a two-photon emission from an atom in a Freedman and Clauser experiment. This process can go through two different intermediate states, realizing two quantum-mechanical paths for biexitonic recombination.
In experiments \cite{Akopian06qdot,Stevenson06source,Kuroda13qdot,Jayakumar13qdot} the photon pair emitted along one path is polarized vertically; along the other, horizontally. Thus recombination of such a biexciton creates an optical Bell state $|\Phi^{(\pm)}\rangle$ introduced in (\ref{BellStates}), provided that the polarization terms are not ``tagged" by either the final (ground) state of the quantum dot, or the optical wavelength. Then violation of Bell’s inequality is possible, and in fact has been observed with a confidence of five standard deviations \cite{Kuroda13qdot}. 

The biexciton recombination process is broadband enough to provide a significant wavelength overlap even if perfect wavelength degeneracy cannot be achieved. This allows one to erase the wavelength distinguishablility by spectral filtering and achieve a polarization-entangled state capable of violating  Bell’s inequality by more than three standard deviations \cite{Akopian06qdot}. Similar mechanisms can lead to polarization-entangled photon pairs emission from impurities in semiconductor \cite{Dotti15qe} and from the hybrid piezoelectric-semiconductor quantum dot systems \cite{Trotta14qdot}. The latter system has been also used to demonstrate Bell’s inequality violation.

Generation of entangled photon pairs by quantum dots is unique in that the pairs themselves have sub-Poissonian statistics, which allows to generate single photon pairs using a pulsed pump. This aspect of the quantum dot entangled light sources was highlighted by Young \textit{et al.} \cite{Young06qdot}, who demonstrated the triggered emission of polarization-entangled photon pairs from the biexciton cascade of a single InAs quantum dot embedded in a GaAs/AlAs planar microcavity. They also showed that quantum dot engineering can reduce the energy gap between the intermediate states, minimizing or removing the need for thorough spectral filtering. Deterministically exciting biexcitons by optical $\pi$-pulse, M{\"u}ller \textit{et al.}  \cite{Muller14qdot} have demonstrated a true ``polarization entangled photon pair on demand" operation with unprecedented anibunching parameter $g^{(2)}(0)< 0.004$ and high entanglement fidelity $(0.81\pm 0.02)$.

\subsection{Parametric down conversion}\label{sec:spdc}

Spontaneous parametric down conversion (SPDC), optical parametric amplification (OPA) and oscillation (OPO) are among the most important sources of non-classical light. All these closely related processes are enabled by the second-order nonlinear response of non-centrosymmetric optical crystals, characterized by quadratic susceptibility $\chi^{(2)}$. This process, originally called parametric scattering or parametric fluorescence, was first observed in 1965 \cite{Giordmaine65spdc} and widely studied later. From the quantum point of view, i.e. in terms of photon pair emission, this process was first discussed in 1969 by Klyshko \cite{Klyshko69PDC}. One year later, the ``simultaneity" of these photons (called the \textit{signal} and \textit{idler}) was observed by Burnham and Weinberg \cite{Burnham70cal}. 

We now know that the reported ``simultaneity" reflected the resolution of the time measurements rather than the physical nature of the biphoton wavefunction. The signal-idler correlation  time is finite, and is closely related to their optical spectra and the group velocity dispersion (GVD) of the parametric nonlinear crystal \cite{strekalov05g1g2}. The temporal correlation function can take on different forms for different types of phase matching \cite{Rubin94TypeII,Dauler99subfs,Valencia02disp,strekalov05g1g2JMO}, with the width ranging over six orders of magnitude: from 14 femtoseconds for free-space SPDC in a very short crystal \cite{Dauler99subfs} to 10-30 nanoseconds for SPDC in a high-finesse optical resonators \cite{Bao08pairs,Fekete13src,Schunk15CsRb,Lenhard15,Scholz09narrow,Chuu12bright,Fortsch13NC,Fortsch15sm}. Shaping this correlation function is an important problem in quantum communications. With SPDC lacking the photon-storage capability available to the atomic sources, this problem is quite challenging. One possible approach is by interferometric tailoring of the SPDC spectra using two or possibly more coherently pumped crystals \cite{Burlkaov97bigpaper,Korystov2001hooks,Iskhakov16bright}. Another approach is based on using a dispersive media \cite{Valencia02disp}. There is also a proposal for leveraging the \textit{temporal ghost imaging} \cite{Setala10TGI}, which is similar to spatial ghost imaging\cite{Pittman95Qimaging} but relies on temporal rather than spatial masks (implemented e.g. by electro-optical modulators) \cite{Sych16TGIxxx,Averchenko16TGIxxx}.

Parametric down conversion has been described in great detail in many books and papers, which spares us the necessity to reproduce all the analysis and derivations here. Let us just list the most fundamental facts.
The energy and momentum conservation for the pump, signal and idler photons impose the phase matching conditions  
\begin{eqnarray}
\omega_p&=&\omega_s +\omega_i,\label{pmom}\\
\vec{k}_p&=&\vec{k}_s +\vec{k}_{i},\label{pmk}
\end{eqnarray}
where the frequencies are related to the wave numbers by the dispersion relations $\omega=ckn(\lambda)$. It is the combination of these three constraints that is responsible for the free-space SPDC light appearing as a set of colorful rings. In most materials, normal chromatic dispersion of the refractive index $n(\lambda)$ prohibits parametric phase matching by making (\ref{pmom}) and (\ref{pmk}) incompatible. However it can be compensated by polarization dispersion in birefringent materials. For example, the pump polarization can be made orthogonal to both signal and idler polarizations, which is known as Type-I PDC configuration. Alternatively, either signal or idler polarization can be parallel to that of the pump in Type-II PDC. Type-0 PDC, when all three fields are polarized in the same plane, can be attained by using various periodical poling techniques which modifies (\ref{pmk}) by adding or subtracting a multiple of the poling structure wave vector $\vec{e}\,2\pi/\Lambda$, where $\Lambda$ is the poling period and $\vec{e}$ is its direction.

A pair of coupled signal and idler modes with photon annihilation operators $a_s$ and $a_i$, respectively, is governed by the evolution operator 
\begin{equation}
\hat{U}(t)=\exp\left\{-\frac{i}{\hbar}\int_0^t\hat{H}_{int}(t)dt\right\},\quad\hat{H}_{int}(t)=i\hbar g(t)(a^\dagger_sa^\dagger_i-a_sa_i). \label{Uparam}
\end{equation} 
This is an approximation assuming that the pump field can be treated classically, i.e. that one can neglect the annihilation of one pump photon for every creation of a signal/idler photon pair. The function $g(t)$ in (\ref{Uparam}) describes parametric interaction:
\begin{equation}
g= 2\pi\sigma(t)(\chi^{(2)}\vdots \vec{e}_p\vec{e}_s\vec{e}_i)\frac{\sqrt{\omega_s\omega_i}}{n_sn_i}, \label{Omega}
\end{equation}
where $(\chi^{(2)}\vdots \vec{e}_p\vec{e}_s\vec{e}_i)$ is the scalar product of the nonlinear susceptibility tensor with the interacting fields unit vectors. The overlap integral 
\begin{equation}
\sigma(t)=\int\psi_s(\vec{r})\psi_i(\vec{r})E_p^*(\vec{r},t) dV \label{ovlp}
\end{equation} 
is calculated for the normalized modes eigenfunctions $||\psi(\vec{r})_{s,i}||=1$ and the pump field envelope $E_p(\vec{r},t)$ . This integral enforces the momentum conservation (\ref{pmk}) for the plane-wave modes.

The time integral $G\equiv\int_0^{T}g(t)\,dt$ is called the parametric gain. Here the interaction time $T$ is determined by the crystal length. The effective interaction length however can be shorter than the crystal length for short pump pulses, when significant longitudinal walk-off between the pump and parametric pulses occurs due to the GVD. Note that depending on the pump phase $G$ may take on negative values, leading to de-amplification. 

\paragraph{Spontaneous parametric down conversion}%%%%%%%%%%%%%%%%%%%%%%

Vacuum-seeded parametric down conversion, or SPDC, is probably the most widely used nonlcassical light source made famous by Bell's inequality violations, early QKD demonstrations, quantum teleportation, and a number of other remarkable achievements PDC has made possible. This process has been realized in low and high gain regimes, in free space, single transverse mode waveguides, and in optical resonators. In the low-gain regime, this process is adequately described by expanding the evolution operator $\hat{U}$ from (\ref{Uparam}) into a power series. The leading term of the expansion represents a two-mode vacuum, next term is a signal-idler photon pair, the third term represents two such pairs, and so on. The amplitudes of these terms form the same power series as for a thermally populated mode \cite{LoudonBook,KlyshkoPhotonsBook,Tapster98SPDCstat}, which determines the peak value of the Glauber correlation function for a weakly populated SPDC mode: $g^{(2)}(0)= 2$. This also allows one to introduce the \textit{effective temperature} for SPDC emission \cite{KlyshkoPhotonsBook}. 

Free-space SPDC provides for a multimode source of spatially-entangled biphotons, which can be used in two-photon imaging discussed in section \ref{sec:aps}. This type of entanglement arises from the momentum conservation (\ref{pmk}). Indeed, even with strictly constrained (e.g., by band-pass filters) optical wavelengths, there are many indistinguishable ways the transverse momentum conservation $\vec{k}^\perp_s+\vec{k}^\perp_i=0$ can be achieved. On the other hand, selecting a single transverse signal mode, and the paired with it idler mode, one obtains a frequency-entangled state (\ref{FreqEnt}).

Type-II SPDC offers an interesting configuration \cite{Kwiat99src} wherein the same pair of spatial modes $A$ and $B$ can be populated by orthogonally polarized signal and idler in both possible permutations, leading to a polarization-entangled state such as $|\Psi^{(+)}\rangle_{AB}$ in (\ref{BellStates}). A closer look shows that this state is also frequency-entangled as in  (\ref{FreqEnt}). Such states that are entangled in more than one degree of freedom at once are called hyperentangled \cite{Barreiro05hyper}.

Polarization entangled photon pairs can also be generated in Type-I SPDC, in a clever configuration of two crystals whose optical axis planes are perpendicular to each other \cite{Kwiat99src}. This configuration provides even more flexibility than the polarization entanglement generation in a Type-II crystal: by varying the phase between the pump field projection on the two crystals' axes (e.g., varying the pump polarization ellipticity), as well as manipulating the polarization and phase of the signal and idler photons between the crystals, one can generate any polarization-entangled state in Hilbert space spanned by the Bell-states basis (\ref{BellStates}), as well as some of mixed states \cite{Kwiat99src}.

Parameters of SPDC biphoton sources such as their wavelengths, bandwidth and pair production rate may vary considerably. Because of accidental generation of multiple photon pairs, ultra high pair rate associated with large $G$ is not always desirable. It is often more important to minimize the chance of accidentally generating a second pair during the measurement. In the limit of very fast measurements it is also important to generate sufficiently few (much less than one on average) photons per \textit{coherence time}, i.e. per longitudinal mode. If this number exceeds unity, then $G>1$ as well, and the power series expansion of the evolution operator $\hat{U}$ does not converge. This means that the already generated parametric photons make a stronger contribution to the further PDC process than the vacuum photons, i.e. we enter the regime of parametric super luminescence. This is accompanied by a transition from thermal (Gaussian) photon number statistics to Poissonian statistics, typical for laser light.

However the parametric light remains non-classical even in the high-gain regime. When the signal and idler are distinguishable, the light is two-mode squeezed, which can be established by measuring the photocurrents difference in the signal and the idler detectors and finding it below the shot noise level. 
When the signal and idler are indistinguishable, we have the squeezed vacuum state such as shown in Fig.~\ref{fig:pq}(d), whose photon-number basis expansion consists of only even terms and $\langle p\rangle=\langle q\rangle=0$. Let us recall that if the signal and idler have the same frequency and the distinguishability is only based on polarization or emission direction, a conversion between two-mode squeezing and squeezed vacuum is trivially accomplished with a polarizing or non-polarizing beam splitter, respectively. In these cases the terms ``two-mode squeezing", ``squeezed vacuum" and even ``two-mode squeezed vacuum" are often used interchangeably. 

\begin{figure}[b]
\centering
\includegraphics*[width=0.5\textwidth]{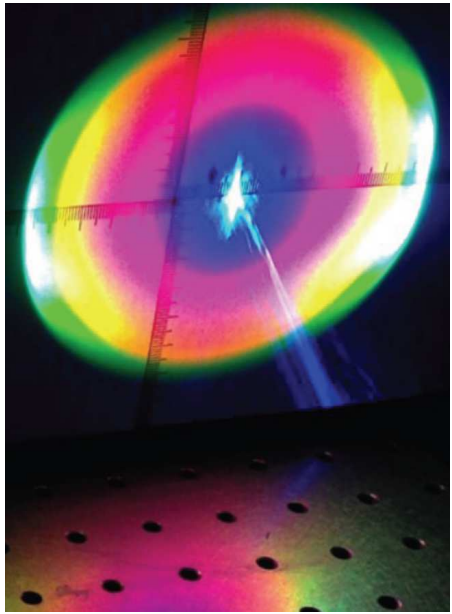}
\caption[]{It is incorrect to think of non-classical light as always faint. Bright parametric light on this photo is a
macroscopic quantum state. Courtesy of M.V. Chekhova.}
\label{fig:bright}       
\end{figure}

Parametric gain determines the mean photon number in a mode $\langle N\rangle=\sinh^2(G)$ as well as the squeezing parameters: $q_{out}=q_{in}\exp(G),\;p_{out}=p_{in}\exp(-G)$. We have used these relations calculating the Wigner function shape in Fig.~\ref{fig:pq}(d). Even in strongly pumped parametric processes, $G$ is typically less than ten. The record value of $G\approx 16$ is reported in \cite{Iskhakov12bright}. But let us not be deceived by these small numbers. Unlike a gain of a common amplifier, parametric gain is exponential, see (\ref{Uparam}), so SPDC with  $G\approx 16$ produces over $10^{13}$ photons per mode. Therefore multimode light generated in parametric down conversion can be quite strong in terms of the optical power, see Fig.~\ref{fig:bright}, but still non-classical.  

Multimode SPDC light is useful for imaging and similar applications. Here the number of modes can be compared to the number of pixels, and directly translates to the spatial resolution. Single mode SPDC, on the other hand, is often desirable for quantum communication applications, when the presence of multiple mutually incoherent modes is equivalent to the loss of the phase information, or decoherence. Spatial and frequency filtering can be employed to purify the SPDC mode structure, but this approach is not power-efficient if the initial source has too many modes. The number of excited transverse modes can be reduced, even to one, by using waveguides instead of bulk crystals. This provides a dramatic benefit over the filtering approach in terms of useful photon pair rate. For example, it is possible to generate and collect about 100,000 photon pairs per second with only 0.1 mW pump \cite{Sanaka01pdc}.

The number of frequency or temporal modes can be controlled by matching the SPDC linewidth, determined by the source length, geometry and GVD, with the transform-limited spectrum of the pump pulse. This can be done e.g. by adjusting the pump pulse duration. 

Combining these two techniques, nearly single-mode parametric sources can be realized \cite{Harder13waveguide}. Let us also mention that the birefringent properties of parametric crystals can make the gain so selective that in the super luminescence regime even free-space parametric sources can approach single-mode operation \cite{Perez15twin}.

Multipartite multiphoton states can be prepared in SPDC process by combining two or more identical coherently pumped sources \cite{Zukowski93multiphot,Zukowski95multiphot}, or by splitting multiphoton states from a single source \cite{Pan2000GHZ,Eibl04W,Radmark09sixphot}. These experiments are difficult because of the thermal statistics of SPDC pairs. Although higher photon-number states are less likely to emerge, they are more likely to cause a detection event with imperfect ($\eta<1$) detectors. Suppressing such events requires limiting the overall photon flux, which leads to very low data rate, typically of the order of 1/second for four-photon measurements and 1/hour for six-photon measurements. 

\paragraph{Optical parametric amplification}%%%%%%%%%%%%%%%%%%%%%%%%%%%%%%%%%%

If a degenerate or non-degenerate parametric process has non-vacuum inputs in the signal and idler modes, it may amplify or de-amplify the input beam(s) depending on the relation between the sum of their phases and the phase of the pump, which determines the sign of $G$. If one of the inputs, e.g. the idler, is in vacuum state for which the phase is not defined, then the signal will always be amplified. On the phase space diagram it will appear as both displacement and quadrature-squeezing \cite{Chekhova15sqz_rev}.  

Like SPDC, OPA is a common technique for generating non-classical light. This technique is most suitable for squeezing coherent light pulses seeding the OPA. A 2 dB \cite{Aytur92sqz} and then 5.8 dB \cite{Kim94sqz} squeezing of 270 ns long pulses in a degenerate Type-II parametric amplifier was demonstrated. A thousand times shorter squeezed vacuum pulses (250 fs, 1.7 dB squeezing) were generated in a Sagnac interferometer configuration using periodically poled lithium niobate crystal \cite{Hirosawa09fs-sqz}.

Continuous wave coherent states can also be used for seeding the OPAs, which allows for precise control of the local oscillator phase. This technique has been used to generate quadrature-squeezed light by injecting fundamental laser light into a degenerate OPA waveguide made from periodically poled KTP and pumped by the second harmonic of the fundamental laser light, reaching 2.2 dB of squeezing \cite{Pysher09sqz}. Realizing a similar process in a monolitic cavity with highly reflective coating on the parametric crystal facets, 6 dB of squeezing has been reached \cite{Breitenbach97sqz}. Using a Type-II OPA in a bow-tie cavity yielded 3.6 dB of polarization squeezing \cite{Lassen07sqz}, which corresponds to
reduction of the quantum uncertainty of the observables associated with the Stokes operators (\ref{StokesOps}).

Often the OPA seed signal itself is generated in another SPDC process taking place in a similar crystal and pumped by the same pump. This configuration is sometimes called a nonlinear interferometer. We have already encountered it discussing the spectroscopy applications in section \ref{sec:aps}. The high mode selectivity of such interferometers has allowed to implement a nearly single-mode squeezed vacuum source without a significant decrease in the output brightness \cite{Perez14sqz,Iskhakov16bright}. It is also possible to cascade more than two OPAs. A system of three OPAs reported in \cite{Yan12sqz} has boosted the two-mode squeezing from 5.3 dB measured after the first OPA to 8.1 dB after the third one.

\paragraph{Parametric processes in cavities}%%%%%%%%%%%%%%%%%%%%%%%%

An amplifier can be turned into an oscillator by providing a positive feedback, e.g. by placing the amplifying media into an optical cavity. Such a setup was used in the first demonstration of parametric squeezing in 1986 by Wu \textit{et al.} \cite{Wu86sqz}. In this experiment, frequency-doubled 1064 nm laser light pumped a degenerate OPO system consisting of a lithium niobate crystal inside a Fabri-Perot resonator. The same fundamental laser light was used as a local oscillator in homodyne detection of the squeezed vacuum. 3.5 dB of squeezing was measured. In 1992 this result was slightly improved to 3.8 dB with a bow-tie cavity \cite{Polzik92sqz}. This configuration was further improved by using periodically poled KTP crystals, which reduced the linear and pump-induced absorption and eliminated the transverse walk-off. 7.2 dB of squeezing was demonstrated in 2006 \cite{Suzuki06sqz}, and 9 dB in 2007 \cite{Takeno07sqz}. Thorough stabilzation of a cavity allowed generation of a narrow-band, 5 dB squeezed vacuum matching the rubidium D1 line \cite{Hetet07sqz}. Using a monolithic cavity boosted the squeezing to 10 dB in 2008 \cite{Vahlbruch08sqz} and to 12.7 dB in 2010 \cite{Eberle10sqz}. Most recently, a new record, 15 dB of squeezing, was reported \cite{Vahlbruch16sqz}.
%12.3 dB with a semi-monolithic OPA cavity \cite{Mehmet11sqz}

The experiments \cite{Wu86sqz,Slusher85sqz,Polzik92sqz,Suzuki06sqz,Takeno07sqz,Vahlbruch08sqz,Eberle10sqz,Vahlbruch16sqz,Hetet07sqz} were carried out below the OPO threshold. This means that the mean photon number per mode was below unity, or in other words, the process was predominantly vacuum-seeded, in contrast to the case of self-sustained oscillations. In this sense a sub-threshold OPO can be compared to a very long crystal in an SPDC experiment. By contrast, operating an OPO \textit{above} the threshold turns it into a laser. This is not an ordinary classical laser, however. A non-degenerate OPO laser emits two beams that are quantum-correlated, or two-mode squeezed. This has been demonstrated already in 1987 by Heidmann \textit{et al.}, who used a Type-II OPO to generate a few milliwatts in each near-degenerate signal and idler beams \cite{Heidmann87sqz}. In a few years the same approach yielded 8.5 dB of two-mode squeezing \cite{Mertz91sqz}, which had remained a squeezing world record perhaps for the longest time. 

The photon number correlation between the signal and idler beams can be used to prepare sub-Poissonian light in either one of these beams. This was demonstrated in 1988 by Tapster et al. \cite{Tapster88feedback} who detected the fluctuations of the signal beam power emitted in a Type-I SPDC process from a KDP crystal in frequency-degenerate but non-collinear configuration, and fed them back to the pump power thereby achieving the photon-number squeezing in the idler beam. A variation of this experiment was performed later with a sub-threshold OPO\cite{Mertz91feedback}, in which case the signal measurement was \textit{fed forward} to a fast intensity modulator placed in the idler beam. In \cite{Mertz91feedback} one can also find an extensive theoretical analysis of both feedforward and feedback techniques applied for preparing sub-Poissonian light in PDC. In section \ref{sec:lasers} we will see how both these techniques can be applied to other types of lasers to generate non-classical light.

This approach received an interesting developement in 2003 \cite{Laurat03sqz}, when instead of actively using the signal power fluctuations in a feedback or feedforward loops, Laurat \textit{et al.} used them for conditioning the signal-idler squeezing measurement. Only those measurements were retained when the fluctuations were below a certain threshold. Thereby a continuous-variables post-selective measurement was implemented, which allowed to observe 7.5 dB of squeezing. 

Discussing the quantum information applications of non-classical light, we have mentioned the importance of making the source narrow-band enough to match the optical transitions widths in gas phase ensemble quantum memories, often implemented with atoms or ions. An OPO provides such an opportunity. Above the threshold, its line can be considerably narrower than the cold cavity linewidth due to the Schawlow-Townes effect. Thus even with modest cavities OPO light can match the narrow atomic transitions. Hald \textit{et al.} used this approach to observe spin squeezing of cold atomic ensemble induced by interaction with squeezed vacuum \cite{Hald99}. Later it was shown that such a spin-squeezed atomic state can regenerate the squeezed vacuum, thereby verifying its storage \cite{Honda08sqz}. 

It is more difficult to achieve narrow-line OPO operation below the threshold. Usually it requires external high-$Q$ filter cavities \cite{Bao08pairs,Scholz09sm,Fekete13src,Lenhard15} or post-selection \cite{Wolfgramm11src} techniques that considerably reduce the signal rate, as well as introduce inevitable losses at the edges of the filter windows. It would be desirable to generate photon pairs directly into a single or a few easily separated modes. This became possible by using WGM micro-resonators.

In WGM resonators light is guided along a smooth optical surface of rotation by continuous total internal reflection, similarly to how sound is guided in their namesake acoustical analogues. WGM resonators defy the postulate of light propagating in a straight line in the most profound way: here the light ray bends at every point. The WGM eigenfunctions inside of a spherical resonator  are 
\begin{equation}
E_{mlq}(r,\theta,\varphi)=E_0j_m(nk_qr)\, P_l^m(\cos\theta)\,e^{im\varphi},\label{FieldSphere}
\end{equation} 
where $(r,\theta,\varphi)$ are usual spherical coordinates, $j_m$ is the spherical Bessel function of order $m$, $P_l^m$ are the associated Legendre Polynomials, and $E_0$ is the amplitude. The eigenvalue $k_q$ for a given radial mode number $q=1,2,3,\ldots$ is found by matching the internal Bessel and external Hankel eigenfunctions according to the boundary condition at the resonator rim $r=R$. For relatively large WGM resonators with small evanescent field the approximation $j_m(nk_qR)=0$ yields quite accurate results. 

\begin{figure}[b]
\centering
\includegraphics*[width=\textwidth]{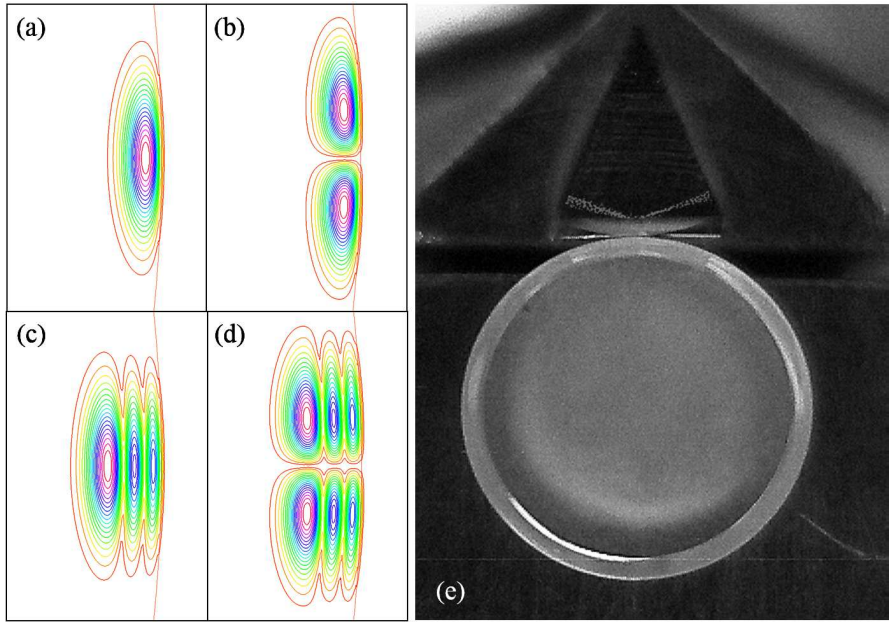}
\caption[]{Intensity distribution in the $(r,\theta)$ cross section of a WGM resonator for the fundamental mode $q=1,p=0$ (a), higher-order modes $q=1,p=1$ (b), $q=3,p=0$ (c), $q=3,p=1$ (d); and the top $(r,\varphi)$ view of a resonator with the coupling prism (e). Optical beams, visible inside the prism because of fluorescence, are focused at the coupling region where the total internal reflection is locally frustrated.}
\label{fig:WGMR}       
\end{figure} 

It is convenient to introduce $p=l-|m|=0,1,2\ldots$ which gives the mode order in the $\theta$ direction, similarly to how $q$ gives it in the radial direction. Intensity distributions in the fundamental and three higher-order WGMs are shown in Fig.~\ref{fig:WGMR}(a)-(d). Coupling of WGM resonators to external optical beams is usually done via frustrated total internal reflection, which is achieved by placing a higher-index waveguide or prism in the evanescent field of the resonator, see Fig.~\ref{fig:WGMR}(e).

More detailed discussion of WGM resonators and their properties can be found in review papers \cite{Matsko06review1,Matsko06review2,Chiasera10WGMrev,Strekalov16rev}. Here we only make two comments regarding WGM resonators that are relevant to our topic. First, the quality factor $Q$ of WGM resonators made from optically nonlinear crystals typically ranges from 10 to 100 millions. For a resonator with 1 mm circumference and 1 $\mu$m wavelength this translates to the finesse ${\cal F}=Q/m\sim 10^{4}-10^{5}$. Limited by absorption of the material, high $Q$ persists within its entire transparency window, which for a good optical crystal may well exceed an octave. Therefore the pump, signal and idler are all high-finesse modes, which increases the nonlinear optical conversion efficiency by a factor of ${\cal F}\,^3\sim 10^{12}-10^{15}$ compared to the same millimeter-long crystal without a cavity. This is a very strong enhancement which allows to seriously discuss the perspectives of doing nonlinear and quantum optics with a few or even single photons, in particular implementing optical quantum logic gates \cite{Sun13switch}.  

The second note concerns the SPDC phase matching. While the formalism (\ref{Uparam}), (\ref{Omega}), (\ref{ovlp}) still applies, the overlap integral (\ref{ovlp}) leads to selection rules that are much less restrictive than the usual phase matching (\ref{pmk}). In fact the angular part of this integral yields the Clebsch-Gordan coefficients, reminding us that in spherical geometry the orbital momenta are conserved, rather than linear momenta. The radial part leads to no strict selection rules, but it favors such combinations when $q_p\approx q_s+q_i$ \cite{Strekalov14SFG}.

SPDC was observed in WGM resonators made from various optically nonlinear crystals and at various pump wavelengths both above \cite{Savchenkov07THz,Fuerst10PDC,fuerst11sqz,Beckmann11,Beckmann12coupling,Werner12BlueOPO,Werner15OPO} and below\cite{Fortsch13NC,Fortsch15sm,Fortsch15jopt,Schunk15CsRb} the OPO threshold which for such resonators can be as low as several microwatts \cite{Fuerst10PDC}. Two-mode squeezing above the threshold was reported by F{\"u}rst \textit{et al.} \cite{fuerst11sqz}. The emitted signal and idler wavelengths can be tuned in a very wide range but at the same time with a great precision using a combination of temperature tuning, pump mode selection and evanescent field manipulation. Adjusting these parameters, Schunk \textit{et al.} have been able to tune the signal wavelength to an atomic transition and observe fluorescence induced by single heralded photons \cite{Schunk15CsRb}. In this experiment both cesium and rubidium D1 transitions were accessed using the same laser and the same resonator with the resonator temperature change by less than 2$^\circ$C.

Narrow linewidth of WGMs leads to a relatively sparse spectrum. Leveraging the selection rules, this can be used for engineering a single-mode parametric light source. Strictly single mode operation attested to by a Glauber correlation function measurement on the signal beam $g^{(2)}(0)=2.01\pm0.07$ was demonstrated by F{\"o}rtsch \textit{et al.} with only minimal spectral filtering \cite{Fortsch15sm}. In this experiment the spectral width of the pulsed pump was transform-limited to approximately 20 MHz, exceeding the signal and idler spectral widths (both equal to the resonator linewidth) by more than a factor of two. Hence even a very careful measurement of the signal frequency would not allow to identify its idler twin photon among the others using relation (\ref{pmom}), and true single-mode regime is achieved.

By the same argument, single-mode operation should not be expected with a CW pump having a linewidth smaller than that of a resonator mode. However an experiment using sub-kHz wide CW laser pumping a WMG resonator with several MHz linewidth \cite{Fortsch13NC} showed surprisingly few (approximately three, where it should be thousands) SPDC modes, consistently with $g^{(2)}(0)\approx 1.3$. Note that in this experiment the ``parasitic" SPDC into a wrong family of signal and idler WGMs has not been filtered out. Such filtering has improved $g^{(2)}(0)$ from 1.5 to 2 in the pulsed light experiment \cite{Fortsch15sm}, see above. Therefore the extra modes observed in \cite{Fortsch13NC} are more likely to be associated with different mode families than with photons distinguishability within a single WGM. 

The apparent paradox is resolved if we contemplate the fact that limitation of the observation time prevents us, even in principle, from performing a frequency measurement of the signal photon with the resolution required to localize the idler photon within a WGM linewidth. In this respect, gating a photon-detection measurement is equivalent to pulsing the pump. In both experiments \cite{Fortsch15sm} and \cite{Fortsch13NC} the measurement time was defined by the resolution of the instrument recording the signal-idler coincidences, 1 ns and 162 ps respectively, much too short for resolving the WGM linewidth. 

Closing this section we would like to make two remarks regarding the cavity-assisted nonlinear optical processes. The first one is  that squeezing can be attained not only in PDC but also in other such processes. For example,  both the second harmonic \cite{Collett85sqz,Sizmann90SHGsqz,Kurz92SHsqz} and fundamental pump wavelength \cite{Drummond81SH,Pereira88SHGpumpSQZ} in the frequency-doubling processes may be squeezed. But in such processes the amount of squeezing is inherently limited and most likely does not present significant practical interest. The second remark is that the parametric down-conversion
near degeneracy may populate multiple pairs of quantum-correlated signal and idler modes  \cite{Hage10multipartite,Pysher11ent_comb}, leading to an optical comb. Such quantum-correlated optical combs may be used for creating multipartite entangled states, highly desired in many quantum information applications, e.g. in linear quantum computing. Finally, we would like to point out that WGM is not the only type of the optically nonlinear monolithic resonators based on total internal reflection. OPO based on square-shaped monolithic resonators has been recently implemented to generate 2.6 dB of vacuum squeezing \cite{Brieussel16sqz}.

\subsection{Kerr nonlinearity in fibers and resonators}\label{sec:Kerr}

\begin{figure}[htb]
\centering
\includegraphics*[width=0.5\textwidth]{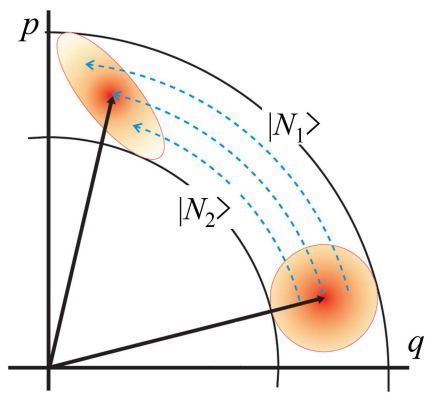}
\caption[]{Illustration of an input coherent state squeezing via self phase modulation.}
\label{fig:Kerr}       
\end{figure}

A monochromatic wave propagating in Kerr media experiences self-phase modulation (SPM) that can be described by the Kerr Hamiltonian
\begin{equation}
H = \hbar K a^\dagger a a^\dagger a\label{HKerr}
\end{equation}
and by the associated time evolution operator \cite{GerryKnight}. If the nonlinear phase shift is small enough this interaction can be approximated by a dependence of the index of refraction $n$ on the intensity $I$ \cite{Chiao64Kerr}:
\begin{equation}
n=n_0+n_2I. \label{n2}
\end{equation}
Relation (\ref{n2}) is applicable to classical and quantum fluctuations of intensity. Expanding e.g. a coherent state in the photon-number basis we observe that the SPM advances a higher-number state $|N_1\rangle$ further in the phase space than a lower-number state  $|N_2\rangle$. As a result, a characteristic shearing of the Wigner function occurs, as illustrated in Fig.~\ref{fig:Kerr}, eventually leading to a crescent shape similar to Fig.~\ref{fig:pq}(f) and indicating the number-phase squeezing \cite{Kitagawa86sqz} . The direction of shearing is opposite for materials with self-focusing ($n_2>0$) and self-defocusing ($n_2<0$). Note that SPM broadens the optical spectrum, leading to generation of frequency-shifted fields, but preserves the initial field energy. This process can be also described as degenerate four-wave mixing; in continuous-spectrum systems there is no clear boundary between these two processes.     

Broad-band Kerr response in transparent dielectrics is much weaker than the resonant Kerr response in atoms, or the quadratic response in optical crystals. However, the Kerr nonlinearity in dielectrics has an important advantage: it is present also in amorphous materials such as fused silica, that can be shaped into long single-mode fibers with very low loss. This advantage allowed Shelby \textit{et al.} \cite{Shelby86sqz} to observe Kerr squeezing in fiber already in 1986, the same year as the first OPO squeezing was reported and a year after the first demonstration of squeezing in a sodium beam. They used 114 m of liquid helium cooled single-mode optical fiber pumped with CW 647 nm laser light. Reflecting the output light off a  single-ended cavity they varied the phase between the pump (also serving as the local oscillator) and the squeezed sideband to observe 0.6 dB of squeezing. Liquid helium had to be used to suppress stimulated Brillouin oscillations and
spontaneous guided acoustic-wave Brillouin scattering (GAWBS), the acousto-optic phenomena presenting the main obstacles to CW Kerr-squeezing in fibers. 

These obstacles can be circumvented by using short pulses and high peak intensities. Because of different power dependence of the Kerr and Brillouin responses this effectively minimizes the latter. Bergman and Haus observed 5 dB of squeezing with 100-ps pulses propagating in a 50 m fiber loop Sagnac interferometer \cite{Bergman91sqz}. Alleviating the problem with GAWBS, short pulses bring about a difficulty of their own: GVD causes them to spread, losing the advantage of high peak power. This problem can be solved using optical solitons. Rosenbluh and Shelby have detected a modest (1.1 dB) squeezing of 200-fs soliton pulses propagating at room temperature in 5 m of optical fiber symmetric Sagnac interferometer \cite{Rosenbluh91sqz}. Asymmetric Sagnac interferometers were later used to produce stronger amplitude squeezing of solitons: 3.9 dB (6.0 dB corrected for losses) with 126-fs pulses \cite{Schmitt98sqz}, and  5.7 dB (6.2 dB corrected for losses) with 182-fs pulses \cite{Krylov98sqz}.

Sagnac loops are convenient because they naturally facilitate a homodyne measurement. However, detecting the photon-number squeezing in a direct measurement is also possible. This was accomplished in a unidirectionally pumped 1.5 km fiber, yielding 2.3 dB (3.7 dB corrected for losses) squeezing of 2.3-ps soliton pulses \cite{Friberg96sqz}. In combination with the propagation length dependent spectral filtering, this technique has lead to even stronger (3.8 dB) squeezing of 130-fs pulses \cite{Spalter98sqz}. Squeezing bandwidth in this experiment is shown to be at least 2 GHz. Even higher bandwidths are theoretically possible. It is furthermore possible to generate mid-infrared time-locked patterns of squeezed vacuum with the amplitude fluctuations varying from below to above the shot noise limit, i.e. from squeezing to anti-squeezing, on the sub-cycle time scale \cite{Riek17sqz}. Observation of this phenomenon is enabled by the sub-cycle electro-optic probing \cite{Moskalenko15vac,Riek15vac}. 

The benefit of squeezing solitons does not come entirely for free: solitonic propagation requires specific input pulse shape and area, which makes the squeezing depend on the pulse energy \cite{Friberg96sqz}. But stabilizing the pulse energy is a much more tractable problem than suppressing GAWBS and managing GVD. And in addition, if the input energy is large enough for the given pulse parameters, the nonlinear dynamical evolution of the the pulse will lead to a soliton solution.

Polarization squeezing can be prepared from quadrature squeezing of two orthogonaly polarized modes by projecting them onto a new polarization basis. Levandovsky \textit{et al.} used for this purpose polarization-maintaining (PM) optical fibers in Sagnac configuration, producing about 1 dB of squeezing \cite{Levandovsky99sqz}. Better results were obtained with a unidirectionally pumped 13.3 m PM fiber \cite{Heersink05sqz}, in which case 130-fs soliton pulses were squeezed to  5.1 dB. This result was later improved to 6.8 dB (10.4 dB corrected for losses), but Raman scattering was found to become a limiting factor at that level \cite{Dong08sqz}.

An interesting approach was taken by Margalit \textit{et al.} \cite{Margalit98Xsqz}, who used \textit{off-diagonal} components of the $\chi^{(3)}$ tensor to cross-phase modulate orthogonal polarizations. In this case linearly polarized 1-nJ 150-fs pulses propagating unidirectionally in a non-PM fiber induced 3 dB of vacuum squeezing in the orthogonal polarization.

Invention of microstructured, hollow-core and photonic crystal fibers opened new opportunities in Kerr squeezing. In  microstructured fibers, light is confined primarily in a thin solid core which concentrates the optical field in a smaller volume and increases the Kerr interaction strength. Furthermore, GVD in such fibers can be engineered by designing the structure around the core. Pumping a microstructured fiber near its zero GVD with 38-fs pulses, Hirosawa \textit{et al.} \cite{Hirosawa05sqz} observed a spectrally broadened optical signal with up to 4.6 dB (10.3 dB corrected for losses) squeezing for some sidebands. Milanovic \textit{et al.} \cite{Milanovic10sqz} observed 3.9 dB of squeezing and a reduction of excess noise, i.e. an increase in purity, as compared to standard fiber squeezing experiments. Four wave mixing in microstructured fibers has been also used to create pulsed photon pairs at a rate rivaling the best SPDC souces \cite{Rarity05fiber,Fan07src}.

\begin{figure}[b]
\centering
\includegraphics*[width=0.7\textwidth]{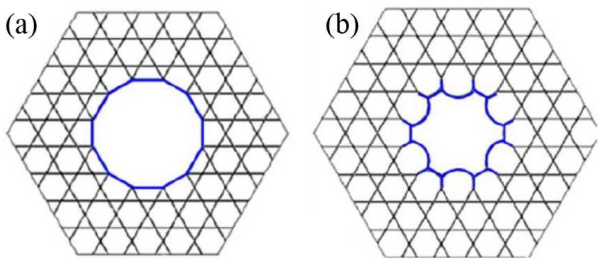}
\caption[]{Kagome fibers with a central channel designed for gas filling: a circular channel contour (a) and a hypocycloidal contour (b) designed to minimize the optical field diameter.} %Reprinted from  \cite{Bradley13Kagome}.
\label{fig:Kagome}       
\end{figure}

Another opportunity lies in combining the benefits of strong Kerr response of atomic transition with the field confinement and GVD engineering accessible in hollow-core optical fibers, in particular those with cross section resembling a traditional Japanese woven basket, which earned them a nickname Kagome fibers, see Fig.~\ref{fig:Kagome}. In Kagome fibers, light propagates mainly inside the central hollow channel, which can be filled with a Kerr media of choice. GVD can still be tailored by designing the fiber microstructure surrounding the channel, but it can furthermore by dynamically fine-tuned by changing the gas pressure, literally inflating the Kagome fiber during the drawing process or even during the measurement \cite{Nold10THG}. At the same time, Brioullin
and Raman processes in the fiber material are virtually avoided. First results have demonstrated
squeezing in fibers filled with high pressure argon \cite{Finger15sqz}
and mercury vapour \cite{Vogl15CLEO_Hg}. Filling Kagome fibers with alkali atom vapors has been proposed \cite{Bradley13Kagome} and attempted, but has not yet led to success because of chemically aggressive properties of such vapors.

\begin{figure}[b]
\centering
\includegraphics*[width=0.9\textwidth]{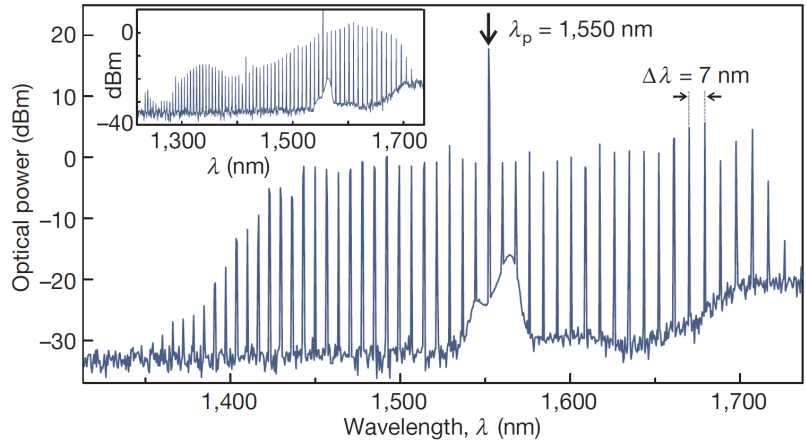}
\caption[]{A Kerr comb generated in a WGM resonator. Reprinted from  \cite{Del'Haye07comb}.}
\label{fig:comb}       
\end{figure}

Extended interaction of strongly confined optical fields can be achieved not only in fibers, but also in resonators. 
In contrast to waveguides, resonators have discrete spectra consisting of nearly-equidistant modes. In this case the SPM, cross-phase modulation (XPM) and four-wave mixing processes are clearly distinct. All these processes play their roles in the formation of Kerr-combs in crystalline WGM resonators \cite{Chembo10comb}, such as shown in Fig.~\ref{fig:comb}. 

WGM combs have been extensively discussed recently, see e.g. \cite{Liang15OPO,Strekalov16rev} and references therein. The aspect that is directly relevant to our discussion is the photon-number correlation between multiple pairs of sidebands placed symmetrically on both sides of  the pump wavelength labeled $\lambda_p$ in  Fig.~\ref{fig:comb}. This correlation arises from the degenerate four-wave mixing (or hyperparametric) process of annihilation of two pump photons and creation a photon pair in two symmetric modes. Below the oscillation threshold this process leads to the generation of entangled photon pairs. A number of
experimental demonstrations of such pairs has emerged recently using on-chip fabricated silicon microring resonators  \cite{Clemmen:2009dn,Azzini:2012rc,Engin:2013qd,Guo:2014db,Grassani:2015zl,Wakabayashi:2015lq,Suo:2015fv}.  The  time-energy entanglement was proved by violating Bell's inequality in \cite{Grassani:2015zl,Wakabayashi:2015lq}, and \cite{Suo:2015fv} has demonstrated time-energy and polarization hyper-entanglement, also confirmed by Bell's inequality violation.

Above the threshold, hyperparametric conversion leads to two-mode squeezing in a multitude of mode pairs. Such squeezing was demonstrated in a microfabricated  Si$_3$N$_4$ ring \cite{Dutt15on-chip_sqz}, which is not strictly speaking a WGM resonator, but is closely related. The free spectral range of this resonator $\Delta\lambda$ was large enough to allow selection of a single pair of squeezed modes by spectral filtering. These modes were found to be squeezed at the level of 1.7 dB (5 dB corrected for losses). Broadband quadrature squeezing in a similar resonator has been theoretically predicted \cite{Hoff15sqz}. 

Closing this section, we would like to mention that interaction of light with mechanical vibrations is not always harmful for preparation of non-classical light as in the case of GAWBS. It can be used to one's advantage. Recently, it was shown
that squeezed light can be created by coupling light with a mechanical oscillator. Here the radiation pressure quantum fluctuations induce the resonator motion which in turn imparts a phase shift to the laser light. Intensity-dependent
phase shift leads to optical squeezing in close analogy to the Kerr effect. In this way squeezing of 1.7 dB was demonstrated in a bulk cavity setup containing a thin partially transparent mechanical membrane \cite{Purdy13sqz,Painter13sqz}.
 
\subsection{Lasers and other feedback systems}\label{sec:lasers}

Laser light is commonly believed to be the best real-world approximation of a coherent state of an optical mode. However this is not always the case. The nonlinear response of a laser cavity can lead to sub-Poisson statistics of the emitted light, i.e. photon-number squeezing illustrated in Fig.~\ref{fig:pq}(f). To understand the physical mechanisms of intensity fluctuation suppression in lasers, consider an experiment with a vacuum tube filled with mercury vapor, carried out in 1985 \cite{Teich85sub}. In this experiment, a constant current flowing through the tube caused the fluorescence with the photon rate fluctuations below the shot noise. While the electrons emitted from the cathode have Poisson statistics, their flow through the vacuum tube is regulated by both the anode potential and the space charge of the electron flow. If the current increases, so does the negative space charge, which leads to the current fluctuation suppression. In other words, the space charge acts as a compressible buffer, smoothing out these fluctuations below the classical limit, which is reflected in the emitted photons statistics. This is the same mechanism, which allowed Schottky and Spehnke \cite{Schottky37} to observe a sub shot noise electron current in a vacuum tube in 1937.

A similar mechanism is present in semiconductor lasers operating in the constant-current (but not in the constant-voltage) regime, where the junction voltage provides a negative feedback regulating the current in the region of recombination \cite{Yamamoto87sqz}. This experiment was carried out using laser diodes at room temperature \cite{Machida87sqz} and at 77 K \cite{Machida88sqz}. In both cases approximately 1.7 dB amplitude squeezing (corrected for detectors efficiency) was detected in a very broad frequency range. Evidently, the squeezing measurement in these experiments was impeded by low collection efficiency. Improving this efficiency by ``face-to-face" coupling of the laser diode and the photo diode, and cooling the assembly down to 66 K, the same group was able to boost the squeezing to 8.3 dB  \cite{Richardson91sqz}. Considering the 89\% quantum efficiency of the photodiode, this corresponds to 14 dB inferred squeezing. However, neither this nor other groups were later able to reproduce this large squeezing in a semiconductor laser, showing that there must be parameters not well understood and controlled in the initial experiment. Nevertheless their experiment initiated work in other groups, which eventually led to a better understanding.

Although the space charge model gives a qualitative understanding of the phenomenon, it does not capture many important details. In 1995, Marin \textit{et. al.} conceded that ``the very mechanisms capable of explaining why some laser diodes and not others were generating sub-shot-noise light remained unclear" \cite{Marin95sqz}. They came to the conclusion that one of these mechanisms is the cross-talk between the main mode and other weakly excited modes, which should lead to their anti-correlation, i.e. two-mode or even multipartite squeezing. Later, the same group developed a theoretical understanding by identifying two excess noise sources, the Petermann excess noise and the leakage current noise, to explain the limitations of the squeezing observed \cite{Maurin05laser}. 

Another relevant factor is the optical injection into the laser cavity. The effects of an external laser injection at 10 K  \cite{Wang93sqz} and self-injection at room temperature \cite{Freeman93sqz} were studied in quantum-well lasers. Over 3 dB \cite{Wang93sqz} and 1.8 dB \cite{Freeman93sqz} photon-number squeezing was observed. A weak squeezing in a free-running quantum-well laser was also observed at room temperature \cite{Wolfl02sqz,Uemukai05sqz}.

The negative feedback suppressing the current (and hence the optical power) fluctuations does not necessarily have to be facilitated by the laser cavity. In section \ref{sec:spdc} we already discussed an example of the electronic feedback derived from the signal measurement to control the idler photon statistics in PDC. 
A similar technique was applied to a semiconductor laser in 1986 by Yamamoto \textit{et. al.} \cite{Yamamoto86sqz}. Because the laser beam lacks a quantum-correlated twin, Yamamoto employed a XPM-based quantum nondemolition (QND) measurement to monitor the output laser power. The power fluctuations of the laser beam were imprinted onto the phase of a probe beam, recovered in a heterodyne measurement, and fed back to the laser current. As a result, the amplitude squeezing ranging from 5 dB at 16 MHz to 10 dB below 2 MHz was observed.

It might seem that a linear beam splitter could provide a simpler alternative to a QND measurement in preparation of non-classical light with the feedback technique. This approach indeed leads to a very interesting field dynamics known as \textit{squashing} \cite{Buchler99squash}. The term ``squashing" pertains to the fields propagating inside the loop, and is fundamentlly different from squeezing. The most remarkable property of the in-loop squashed optical field, theoretically shown by Shapiro \textit{et al.} \cite{Shapiro87feedback}, is that such a field does not obey the usual commutation relations. Therefore it is not subject to Heisenberg uncertainty principle, and its photon-number uncertainty can be reduced below the classical limit without the phase noise penalty \cite{Buchler99squash}.  
It is worth noting that not only a state of an optical mode, but also a motional state of a trapped ion can be squashed in a feedback loop \cite{Mancini2000squash}.

In the context of nonclassical light applications, the possibility of generating optical fields not constrained by the  Heisenberg uncertainty relations appears too good to be true. And indeed, it has been shown that out-coupling the squashed field from the loop destroys its remarkable properties \cite{Shapiro87feedback}. In fact, it has been pointed out \cite{Buchler99squash} that even fully characterizing these properties, which is only possible within the loop, is a highly nontrivial experimental problem that requires a QND measurement. Therefore using the electronic feedback systems for generating non-classical light has not attracted much of practical interest. Using feedforward, on the other hand, is quite common in commercial optical devices known as ``noise eaters" that can suppress power fluctuations within the classical limit.

It would seem that diode lasers offer the most robust and easily scalable technology for generating non-classical (photon-number squeezed, or sub-Poissonian) light. They have also shown a promise in generating strongly squeezed states \cite{Richardson91sqz}. However the interest to this field apparently waned in the first decade of the 21$^{\rm st}$ century. The reason for this skepticism could be that the discovery of the excess noise sources by Maurin et al. \cite{Maurin05laser} made it clear that it is difficult to fabricate a laser that would \textit{predictably} generate strong squeezing. If this is the case, a new advance in the field may be expected from improving the semiconductor technology. 

\section{Final remarks}\label{sec:concl}

Non-classical light has played an important role in development of quantum theory, starting form the early tests of local realism performed with entangled photons in 1972 \cite{Freedman72Bell}. Following this pioneering experiment, many striking quantum phenomena have been discovered via non-classical optics research. Fluctuations of the optical field intensity have been suppressed below the shot-noise limit, which in classical notation requires negative probabilities. The concept of a \textit{biphoton}, and later of a multipartite entangled state, was proven to be tangible. Thus physicists gained hands-on experience with a system that may consist of space-like separated parts and yet constitute a single physical entity. Experimental \textit{quantum teleportation} has been made possible with such systems.

Not only fundamental, but also applied science and technology have a lot to benefit from non-classical light. Sub shot noise characteristics of the squeezed light directly points to one group of such applications: high resolution metrology. Optical phase in an interferometer, optical beam displacement, sub-wavelength image discerning and recording are just a few topics from this group. Information encoded in non-orthogonal single photon states or in any other non-orthogonal pure quantum states is protected from copying by fundamental laws of physics, which gives rise to another  large group of applications concerned with information security. Furthermore, this information can be processed using mind-blowing quantum logic operations (such as e.g. a $\sqrt{\rm NOT}$ gate) allowing, in perspective, to realize a quantum computer and the quantum internet.

But how is this wonderful non-classical light generated? The purpose of this chapter has been to provide a brief introductory tour over the most common sources of quantum light.
The variety of physical systems capable of generating non-classical light is very broad. We encountered atomic beams, vapor cells, laser-cooled atomic clouds and even individual trapped atoms or ions; optical crystals and fibers; semiconductor nanoparticles and diode lasers.  

With such a great variety of physical systems to discuss, we did not have an opportunity to provide much of detail regarding each system and its performance. Instead, we rely on references that are strategically placed so that an interested reader would be able to easily ``zoom in" on any part of our review by downloading the appropriate publications.

Despite the great diversity of the quantum light sources, a few common properties can be summarized that are important for the majority of non-classical light applications. They are the following.
\begin{itemize}
\item{\it Optical nonlinearity.} This is a driving mechanism for generating non-classical light. Strongly nonlinear optical systems require less pump power and as a consequence are less noisy and more technologically acceptable. Resonant nonlinearity of natural or artificial atoms, and broad-band nonlinearity of laser gain media are two examples that may  surpass other systems by far.
\item{\it Optical loss.} When photons are randomly removed from the system, statistics of the remaining photons becomes more and more Poissonian. For many (but not all) quantum states this results in diminishing their non-classical characteristics, such as e. g. squeezing. For quantum states with zero displacement in phase space such as Fock-states, squeezed vacuum states, cat states (i.e. superpositions of coherent states) the statistical loss of a photon on average is enough to largely reduce the non-classical property \cite{Caldeira85damping,Leuchs05dissipation}. Good examples of low-loss systems are optical nonlinear crystals and fibers. 
\item{\it Mode structure.} While imaging applications require multiple transverse modes, the applications concerned with sensing and  information processing may require strictly single-mode light. Bulk nonlinear crystals are natural sources for multimode light; on the other hand wavequides, fibers and optical cavities can be used to achieve single-mode operation.
\item{\it Wavelength and bandwidth.} Using non-classical light in conjunction with ``massive" qubits, as suggested by the quantum network paradigm, requires matching their central wavelengths and bandwidths. Therefore these source parameters either need to be precisely engineered, which is possible with atom-based sources, or tunable. Wavelength tuning is readily available in bulk crystals, but their emission is usually very broad-band. Fine-tuning of parametric light to atomic transitions in both central wavelength and bandwidth has been achieved with sub- or above-threshold OPOs.
\item{\it Practical utility.} It is generally desirable to avoid cryogenic temperatures and other stringent environment requirements. Unfortunately, many of quantum dot, quantum well and trapped atoms sources of non-classical light fail this requirement.  Therefore, progress may come via two different routes: (1) improving room temperature systems, or (2) developing compact sources and low cost cryogenic fridges. 
\end{itemize}
In general, we see that progress in quantum optics comes from developing: (1) light sources, (2) light confinement strategies, (3) materials with strong optically non-linear response. For a particular goal, one achieves the best result if one optimizes the combination of items from these three categories. We have already discussed one such example, a hollow-core fiber filled with atom media, and it appears plausible that more such examples may emerge in the future.

We thank Drs. M. Raymer and M. Gurioli for valuable comments. 
D. V. S. would like to thank the Alexander von Humboldt Foundation for sponsoring his collaboration with the Max Plank Institute for the physics of light in Erlangen.

\vspace*{0.4in}

\textbf{\large Appendix. Is coherent light quantum?}

\vspace*{0.2in}

\noindent Let us consider the following series of thought experiments. The toolbox we need contains a source of laser light, a beam splitter, two time resolving detectors of high bandwidth, and electronic equipment to analyze the detector signals. In the first experiment (1) measuring the intensity correlations after splitting the laser light with the beam splitter yields a $g^{(2)}(\tau)$ which is independent of time $\tau$. This can be described by a classical model, namely classical light fields without fluctuations - fine. Now the second experiment (2) is to measure the intensity of the laser light as a function of time. The result is a fluctuating detector signal (corresponding to the Poisson statistics of the photons in a quantum language). A classical model can also describe this. This time it is a
model in which the classical electric fields fluctuate - this is also fine, but note that the models required are not compatible.

You may not be satisfied and argue that the fluctuation observed in experiment (2) may well come from the detectors themselves contributing noise. This would average out in experiment (1) because the noises introduced by the two detectors are of course not correlated. But suppose the lab next door happens to have amplitude squeezed light, with intensity fluctuations suppressed by 15 dB below the shot noise. Measuring the squeezed light intensity noise you convince yourself easily that the detector does not introduce enough noise to explain experiment (2). Note that this test should convince you even if you have no clue what the squeezed light is. 

But you do not want to give up so easily and you say ``what if a classically noisy light field enters the second input port, uncorrelated with the laser light but likewise modeled by classical stochastic fluctuations?". And you are right, this more involved classical model would explain both experiments (1) and (2) - yet there is (3) a third experiment we can do. We can check the intensity of the light arriving at this second input port of the beam splitter and no matter how sensitive the intensity measuring detectors are they will detect no signal. But this is not compatible with a classical model: classical fluctuations always lead to measurable intensity noise. 

We conclude by noting that obviously coherent states are non-classical because there is no single classical stochastic model which describes all possible experiments with laser light. But as we have seen it is tedious to go through these arguments, and no simple measure of non-classicality was found so far qualifying a coherent state as non-classical. Nevertheless, the non-classical nature of a coherent state is used in some quantum protocols.

It is interesting to note that there is a much different scenario in which experiments with coherent states cannot be described classically without field quantization, i.e. with semi classical theory. Coherent states lead e.g. to a revival of Rabi oscillations in their interaction with an atom in the Jaynes Cummings model. This effect can only be properly described when properly accounting for the quantization of the electromagnetic field \cite{Eberly80revival,Rempe87revival}. Thus the hypothesis is that for any pure quantum state it is always possible to find experimental scenarios, which can only be properly described using  field quantization. Let us furthermore note that also thermal states, i.e. mixed quantum states, can still be somewhat nonclassical in nature if the classical excess noise is not too much larger than the underlying quantum uncertainty.

%\bibliographystyle{myunsrt}
%\bibliography{./../bibliography}

\begin{thebibliography}{100}

\bibitem{LoudonBook}
R.~Loudon.
\newblock {\em {The quantum theory of light}}.
\newblock Oxford university press, 2000.

\bibitem{Klyshko98concepts}
D.~N. Klyshko.
\newblock Basic quantum mechanical concepts from the operational viewpoint.
\newblock {\em Physics-Uspekhi}, 41:885--922, Sep 1998.

\bibitem{Walls79review}
D.~F. Walls.
\newblock Evidence for the quantum nature of light.
\newblock {\em Nature}, 280:451--454, Aug 1979.

\bibitem{Paul82rev}
H.~Paul.
\newblock Photon antibunching.
\newblock {\em Rev. Mod. Phys.}, 54:1061--1102, Oct 1982.

\bibitem{Leuchs1986Ch26}
G.~Leuchs.
\newblock Photon statistics, antibunching and squeezed states.
\newblock In G.~T. Moore and M.~O. Scully, editors, {\em Frontiers of
  Nonequilibrium Statistical Physics}. Springer US, 1986.

\bibitem{Klyshko96rev_nonclas}
D.~N. Klyshko.
\newblock The nonclassical light.
\newblock {\em Physics - Uspekhi}, 39:573--596, 1996.

\bibitem{Kimble77antibunch}
H.~J. Kimble, M.~Dagenais, and L.~Mandel.
\newblock Photon antibunching in resonance fluorescence.
\newblock {\em Phys. Rev. Lett.}, 39, Sep 1997.

\bibitem{Klyshko94aspects}
D.~N. Klyshko.
\newblock Quantum optics: quantum, classical, and metaphysical aspects.
\newblock {\em Physics - Uspekhi}, 37:1097--1123, 1994.

\bibitem{Bell64}
J.~S. Bell.
\newblock On the einstein podolsky rosen paradox.
\newblock {\em Physics}, 1:195--200, 1964.

\bibitem{chsh1969}
J.~F. Clauser, M.~A. Horne, A.~Shimony, and R.~A. Holt.
\newblock Proposed experiment to test local hidden-variables theories.
\newblock {\em Phys. Rev. Lett.}, 23:880--884, Oct 1969.

\bibitem{Clauser74bell}
J.~F. Clauser and M.~A. Horne.
\newblock Experimental consequences of objective local theories.
\newblock {\em Phys. Rev. D}, 10:526--535, Jul 1974.

\bibitem{Clauser78bell}
J.~F. Clauser and A.~Shimony.
\newblock Bell's theorem: experimental tests and implications.
\newblock {\em Rep. Prog. Phys.}, 41:1881--1927, 1978.

\bibitem{Cerf97entropy}
N.~J. Cerf and C.~Adami.
\newblock Negative entropy and information in quantum mechanics.
\newblock {\em Phys. Rev. Lett.}, 79:5194--5197, Dec 1997.

\bibitem{nielsen2010quantum}
M.~A. Nielsen and I.~L. Chuang.
\newblock {\em Quantum computation and quantum information}.
\newblock Cambridge university press, 2010.

\bibitem{SchleichBk}
W.~P. Schleich.
\newblock {\em Quantum Optics in Phase Space}.
\newblock Wiley-VCH Verlag Berlin GmbH, Berlin, 2001.

\bibitem{Hillery84rev}
M.~Hillery, R.~F. O'Connell, M.~O. Scully, and E.~P. Wigner.
\newblock Distribution functions in physics: fundamentals.
\newblock {\em Phys. Rep.}, 106:121--167, 1984.

\bibitem{Hillery89noise}
M.~Hillery.
\newblock Total noise and nonclassical states.
\newblock {\em Phys. Rev. A}, 39:2994--3002, Mar 1989.

\bibitem{Lee90nonclas}
C.~T. Lee.
\newblock Higher-order criteria for nonclassical effects in photon statistics.
\newblock {\em Phys. Rev. A}, 41:1721--1723, Feb 1990.

\bibitem{Klyshko96nonclas}
D.~N. Klyshko.
\newblock Observable signs of nonclassical light.
\newblock {\em Phys. Lett. A}, 213:7, Apr 1996.

\bibitem{Yurke86cat}
B.~Yurke and D.~Stoler.
\newblock Generating quantum mechanical superpositions of macroscopically
  distinguishable states via amplitude dispersion.
\newblock {\em Phys. Rev. Lett.}, 57:13--16, Jul 1986.

\bibitem{Vlastakis13cat}
B.~Vlastakis, G.~Kirchmair, Z.~Leghtas, S.~E. Nigg, L.~Frunzio, S.~M. Girvin,
  M.~Mirrahimi, M.~H. Devoret, and R.~J. Schoelkopf.
\newblock Deterministically encoding quantum information using 100-photon
  schr{\"o}dinger cat states.
\newblock {\em Science}, 342:607--610, Nov 2013.

\bibitem{Kwiat98hyper}
P.~G. Kwiat and H.~Weinfurter.
\newblock Embedded bell-state analysis.
\newblock {\em Phys. Rev. A}, 58:R2623--R2626, Oct 1998.

\bibitem{Dowling2008noon}
J.~Dowling.
\newblock Quantum optical metrology - the lowdown on hing-noon states.
\newblock {\em Contemp. Phys.}, 49:125--143, 2008.

\bibitem{Afek10noon}
I.~Afek, O.~Ambar, and Y.~Silberberg.
\newblock High-noon states by mixing quantum and classical light.
\newblock {\em Science}, 328:879--881, May 2010.

\bibitem{Zhang07quad_ent}
Y.~Zhang, T.~Furuta, R.~Okubo, K.~Takahashi, and T.~Hirano.
\newblock Experimental generation of broadband quadrature entanglement using
  laser pulses.
\newblock {\em Phys. Rev. A}, 76:012314, Jul 2007.

\bibitem{Yoshino07quad_ent}
K.-i. Yoshino, T.~Aoki, and A.~Furusawa.
\newblock Generation of continuous-wave broadband entangled beams using
  periodically poled lithium niobate waveguides.
\newblock {\em Appl. Phys. Lett.}, 90:041111, Jan 2007.

\bibitem{Andersen10qi}
U.~L. Andersen, G.~Leuchs, and C.~Silberhorn.
\newblock Continuous-variable quantum information processing.
\newblock {\em Las. \& Phot. Rev.}, 4:337--354, Jan 2010.

\bibitem{Bowen03teleport}
W.~P. Bowen, N.~Treps, B.~C. Buchler, R.~Schnabel, T.~C. Ralph, H.-A. Bachor,
  T.~Symul, and P.~K. Lam.
\newblock Experimental investigation of continuous-variable quantum
  teleportation.
\newblock {\em Phys. Rev. A}, 67:032302, Mar 2003.

\bibitem{Chekhova15sqz_rev}
M.~V. Chekhova, G.~Leuchs, and M.~\.{Z}ukowski.
\newblock Bright squeezed vacuum: Entanglement of macroscopic light beams.
\newblock {\em Opt. Comm.}, 337:27--43, Feb 2015.

\bibitem{Collett85sqz}
M.~J. Collett and D.~F. Walls.
\newblock Squeezing spectra for nonlinear optical systems.
\newblock {\em Phys. Rev. A}, 32:2887--2892, Nov 1985.

\bibitem{Reid88sqz}
M.~D. Reid and P.~D. Drummond.
\newblock Quantum correlations of phase in nondegenerate parametric
  oscillation.
\newblock {\em Phys. Rev. Lett.}, 60:2731--2733, Jun 1988.

\bibitem{Fabre89opo}
C.~Fabre, E.~Giacobino, A.~Heidmann, and S.~Reynaud.
\newblock Noise characteristics of a non-degenerate optical parametric
  oscillator - application to quantum noise reduction.
\newblock {\em J.de Phys.}, 50:1209--1225, Jan 1989.

\bibitem{fuerst11sqz}
J.~U. F{\"u}rst, D.~V. Strekalov, D.~Elser, A.~Aiello, U.~L. Andersen,
  C.~Marquardt, and G.~Leuchs.
\newblock Quantum light from a whispering-gallery-mode disk resonator.
\newblock {\em Phys. Rev. Lett.}, 106:113901, Mar 2011.

\bibitem{Yurke86interf}
B.~Yurke, S.~L. McCall, and J.~R. Klauder.
\newblock Su(2) and su(1,1) interferometers.
\newblock {\em Phys. Rev. A}, 33:4033--4054, Jun 1986.

\bibitem{Campos89BS}
R.~A. Campos, B.~E.~A. Saleh, and M.~C. Teich.
\newblock Quantum-mechanical lossless beam splitter: Su(2) symmetry and photon
  statistics.
\newblock {\em Phys. Rev. A}, 40:1371--1384, Aug 1989.

\bibitem{Spasibko14BS}
K.~Y. Spasibko, F.~T{\"o}ppel, T.~S. Iskhakov, M.~Stobińska, M.~V. Chekhova,
  and G.~Leuchs.
\newblock Interference of macroscopic beams on a beam splitter: phase
  uncertainty converted into photon-number uncertainty.
\newblock {\em New J. Phys.}, 16:013025, Jan 2014.

\bibitem{Greenberger90GHZ}
D.~M. Greenberger, M.~A. Horne, A.~Shimony, and A.~Zeilinger.
\newblock Bell's theorem without inequalities.
\newblock {\em Am. J. Phys.}, 58:1131--1143, 1990.

\bibitem{Pan2000GHZ}
J.-W. Pan, D.~Bouwmeester, M.~Daniell, H.~Weinfurter, and A.~Zeilinger.
\newblock Experimental test of quantum nonlocality in three-photon
  greenberger-horne-zeilinger entanglement.
\newblock {\em Nature}, 403:515--519, Feb 2000.

\bibitem{Dur2000W}
W.~Dur, G.~Vidal, and J.~I. Cirac.
\newblock Three qubits can be entangled in two inequivalent ways.
\newblock {\em Phys. Rev. A}, 62:062314, Nov 2000.

\bibitem{Eibl04W}
M.~Eibl, N.~Kiesel, M.~Bourennane, C.~Kurtsiefer, and H.~Weinfurter.
\newblock Experimental realization of a three-qubit entangled w state.
\newblock {\em Phys. Rev. Lett.}, 92:077901, Feb 2004.

\bibitem{Wen10resolution}
J.~Wen, S.~Du, and M.~Xiao.
\newblock Improving spatial resolution in quantum imaging beyond the rayleigh
  diffraction limit using multiphoton w entangled states.
\newblock {\em Phys. Lett. A}, 374:3908--3911, Aug 2010.

\bibitem{Briegel01cluster}
H.~J. Briegel and R.~Raussendorf.
\newblock Persistent entanglement in arrays of interacting particles.
\newblock {\em Phys. Rev. Lett.}, 86:910--913, Jan 2001.

\bibitem{Hein04graph}
M.~Hein, J.~Eisert, and H.~J. Briegel.
\newblock Multiparty entanglement in graph states.
\newblock {\em Phys. Rev. A}, 69:062311, Jun 2004.

\bibitem{Smolin01}
J.~A. Smolin.
\newblock Four-party unlockable bound entangled state.
\newblock {\em Phys. Rev. A}, 63:032306, Mar 2001.

\bibitem{BP2000qkd}
H.~Bechmann-Pasquinucci and W.~Tittel.
\newblock Quantum cryptography using larger alphabets.
\newblock {\em Phys. Rev. A}, 61:062308, May 2000.

\bibitem{Walborn06qkd}
S.~P. Walborn, D.~S. Lemelle, M.~P. Almeida, and P.~H.~S. Ribeiro.
\newblock Quantum key distribution with higher-order alphabets using spatially
  encoded qudits.
\newblock {\em Phys. Rev. Lett.}, 96:090501, Mar 2006.

\bibitem{Dixon12qkd}
P.~B. Dixon, G.~A. Howland, J.~Schneeloch, and J.~C. Howell.
\newblock Quantum mutual information capacity for high-dimensional entangled
  states.
\newblock {\em Phys. Rev. Lett.}, 108:143603, Apr 2012.

\bibitem{Wasilewski06sqz}
W.~Wasilewski, A.~I. Lvovsky, K.~Banaszek, and C.~Radzewicz.
\newblock Pulsed squeezed light: Simultaneous squeezing of multiple modes.
\newblock {\em Phys. Rev. A}, 73:063819, Jun 2006.

\bibitem{Collins02Bell}
D.~Collins, N.~Gisin, N.~Linden, S.~Massar, and S.~Popescu.
\newblock Bell inequalities for arbitrarily high-dimensional systems.
\newblock {\em Phys. Rev. Lett.}, 88:040404, Jan 2002.

\bibitem{Lo16entangl}
H.-P. Lo, C.-M. Li, A.~Yabushita, Y.-N. Chen, C.-W. Luo, and T.~Kobayashi.
\newblock Experimental violation of bell inequalities for multi-dimensional
  systems.
\newblock {\em Sci. Rep.}, 6:22088, Jan 2016.

\bibitem{Dada11Bell}
A.~C. Dada, J.~Leach, G.~S. Buller, M.~J. Padgett, and E.~Andersson.
\newblock Experimental high-dimensional two-photon entanglement and violations
  of generalized bell inequalities.
\newblock {\em Nat. Phys.}, 7:677--680, Sep 2011.

\bibitem{Mair01OAM}
A.~Mair, A.~Vaziri, G.~Weihs, and A.~Zeilinger.
\newblock Entanglement of the orbital angular momentum states of photons.
\newblock {\em Nature}, 412:313--316, Jul 2001.

\bibitem{Krenn14OAM}
M.~Krenn, M.~Huber, R.~Fickler, R.~Lapkiewicz, S.~Ramelow, and A.~Zeilinger.
\newblock Generation and confirmation of a (100 x 100)-dimensional entangled
  quantum system.
\newblock {\em PNAS}, 111:6243--6247, Apr 2014.

\bibitem{Hiesmayr16OAM}
B.~C. Hiesmayr, M.~J.~A. de~Dood, and W.~L{\"o}ffler.
\newblock Observation of four-photon orbital angular momentum entanglement.
\newblock {\em Phys. Rev. Lett.}, 116:073601, Feb 2016.

\bibitem{Mitchell14_100part}
M.~W. Mitchell and F.~A. Beduini.
\newblock Extreme spin squeezing for photons.
\newblock {\em New J. Phys.}, 16:073027, Jul 2014.

\bibitem{Horodecki09quant}
R.~Horodecki, P.~Horodecki, M.~Horodecki, and K.~Horodecki.
\newblock Quantum entanglement.
\newblock {\em Rev. Mod. Phys.}, 81:865--942, Apr 2009.

\bibitem{Vidal02negativity}
G.~Vidal and R.~F. Werner.
\newblock Computable measure of entanglement.
\newblock {\em Phys. Rev. A}, 65:032314, Mar 2002.

\bibitem{Wootters98concur}
W.~K. Wootters.
\newblock Entanglement of formation of an arbitrary state of two qubits.
\newblock {\em Phys. Rev. Lett.}, 80:2245--2248, Mar 1998.

\bibitem{Hildebrand07concur}
R.~c.
\newblock Concurrence revisited.
\newblock {\em J. Math. Phys.}, 48:102108--102108, Oct 2007.

\bibitem{Zeldovich69cal}
B.~Y. Zel’dovich and D.~N. Klyshko.
\newblock Statistics of field in parametric luminescence.
\newblock {\em Sov. Phys. JETP Lett.}, 9:40–44, 1969.

\bibitem{Burnham70cal}
D.~C. Burnham and D.~L. Weinberg.
\newblock Observation of simultaneity in parametric production of optical
  photon pairs.
\newblock {\em Phys. Rev. Lett.}, 25:84--87, Jul 1970.

\bibitem{Klyshko80calibration}
D.~N. Klyshko.
\newblock Use of two-photon light for absolute calibration of photoelectric
  detectors.
\newblock {\em Quant. El.}, 7:1932--1940, Sep 1980.

\bibitem{Polyakov07cal}
S.~V. Polyakov and A.~L. Migdall.
\newblock High accuracy verification of a correlatedphoton- based method for
  determining photoncounting detection efficiency.
\newblock {\em Opt. Expr.}, 15:1390--1407, Jan 2007.

\bibitem{Ware07cal}
M.~Ware, A.~Migdall, J.~Bienfang, and S.~Polyakov.
\newblock Calibrating photon-counting detectors to high accuracy: background
  and deadtime issues.
\newblock {\em J. Mod. Opt.}, 54:361--372, Jan 2007.

\bibitem{Czitrovszky2000cal}
A.~Czitrovszky, A.~Sergienko, P.~Jani, and A.~Nagy.
\newblock Measurement of quantum efficiency using correlated photon pairs and a
  single-detector technique.
\newblock {\em Metrologia}, 37:617--620, 2000.

\bibitem{Lebedev08cal}
M.~V. Lebedev, A.~A. Shchekin, and O.~V. Misochko.
\newblock Two-electron pulses of a photomultiplier and two-photon photoeffect.
\newblock {\em Q. El.}, 38:710--723, Aug 2008.

\bibitem{Brida06cal}
G.~Brida, M.~Genovese, I.~Ruo-Berchera, M.~Chekhova, and A.~Penin.
\newblock Possibility of absolute calibration of analog detectors by using
  parametric downconversion: a systematic study.
\newblock {\em JOSA B}, 23:2185--2193, Oct 2006.

\bibitem{Vahlbruch16sqz}
H.~Vahlbruch, M.~Mehmet, K.~Danzmann, and R.~Schnabel.
\newblock Detection of 15 db squeezed states of light and their application for
  the absolute calibration of photoelectric quantum efficiency.
\newblock {\em Phys. Rev. Lett.}, 117:110801, Sep 2016.

\bibitem{Brida10cal}
G.~Brida, I.~P. Degiovanni, M.~Genovese, M.~L. Rastello, and I.~Ruo-Berchera.
\newblock Detection of multimode spatial correlation in pdc and application to
  the absolute calibration of a ccd camera.
\newblock {\em Opt. Expr.}, 18:20572--20584, Sep 2010.

\bibitem{KlyshkoPhotonsBook}
D.~N. Klyshko.
\newblock {\em {Photons and Nonlinear optics}}.
\newblock Taylor and Francis, New York, NY USA, 1988.

\bibitem{Klyshko87bright}
D.~N. Klyshko and A.~N. Penin.
\newblock The prospects of quantum photometry.
\newblock {\em Sov. Phys. Usp.}, 30:716--723, 1987.

\bibitem{Xiao87sqz}
M.~Xiao, L.-A. Wu, and H.~J. Kimble.
\newblock Precision measurement beyond the shot-noise limit.
\newblock {\em Phys. Rev. Lett.}, 59:278--281, Jul 1987.

\bibitem{Grangier87sqz}
P.~Grangier, R.~E. Slusher, B.~Yurke, and A.~LaPorta.
\newblock Squeezed-light- enhanced polarization interferometer.
\newblock {\em Phys. Rev. Lett.}, 59:2153--2156, Nov 1987.

\bibitem{LIGO11sqz}
T.~L.~S. Collaboration.
\newblock A gravitational wave observatory operating beyond the quantum
  shot-noise limit.
\newblock {\em Nat. Phys.}, 7:962--965, Jan 2011.

\bibitem{Polzik92sqz}
E.~S. Polzik, J.~Carri, and H.~J. Kimble.
\newblock Spectroscopy with squeezed light.
\newblock {\em Phys. Rev. Lett.}, 68:3020--3023, May 1992.

\bibitem{Ribeiro97spectrscopy}
P.~H.~S. Ribeiro, C.~Schwob, A.~Maitre, and C.~Fabre.
\newblock Sub-shot-noise high-sensitivity spectroscopy with optical parametric
  oscillator twin beams.
\newblock {\em Opt. Lett.}, 22:1893--1895, Dec 1997.

\bibitem{Taylor13sqz}
M.~A. Taylor, J.~Janousek, V.~Daria, J.~Knittel, B.~Hage, H.-A. Bachor, and
  W.~P. Bowen.
\newblock Biological measurement beyond the quantum limit.
\newblock {\em Nat. Phot.}, 7:229--233, Mar 2013.

\bibitem{Gea-Banacloche89two-phot}
J.~Gea-Banacloche.
\newblock Two-photon absorption of nonclassical light.
\newblock {\em Phys. Rev. Lett.}, 62:1603--1606, Apr 1989.

\bibitem{Javanainen90two-phot}
J.~Javanainen and P.~L. Gould.
\newblock Linear intensity dependence of a two-photon transition rate.
\newblock {\em Phys. Rev. A}, 41:5088--5091, May 1990.

\bibitem{Dayan07TPA}
B.~Dayan.
\newblock Theory of two-photon interactions with broadband down-converted light
  and entangled photons.
\newblock {\em Phys. Rev. A}, 76:043813, Oct 2007.

\bibitem{Georgiades95twophot}
N.~P. Georgiades, E.~S. Polzik, K.~Edamatsu, H.~J. Kimble, and A.~S. Parkins.
\newblock Nonclassical excitation for atoms in a squeezed vacuum.
\newblock {\em Phys. Rev. Lett.}, 75:3426--3429, Nov 1995.

\bibitem{Dayan04TPA}
B.~Dayan, A.~Pe'er, A.~A. Friesem, and Y.~Silberberg.
\newblock Two photon absorption and coherent control with broadband
  down-converted light.
\newblock {\em Phys. Rev. Lett.}, 93:023005, Jul 2004.

\bibitem{Boitier09two-phot}
F.~Boitier, A.~Godard, E.~Rosencher, and C.~Fabre.
\newblock Measuring photon bunching at ultrashort timescale by two-photon
  absorption in semiconductors.
\newblock {\em Nat. Phys.}, 5:267--270, Apr 2009.

\bibitem{Korystov2001hooks}
D.~Y. Korystov, S.~P. Kulik, and A.~N. Penin.
\newblock Rozhdestvenski hooks in two-photon parametric light scattering.
\newblock {\em JETP Lett.}, 73:214--218, 2001.

\bibitem{Boto00litho}
A.~N. Boto, P.~Kok, D.~S. Abrams, S.~L. Braunstein, C.~P. Williams, and J.~P.
  Dowling.
\newblock Quantum interferometric optical lithography: Exploiting entanglement
  to beat the diffraction limit.
\newblock {\em Phys. Rev. Lett.}, 85:2733--2736, Sep 2000.

\bibitem{Peer04litho}
A.~Pe'er, B.~Dayan, M.~Vucelja, Y.~Silberberg, and A.~A. Friesem.
\newblock Quantum lithography by coherent control of classical light pulses.
\newblock {\em Opt. Expr.}, 12:6600--6605, Dec 2004.

\bibitem{Nagasako01opa}
E.~M. Nagasako, S.~J. Bentley, R.~W. Boyd, and G.~S. Agarwal.
\newblock Nonclassical two-photon interferometry and lithography with high-gain
  parametric amplifiers.
\newblock {\em Phys. Rev. A}, 64:043802, Oct 2001.

\bibitem{Dayan05pdc-sfg}
B.~Dayan, A.~Pe'er, A.~A. Friesem, and Y.~Silberberg.
\newblock Nonlinear interactions with an ultrahigh flux of broadband entangled
  photons.
\newblock {\em Phys. Rev. Lett.}, 94:043602, Feb 2005.

\bibitem{Pittman95Qimaging}
T.~B. Pittman, Y.~H. Shih, D.~V. Strekalov, and A.~V. Sergienko.
\newblock Optical imaging by means of two-photon quantum entanglement.
\newblock {\em Phys. Rev. A}, 52:R3429--R3432, Nov 1995.

\bibitem{Nasr03QOCT}
M.~B. Nasr, B.~E.~A. Saleh, A.~V. Sergienko, and M.~C. Teich.
\newblock Demonstration of dispersion-canceled quantum-optical coherence
  tomography.
\newblock {\em Phys. Rev. Lett.}, 91:083601, Aug 2003.

\bibitem{Treps03pointer}
N.~Treps, N.~Grosse, W.~P. Bowen, C.~Fabre, H.-A. Bachor, and P.~K. Lam.
\newblock A quantum laser pointer.
\newblock {\em Science}, 301:940--943, Aug 2003.

\bibitem{Feynman82qc}
R.~P. Feynman.
\newblock Simulating physics with computers.
\newblock {\em Int. J. Theor. Phys.}, 21:467--488, 1982.

\bibitem{Shor97}
P.~W. Shor.
\newblock Polynomial-time algorithms for prime factorization and discrete
  logarithms on a quantum computer.
\newblock {\em SIAM J. Comp.}, 26:1484--1509, 1997.

\bibitem{Clauser96factor}
J.~F. Clauser and J.~P. Dowling.
\newblock Factoring integers with young's n-slit interferometer.
\newblock {\em Phys. Rev. A}, 53:4587--4590, Jun 1996.

\bibitem{Franson04QZB}
J.~D. Franson, B.~C. Jacobs, and T.~B. Pittman.
\newblock Quantum computing using single photons and the zeno effect.
\newblock {\em Phys. Rev. A}, 70:062302, Dec 2004.

\bibitem{Franson07WGM_QZB}
J.~D. Franson, T.~B. Pittman, and B.~C. Jacobs.
\newblock Zeno logic gates using microcavities.
\newblock {\em JOSA B}, 24:209--213, Feb 2007.

\bibitem{Clader13switch}
B.~D. Clader, S.~M. Hendrickson, R.~M. Camacho, and B.~C. Jacobs.
\newblock All-optical microdisk switch using eit.
\newblock {\em Opt. Expr.}, 21:6169--6179, Mar 2013.

\bibitem{Huang10QZB}
Y.-P. Huang, J.~B. Altepeter, and P.~Kumar.
\newblock Interaction-free all-optical switching via the quantum zeno effect.
\newblock {\em Phys. Rev. A}, 82:063826, Dec 2010.

\bibitem{Huang10switch_prop}
Y.-P. Huang and P.~Kumar.
\newblock Interaction-free all-optical switching in chi$^{(2)}$ microdisks for
  quantum applications.
\newblock {\em Opt. Lett.}, 35:2376--2378, Jul 2010.

\bibitem{Sun13switch}
Y.-Z. Sun, Y.-P. Huang, and P.~Kumar.
\newblock Photonic nonlinearities via quantum zeno blockade.
\newblock {\em Phys. Rev. Lett.}, 110:223901, May 2013.

\bibitem{Hendrickson13switch}
S.~M. Hendrickson, C.~N. Weiler, R.~M. Camacho, P.~T. Rakich, A.~I. Young,
  M.~J. Shaw, T.~B. Pittman, J.~D. Franson, and B.~C. Jacobs.
\newblock All-optical-switching demonstration using two-photon absorption and
  the zeno effect.
\newblock {\em Phys. Rev. A}, 87:23808, Feb 2013.

\bibitem{Strekalov14switch}
D.~V. Strekalov, A.~S. Kowligy, Y.-P. Huang, and P.~Kumar.
\newblock Progress towards interaction-free all-optical devices.
\newblock {\em Phys. Rev. A}, 89:063820, Jun 2014.

\bibitem{Kimble08internet}
H.~J. Kimble.
\newblock The quantum internet.
\newblock {\em Nature}, 453:1023--1030, Jun 2008.

\bibitem{Aoki09switch}
T.~Aoki, A.~S. Parkins, D.~J. Alton, C.~A. Regal, B.~Dayan, E.~Ostby, K.~J.
  Vahala, and H.~J. Kimble.
\newblock Efficient routing of single photons by one atom and a microtoroidal
  cavity.
\newblock {\em Phys. Rev. Lett.}, 102:083601, Feb 2009.

\bibitem{Specht11qmem}
H.~P. Specht, C.~N{\"o}lleke, A.~Reiserer, M.~Uphoff, E.~Figueroa, S.~Ritter,
  and G.~Rempe.
\newblock A single-atom quantum memory.
\newblock {\em Nature}, 473:190--193, May 2011.

\bibitem{Ourjoumtsev11atom}
A.~Ourjoumtsev, A.~Kubanek, M.~Koch, C.~Sames, P.~W.~H. Pinkse, G.~Rempe, and
  K.~Murr.
\newblock Observation of squeezed light from one atom excited with two photons.
\newblock {\em Nature}, 474:623--626, Jun 2011.

\bibitem{Chen13switch}
W.~Chen, K.~M. Beck, R.~B{\"u}cker, M.~Gullans, M.~D. Lukin, H.~Tanji-Suzuki,
  and V.~Vuleti{\'c}.
\newblock All-optical switch and transistor gated by one stored photon.
\newblock {\em Science}, 341:768--770, Aug 2013.

\bibitem{Baur14switch}
S.~Baur, D.~Tiarks, G.~Rempe, and S.~D{\"u}rr.
\newblock Single-photon switch based on rydberg blockade.
\newblock {\em Phys. Rev. Lett.}, 112:073901, Feb 2014.

\bibitem{Shomroni14switch}
X.~Shomroni, S.~Rosenblum, Y.~Lovsky, O.~Bechler, G.~Guendelman, and B.~Dayan.
\newblock All-optical routing of single photons by a one-atom switch controlled
  by a single photon.
\newblock {\em Science}, 345:903--906, Aug 2014.

\bibitem{Tiecke14switch}
T.~G. Tiecke, J.~D. Thompson, N.~P. de~Leon, L.~R. Liu, V.~Vuleti{\'c}, and
  M.~D. Lukin.
\newblock Nanophotonic quantum phase switch with a single atom.
\newblock {\em Nature}, 508:241--244, Apr 2014.

\bibitem{Rosenblum15single}
S.~Rosenblum, O.~Bechler, I.~Shomroni, Y.~Lovsky, G.~Guendelman, and B.~Dayan.
\newblock Extraction of a single photon from an optical pulse.
\newblock {\em Nat. Phot.}, 10:19--22, Jan 2016.

\bibitem{Michler2000qdot1}
P.~Michler, A.~Kiraz, C.~Becher, W.~V. Schoenfeld, P.~M. Petroff, L.~Zhang,
  E.~Hu, and A.~Imamoglu.
\newblock A quantum dot single-photon turnstile device.
\newblock {\em Science}, 290:2282--2285, Dec 2000.

\bibitem{Li11nv}
P.-B. Li, S.-Y. Gao, and F.-L. Li.
\newblock Quantum-information transfer with nitrogen-vacancy centers coupled to
  a whispering-gallery microresonator.
\newblock {\em Phys. Rev. A}, 83:054306, May 2011.

\bibitem{Chen12Qgate}
Q.~Chen, W.~L. Yang, and M.~Feng.
\newblock Quantum gate operations in decoherence-free fashion with separate
  nitrogen-vacancy centers coupled to a whispering-gallery mode resonator.
\newblock {\em Eur. Phys. J. D}, 66:238, Sep 2012.

\bibitem{Volz06atom}
J.~Volz, M.~Weber, D.~Schlenk, W.~Rosenfeld, J.~Vrana, K.~Saucke,
  C.~Kurtsiefer, and H.~Weinfurter.
\newblock Observation of entanglement of a single photon with a trapped atom.
\newblock {\em Phys. Rev. Lett.}, 96:030404, Jan 2006.

\bibitem{Beugnon06atom}
J.~Beugnon, M.~P.~A. Jones, J.~Dingjan, B.~Darqui{\'e}, G.~Messin, A.~Browaeys,
  and P.~Grangier.
\newblock Quantum interference between two single photons emitted by
  independently trapped atoms.
\newblock {\em Nature}, 440:779--782, Apr 2006.

\bibitem{Maunz07atom}
P.~Maunz, D.~L. Moehring, S.~Olmschenk, K.~C. Younge, D.~N. Matsukevich, and
  C.~Monroe.
\newblock Quantum interference of photon pairs from two remote trapped atomic
  ions.
\newblock {\em Nat. Phys.}, 3:538--541, Aug 2007.

\bibitem{Leong15HOM}
V.~Leong, S.~Kosen, B.~Srivathsan, G.~K. Gulati, A.~Cer{\`e}, and
  C.~Kurtsiefer.
\newblock Hong-ou-mandel interference between triggered and heralded single
  photons from separate atomic systems.
\newblock {\em Phys. Rev. A}, 91:063829, Jun 2015.

\bibitem{Bao08pairs}
X.-H. Bao, Y.~Qian, J.~Yang, H.~Zhang, Z.-B. Chen, T.~Yang, and J.-W. Pan.
\newblock Generation of narrow-band polarization-entangled photon pairs for
  atomic quantum memories.
\newblock {\em Phys. Rev. Lett.}, 101:190501, Nov 2008.

\bibitem{Fekete13src}
J.~Fekete, D.~Riel{\"a}nder, M.~Cristiani, and H.~de~Riedmatten.
\newblock Ultranarrow-band photon-pair source compatible with solid state
  quantum memories and telecommunication networks.
\newblock {\em Phys. Rev. Lett.}, 110, Jan 2013.

\bibitem{Schunk15CsRb}
G.~Schunk, U.~Vogl, D.~V. Strekalov, M.~F{\"o}rtsch, F.~Sedlmeir, H.~G.~L.
  Schwefel, M.~G{\"o}belt, S.~Christiansen, G.~Leuchs, and C.~Marquardt.
\newblock Interfacing transitions of different alkali atoms and telecom bands
  using one narrowband photon pair source.
\newblock {\em Optica}, 2:773--778, Sep 2015.

\bibitem{Lenhard15}
A.~Lenhard, M.~Bock, C.~Becher, S.~Kucera, J.~Brito, P.~Eich, P.~M{\"u}ller,
  and J.~Eschner.
\newblock Telecom-heralded single-photon absorption by a single atom.
\newblock {\em Phys. Rev. A}, 92:063827, Dec 2015.

\bibitem{Schunk16CsRb}
G.~Schunk, U.~Vogl, F.~Sedlmeir, D.~V. Strekalov, A.~Otterpohl, V.~Averchenko,
  H.~G.~L. Schwefel, G.~Leuchs, and C.~Marquardt.
\newblock Frequency tuning of single photons from a whispering-gallery mode
  resonator to mhz-wide transitions.
\newblock {\em J. Mod. Opt.}, 63:2058--2073, Jan 2016.

\bibitem{Fortsch15sm}
M.~F{\"o}rtsch, G.~Schunk, J.~U. F{\"u}rst, D.~Strekalov, T.~Gerrits, M.~J.
  Stevens, F.~Sedlmeir, H.~G.~L. Schwefel, S.~W. Nam, G.~Leuchs, and
  C.~Marquardt.
\newblock Highly efficient generation of single-mode photon pairs from a
  crystalline whispering-gallery-mode resonator source.
\newblock {\em Phys. Rev. A}, 91:023812, Feb 2015.

\bibitem{Luo15sm}
K.-H. Luo, H.~Herrmann, S.~Krapick, B.~Brecht, R.~Ricken, V.~Quiring, H.~Suche,
  W.~Sohler, and C.~Silberhorn.
\newblock Direct generation of genuine single-longitudinal-mode narrowband
  photon pairs.
\newblock {\em New J. Phys.}, 17:073039, Jul 2015.

\bibitem{Knill01linearqc}
E.~Knill, R.~Laflamme, and G.~J. Milburn.
\newblock A scheme for efficient quantum computation with linear optics.
\newblock {\em Nature}, 409:46--52, Jan 2001.

\bibitem{Bogdanov04qutrit}
Y.~I. Bogdanov, M.~V. Chekhova, L.~A. Krivitsky, S.~P. Kulik, A.~N. Penin,
  A.~A. Zhukov, L.~C. Kwek, C.~H. Oh, and M.~K. Tey.
\newblock Statistical reconstruction of qutrits.
\newblock {\em Phys. Rev. A}, 70:042303, Oct 2004.

\bibitem{Lanyon08qutrit}
B.~P. Lanyon, T.~J. Weinhold, N.~K. Langford, J.~L. O'Brien, K.~J. Resch,
  A.~Gilchrist, and A.~G. White.
\newblock Manipulating biphotonic qutrits.
\newblock {\em Phys. Rev. Lett.}, 100:060504, Feb 2008.

\bibitem{Bogdanov06ququart}
Y.~I. Bogdanov, E.~V. Moreva, G.~A. Maslennikov, R.~F. Galeev, S.~S. Straupe,
  and S.~P. Kulik.
\newblock Polarization states of four-dimensional systems based on biphotons.
\newblock {\em Phys. Rev. A}, 73:063810, Jun 2006.

\bibitem{Luo15ququart}
M.-X. Luo, Y.~Deng, H.-R. Li, and S.-Y. Ma.
\newblock Photonic ququart logic assisted by the cavity-qed system.
\newblock {\em Sci. Rep.}, 5:13255, Jan 2015.

\bibitem{Wootters82noclone}
W.~K. Wootters and W.~H. Zurek.
\newblock A single quantum cannot be cloned.
\newblock {\em Nature}, 299:802--803, Oct 1982.

\bibitem{BB84}
C.~H. Bennett and G.~Brassard.
\newblock Quantum cryptography: Public key distribution and coin tossing.
\newblock In {\em Proceedings of IEEE International Conference on Computers,
  Systems and Signal Processing}, volume 175, page~8, 1984.

\bibitem{Braunstein05rev}
S.~L. Braunstein and P.~van Loock.
\newblock Quantum information with continuous variables.
\newblock {\em Rev. Mod. Phys.}, 77:513--577, Apr 2005.

\bibitem{Crepeau88}
C.~Cr{\`e}peau and J.~Kilian.
\newblock Achieving oblivious transfer using weakened security assumptions.
\newblock In {\em Foundations of Computer Science, 1988., 29th Annual Symposium
  on}, pages 42 -- 52, 1988.

\bibitem{Lunghi13commit}
T.~Lunghi, J.~Kaniewski, F.~Bussi{\`e}res, R.~Houlmann, M.~Tomamichel, A.~Kent,
  N.~Gisin, S.~Wehner, and H.~Zbinden.
\newblock Experimental bit commitment based on quantum communication and
  special relativity.
\newblock {\em Phys. Rev. Lett.}, 111:180504, Nov 2013.

\bibitem{Croal16qsign}
C.~Croal, C.~Peuntinger, B.~Heim, I.~Khan, C.~Marquardt, G.~Leuchs, P.~Wallden,
  E.~Andersson, and N.~Korolkova.
\newblock Free-space quantum signatures using heterodyne measurements.
\newblock {\em Phys. Rev. Lett.}, 117:100503, Sep 2016.

\bibitem{Freedman72Bell}
S.~J. Freedman and J.~F. Clauser.
\newblock Experimental test of local hidden-variable theories.
\newblock {\em Phys. Rev. Lett.}, 28:938--941, Apr 1972.

\bibitem{Brannan14qbit}
T.~Brannan, Z.~Qin, A.~MacRae, and A.~I. Lvovsky.
\newblock Generation and tomography of arbitrary optical qubits using transient
  collective atomic excitations.
\newblock {\em Opt. Lett.}, 39:5447--5450, Sep 2014.

\bibitem{Slusher85sqz}
R.~E. Slusher, L.~W. Hollberg, B.~Yurke, J.~C. Mertz, and J.~F. Valleys.
\newblock Observation of squeezed states generated by four-wave mixing in an
  optical cavity.
\newblock {\em Phys. Rev. Lett.}, 55:2409--2412, Nov 1985.

\bibitem{Lambrecht96sqz}
A.~Lambrecht, T.~Coudreau, A.~M. Steinberg, and E.~Giacobino.
\newblock Squeezing with cold atoms.
\newblock {\em Europhys. Lett.}, 36:93--98, Oct 1996.

\bibitem{McCormick07sqz}
C.~F. McCormick, V.~Boyer, E.~Arimondo, and P.~D. Lett.
\newblock Strong relative intensity squeezing by four-wave mixing in rubidium
  vapor.
\newblock {\em Opt. Lett.}, 32:178--180, Jan 2007.

\bibitem{Corzo11sqz}
N.~Corzo, A.~M. Marino, K.~M. Jones, and P.~D. Lett.
\newblock Multi-spatial-mode single-beam quadrature squeezed states of light
  from four-wave mixing in hot rubidium vapor.
\newblock {\em Opt. Expr.}, 19:21358--21369, Oct 2011.

\bibitem{Boyer08twin}
V.~Boyer, A.~M. Marino, and P.~D. Lett.
\newblock Generation of spatially broadband twin beams for quantum imaging.
\newblock {\em Phys. Rev. Lett.}, 100:143601, Apr 2008.

\bibitem{Balic05atom}
V.~Bali{\'c}, D.~A. Braje, P.~Kolchin, G.~Y. Yin, and S.~E. Harris.
\newblock Generation of paired photons with controllable waveforms.
\newblock {\em Phys. Rev. Lett.}, 94:183601, May 2005.

\bibitem{Chou04heralded}
C.~W. Chou, S.~V. Polyakov, A.~Kuzmich, and H.~J. Kimble.
\newblock Single-photon generation from stored excitation in an atomic
  ensemble.
\newblock {\em Phys. Rev. Lett.}, 92:213601, May 2004.

\bibitem{Polyakov04heralded}
S.~V. Polyakov, C.~W. Chou, D.~Felinto, and H.~J. Kimble.
\newblock Temporal dynamics of photon pairs generated by an atomic ensemble.
\newblock {\em Phys. Rev. Lett.}, 93:263601, Dec 2004.

\bibitem{Eisaman04shaping}
M.~D. Eisaman, L.~Childress, A.~Andr{\'e}, F.~Massou, A.~S. Zibrov, and M.~D.
  Lukin.
\newblock Shaping quantum pulses of light via coherent atomic memory.
\newblock {\em Phys. Rev. Lett.}, 93:233602, Dec 2004.

\bibitem{Chen10atom}
J.~F. Chen, S.~Zhang, H.~Yan, M.~M.~T. Loy, G.~K.~L. Wong, and S.~Du.
\newblock Shaping biphoton temporal waveforms with modulated classical fields.
\newblock {\em Phys. Rev. Lett.}, 104:183604, May 2010.

\bibitem{Matsko02sqz}
A.~B. Matsko, I.~Novikova, G.~R. Welch, D.~Budker, D.~F. Kimball, and S.~M.
  Rochester.
\newblock Vacuum squeezing in atomic media via self-rotation.
\newblock {\em Phys. Rev. A}, 66:043815, Oct 2002.

\bibitem{Barreiro11sqz}
S.~Barreiro, P.~Valente, H.~Failache, and A.~Lezama.
\newblock Polarization squeezing of light by single passage through an atomic
  vapor.
\newblock {\em Phys. Rev. A}, 84:033851, Sep 2011.

\bibitem{Ries03sqz}
J.~Ries, B.~Brezger, and A.~I. Lvovsky.
\newblock Experimental vacuum squeezing in rubidium vapor via self-rotation.
\newblock {\em Phys. Rev. A}, 68:025801, Aug 2003.

\bibitem{Birnbaum05blockade}
K.~M. Birnbaum, A.~Boca, R.~Miller, A.~D. Boozer, T.~E. Northup, and H.~J.
  Kimble.
\newblock Photon blockade in an optical cavity with one trapped atom.
\newblock {\em Nature}, 436:87--90, Jul 2005.

\bibitem{Dayan08turnstile}
B.~Dayan, A.~S. Parkins, T.~Aoki, E.~P. Ostby, K.~J. Vahala, and H.~J. Kimble.
\newblock A photon turnstile dynamically regulated by one atom.
\newblock {\em Science}, 319:1062--1065, Feb 2008.

\bibitem{Munoz14bundle}
C.~S. Mu{\~n}oz, E.~del Valle, A.~G. Tudela, K.~M{\"u}ller, S.~Lichtmannecker,
  M.~Kaniber, C.~Tejedor, J.~J. Finley, and F.~P. Laussy.
\newblock Emitters of n-photon bundles.
\newblock {\em Nat. Phot.}, 8:550--555, Jan 2014.

\bibitem{Strekalov14bundle}
D.~V. Strekalov.
\newblock A bundle of photons, please.
\newblock {\em Nat. Phot.}, 8:500--501, Jul 2014.

\bibitem{Basche92antibunch}
T.~Basche, W.~E. Moerner, M.~Orrit, and H.~Talon.
\newblock Photon antihunching in the fluorescence of a single dye molecule
  trapped in a solid.
\newblock {\em Phys. Rev. Lett.}, 69:1516--1519, Sep 1992.

\bibitem{Brunel99src}
C.~Brunel, B.~Lounis, P.~Tamarat, and M.~Orrit.
\newblock Triggered source of single photons based on controlled single
  molecule fluorescence.
\newblock {\em Phys. Rev. Lett.}, 83:2722--2725, Oct 1999.

\bibitem{Lounis2000single}
B.~Lounis and W.~E. Moerner.
\newblock Single photons on demand from a singlemolecule at room temperature.
\newblock {\em Nature}, 407:491--493, Sep 2000.

\bibitem{Lounis05src}
B.~Lounis and M.~Orrit.
\newblock Single-photon sources.
\newblock {\em Rep. Prog. Phys.}, 68:1129--1179, May 2005.

\bibitem{Buckley12qd_rev}
S.~Buckley, K.~Rivoire, and J.~Vu{\v c}kovi{\'c}.
\newblock Engineered quantum dot single-photon sources.
\newblock {\em Rep. Prog. Phys.}, 75:126503, Dec 2012.

\bibitem{Michler2000qdot}
P.~Michler, A.~Imamoglu, M.~D. Maso, P.~J. Carson, G.~F. Strouse, and S.~K.
  Buratto.
\newblock Quantum correlation among photons from a single quantum dot at room
  temperature.
\newblock {\em Nature}, 406:968--970, Aug 2000.

\bibitem{Bounouar12qd}
S.~Bounouar, M.~Elouneg-Jamroz, M.~d. Hertog, C.~Morchutt, E.~Bellet-Amalric,
  R.~Andr{\'e}, C.~Bougerol, Y.~Genuist, J.-P. Poizat, S.~Tatarenko, and
  K.~Kheng.
\newblock Ultrafast room temperature single-photon source from nanowire-quantum
  dots.
\newblock {\em Nano Lett.}, 12:2977--2981, Jun 2012.

\bibitem{Holmes14qd}
M.~J. Holmes, K.~Choi, S.~Kako, M.~Arita, and Y.~Arakawa.
\newblock Room-temperature triggered single photon emission from a iii-nitride
  site-controlled nanowire quantum dot.
\newblock {\em Nano Lett.}, 14:982--986, Feb 2014.

\bibitem{Hogele08nanotube}
A.~H{\"o}gele, C.~Galland, M.~Winger, and A.~Imamoǧlu.
\newblock Photon antibunching in the photoluminescence spectra of a single
  carbon nanotube.
\newblock {\em Phys. Rev. Lett.}, 100:217401, May 2008.

\bibitem{Schietinger08nv}
S.~Schietinger, T.~Schr\"{o}der, and O.~Benson.
\newblock One-by-one coupling of single defect centers in nanodiamonds to
  high-q modes of an optical microresonator.
\newblock {\em Nano Lett.}, 8(11):3911--3915, November 2008.

\bibitem{Babinec10diamond}
T.~M. Babinec, B.~J.~M. Hausmann, M.~Khan, Y.~Zhang, J.~R. Maze, P.~R. Hemmer,
  and M.~Loncar.
\newblock A diamond nanowire single-photon source.
\newblock {\em Nat. Nanotech.}, 5:195--199, Mar 2010.

\bibitem{Schulte15sqz}
C.~H.~H. Schulte, J.~Hansom, A.~E. Jones, C.~Matthiesen, C.~Le~Gall, and
  M.~Atat{\"u}re.
\newblock Quadrature squeezed photons from a two-level system.
\newblock {\em Nature}, 525:222--225, Sep 2015.

\bibitem{Press07qdot}
D.~Press, S.~G{\"o}tzinger, S.~Reitzenstein, C.~Hofmann, A.~L{\"o}ffler,
  M.~Kamp, A.~Forchel, and Y.~Yamamoto.
\newblock Photon antibunching from a single quantum-dot-microcavity system in
  the strong coupling regime.
\newblock {\em Phys. Rev. Lett.}, 98:117402, Mar 2007.

\bibitem{Strauf07single}
S.~Strauf, N.~G. Stoltz, M.~T. Rakher, L.~A. Coldren, P.~M. Petroff, and
  D.~Bouwmeester.
\newblock High-frequency single-photon source with polarization control.
\newblock {\em Nat. Phot.}, 1:704--708, Dec 2007.

\bibitem{Peter05qdot}
E.~Peter, P.~Senellart, D.~Martrou, A.~Lema{\^i}tre, J.~Hours, J.~M.
  G{\'e}rard, and J.~Bloch.
\newblock Exciton-photon strong-coupling regime for a single quantum dot
  embedded in a microcavity.
\newblock {\em Phys. Rev. Lett.}, 95:067401, Aug 2005.

\bibitem{Srinivasan07qdot}
K.~Srinivasan and O.~Painter.
\newblock Linear and nonlinear optical spectroscopy of a strongly coupled
  microdisk-quantum dot system.
\newblock {\em Nature}, 450:862--865, Dec 2007.

\bibitem{Makhonin14qdot}
M.~N. Makhonin, J.~E. Dixon, R.~J. Coles, B.~Royall, I.~J. Luxmoore, E.~Clarke,
  M.~Hugues, M.~S. Skolnick, and A.~M. Fox.
\newblock Waveguide coupled resonance fluorescence from on-chip quantum
  emitter.
\newblock {\em Nano Lett.}, 14:6997--7002, Dec 2014.

\bibitem{Akopian06qdot}
N.~Akopian, N.~H. Lindner, E.~Poem, Y.~Berlatzky, J.~Avron, D.~Gershoni, B.~D.
  Gerardot, and P.~M. Petroff.
\newblock Entangled photon pairs from semiconductor quantum dots.
\newblock {\em Phys. Rev. Lett.}, 96:130501, Apr 2006.

\bibitem{Stevenson06source}
R.~M. Stevenson, R.~J. Young, P.~Atkinson, K.~Cooper, D.~A. Ritchie, and A.~J.
  Shields.
\newblock A semiconductor source of triggered entangled photon pairs.
\newblock {\em Nature}, 439:179--182, Jan 2006.

\bibitem{Kuroda13qdot}
T.~Kuroda, T.~Mano, N.~Ha, H.~Nakajima, H.~Kumano, B.~Urbaszek, M.~Jo,
  M.~Abbarchi, Y.~Sakuma, K.~Sakoda, I.~Suemune, X.~Marie, and T.~Amand.
\newblock Symmetric quantum dots as efficient sources of highly entangled
  photons: Violation of bell's inequality without spectral and temporal
  filtering.
\newblock {\em Phys. Rev. B}, 88:041306, Jul 2013.

\bibitem{Jayakumar13qdot}
H.~Jayakumar, A.~Predojevi{\'c}, T.~Huber, T.~Kauten, G.~S. Solomon, and
  G.~Weihs.
\newblock Deterministic photon pairs and coherent optical control of a single
  quantum dot.
\newblock {\em Phys. Rev. Lett.}, 110:135505, Mar 2013.

\bibitem{Dotti15qe}
N.~Dotti, F.~Sarti, S.~Bietti, A.~Azarov, A.~Kuznetsov, F.~Biccari,
  A.~Vinattieri, S.~Sanguinetti, M.~Abbarchi, and M.~Gurioli.
\newblock Germanium-based quantum emitters towards a time-reordering
  entanglement scheme with degenerate exciton and biexciton states.
\newblock {\em Phys. Rev. B}, 91:205316, May 2015.

\bibitem{Trotta14qdot}
R.~Trotta, J.~S. Wildmann, E.~Zallo, O.~G. Schmidt, and A.~Rastelli.
\newblock Highly entangled photons from hybrid piezoelectric-semiconductor
  quantum dot devices.
\newblock {\em Nano Lett.}, 14:3439--3444, Jun 2014.

\bibitem{Young06qdot}
R.~J. Young, R.~M. Stevenson, P.~Atkinson, K.~Cooper, D.~A. Ritchie, and A.~J.
  Shields.
\newblock Improved fidelity of triggered entangled photons from single quantum
  dots.
\newblock {\em New J. Phys.}, 8:29, Feb 2006.

\bibitem{Muller14qdot}
M.~M{\"u}ller, S.~Bounouar, K.~D. J{\"o}ns, M.~Gl{\"a}ssl, and P.~Michler.
\newblock On-demand generation of indistinguishable polarization-entangled
  photon pairs.
\newblock {\em Nat. Phot.}, 8:224--228, Mar 2014.

\bibitem{Giordmaine65spdc}
J.~A. Giordmaine and R.~C. Miller.
\newblock Tunable coherent parametric oscillation in linbo3 at optical
  frequencies.
\newblock {\em Phys. Rev. Lett.}, 14:973--976, Jun 1965.

\bibitem{Klyshko69PDC}
D.~N. Klyshko.
\newblock Scattering of light in a medium with nonlinear polarizability.
\newblock {\em JETP Lett.}, 28:522--526, Mar 1969.

\bibitem{strekalov05g1g2}
D.~Strekalov, A.~B. Matsko, A.~A. Savchenkov, and L.~Maleki.
\newblock Relationship between quantum two-photon correlation and classical
  spectrum of light.
\newblock {\em Phys. Rev. A}, 71:041803, Apr 2005.

\bibitem{Rubin94TypeII}
M.~H. Rubin, D.~N. Klyshko, Y.~H. Shih, and A.~V. Sergienko.
\newblock Theory of two-photon entanglement in type-ii optical parametric
  down-conversion.
\newblock {\em PRA}, 50:5122--5133, Dec 1994.

\bibitem{Dauler99subfs}
E.~Dauler, G.~Jaeger, A.~Muller, A.~Migdall, and A.~Sergienko.
\newblock Tests of a two-photon technique for measuring polarization mode
  dispersion with subfemtosecond precision.
\newblock {\em J. Res. Natl. Inst. Stand. Technol.}, 104:1--10, 1999.

\bibitem{Valencia02disp}
A.~Valencia, M.~V. Chekhova, A.~Trifonov, and Y.~Shih.
\newblock Entangled two-photon wave packet in a dispersive medium.
\newblock {\em Phys. Rev. Lett.}, 88:183601, May 2002.

\bibitem{strekalov05g1g2JMO}
D.~Strekalov, A.~B. Matsko, A.~Savchenkov, and L.~Maleki.
\newblock Quantum-correlation metrology with biphotons: where is the limit?
\newblock {\em J. Mod. Opt.}, 52:2233--2243, Nov 2005.

\bibitem{Scholz09narrow}
M.~Scholz, L.~Koch, and O.~Benson.
\newblock Statistics of narrow-band single photons for quantum memories
  generated by ultrabright cavity-enhanced parametric down-conversion.
\newblock {\em Phys. Rev. Lett.}, 102:63603, Feb 2009.

\bibitem{Chuu12bright}
C.-S. Chuu, G.~Y. Yin, and S.~E. Harris.
\newblock A miniature ultrabright source of temporally long, narrowband
  biphotons.
\newblock {\em Appl. Phys. Lett.}, 101:051108, Aug 2012.

\bibitem{Fortsch13NC}
M.~F{\"o}rtsch, J.~U. F{\"u}rst, C.~Wittmann, D.~Strekalov, A.~Aiello, M.~V.
  Chekhova, C.~Silberhorn, G.~Leuchs, and C.~Marquardt.
\newblock A versatile source of single photons for quantum information
  processing.
\newblock {\em Nat. Comm.}, 4:1818, Jan 2013.

\bibitem{Burlkaov97bigpaper}
A.~V. Burlakov, M.~V. Chekhova, D.~N. Klyshko, S.~P. Kulik, A.~N. Penin, Y.~H.
  Shih, and D.~V. Strekalov.
\newblock Interference effects in spontaneous two-photon parametric scattering
  from two macroscopic regions.
\newblock {\em Phys. Rev. A}, 56:3214--3225, Oct 1997.

\bibitem{Iskhakov16bright}
T.~S. Iskhakov, S.~Lemieux, A.~Perez, R.~W. Boyd, G.~Leuchs, and M.~V.
  Chekhova.
\newblock Nonlinear interferometer for tailoring the frequency spectrum of
  bright squeezed vacuum.
\newblock {\em J. Mod. Opt.}, 63:64--70, Jan 2016.

\bibitem{Setala10TGI}
T.~Set{\"a}l{\"a}, T.~Shirai, and A.~T. Friberg.
\newblock Fractional fourier transform in temporal ghost imaging with classical
  light.
\newblock {\em Phys. Rev. A}, 82:043813, Oct 2010.

\bibitem{Sych16TGIxxx}
D.~Sych, V.~Averchenko, and G.~Leuchs.
\newblock Shaping a single photon without interacting with it.
\newblock {\em arXiv:1605.00023v2}, Apr 2016.

\bibitem{Averchenko16TGIxxx}
V.~Averchenko, D.~Sych, and G.~Leuchs.
\newblock Heralded temporal shaping of single photons enabled by entanglement.
\newblock {\em arXiv:1610.03794v1}, Oct 2016.

\bibitem{Tapster98SPDCstat}
P.~R. Tapster and J.~G. Rarity.
\newblock Photon statistics of pulsed parametric light.
\newblock {\em J. Mod. Opt.}, 45:595--604, 1998.

\bibitem{Kwiat99src}
P.~G. Kwiat, E.~Waks, A.~G. White, I.~Appelbaum, and P.~H. Eberhard.
\newblock Ultrabright source of polarization-entangled photons.
\newblock {\em Phys. Rev. A}, 60:R773--R776, Aug 1999.

\bibitem{Barreiro05hyper}
J.~T. Barreiro, N.~K. Langford, N.~A. Peters, and P.~G. Kwiat.
\newblock Generation of hyperentangled photon pairs.
\newblock {\em Phys. Rev. Lett.}, 95:260501, Dec 2005.

\bibitem{Iskhakov12bright}
T.~S. Iskhakov, A.~M. P{\'e}rez, K.~Y. Spasibko, M.~V. Chekhova, and G.~Leuchs.
\newblock Superbunched bright squeezed vacuum state.
\newblock {\em Opt. Lett.}, 37:1919--1921, Jun 2012.

\bibitem{Sanaka01pdc}
K.~Sanaka, K.~Kawahara, and T.~Kuga.
\newblock New high-efficiency source of photon pairs for engineering quantum
  entanglement.
\newblock {\em Phys. Rev. Lett.}, 86:5620--5623, Jun 2001.

\bibitem{Harder13waveguide}
G.~Harder, V.~Ansari, B.~Brecht, T.~Dirmeier, C.~Marquardt, and C.~Silberhorn.
\newblock An optimized photon pair source for quantum circuits.
\newblock {\em Opt Express}, 21:13975--13985, Jun 2013.

\bibitem{Perez15twin}
A.~M. P{\'e}rez, K.~Y. Spasibko, P.~R. Sharapova, O.~V. Tikhonova, G.~Leuchs,
  and M.~V. Chekhova.
\newblock Giant narrowband twin-beam generation along the pump-energy
  propagation direction.
\newblock {\em Nat. Comm.}, 6:7707, Jan 2015.

\bibitem{Zukowski93multiphot}
M.~Zukowski, A.~Zeilinger, M.~A. Horne, and A.~K. Ekert.
\newblock Event-ready-detectors bell experiment via entanglement swapping.
\newblock {\em Phys. Rev. Lett.}, 71:4287--4290, Dec 1993.

\bibitem{Zukowski95multiphot}
M.~Zukowski, A.~Zeilinger, and H.~Weinfurter.
\newblock Entangling photons radiated by independent pulsed sources.
\newblock {\em Ann. NY Acad. Sci.}, 755:91, 1995.

\bibitem{Radmark09sixphot}
M.~R{\aa}dmark, M.~Zukowski, and M.~Bourennane.
\newblock Experimental test of fidelity limits in six-photon interferometry and
  of rotational invariance properties of the photonic six-qubit entanglement
  singlet state.
\newblock {\em Phys. Rev. Lett.}, 103:150501, Oct 2009.

\bibitem{Aytur92sqz}
O.~Aytur and P.~Kumar.
\newblock Squeezed-light generation with a mode-locked q-switched laser and
  detection by using a matched local oscillator.
\newblock {\em Opt. Lett.}, 17:529--531, Apr 1992.

\bibitem{Kim94sqz}
C.~Kim and P.~Kumar.
\newblock Quadrature-squeezed light detection using a self-generated matched
  local oscillator.
\newblock {\em Phys. Rev. Lett.}, 73:1605--1608, Sep 1994.

\bibitem{Hirosawa09fs-sqz}
K.~Hirosawa, Y.~Ito, H.~Ushio, H.~Nakagome, and F.~Kannari.
\newblock Generation of squeezed vacuum pulses using cascaded second-order
  optical nonlinearity of periodically poled lithium niobate in a sagnac
  interferometer.
\newblock {\em Phys. Rev. A}, 80:043832, Oct 2009.

\bibitem{Pysher09sqz}
M.~Pysher, R.~Bloomer, C.~M. Kaleva, T.~D. Roberts, B.~Philip, and O.~Pfister.
\newblock Broadband amplitude squeezing in a periodically poled ktiopo4
  waveguide.
\newblock {\em Opt. Lett.}, 34:256--258, Feb 2009.

\bibitem{Breitenbach97sqz}
G.~Breitenbach, S.~Schiller, and J.~Mlynek.
\newblock Measurement of the quantum states of squeezed light.
\newblock {\em Nature}, 387:471--475, 1997.

\bibitem{Lassen07sqz}
M.~Lassen, M.~Sabuncu, P.~Buchhave, and U.~L. Andersen.
\newblock Generation of polarization squeezing with periodically poled ktp at
  1064 nm.
\newblock {\em Opt. Expr.}, 15:5077--5082, Apr 2007.

\bibitem{Perez14sqz}
A.~M. P{\'e}rez, T.~S. Iskhakov, P.~Sharapova, S.~Lemieux, O.~V. Tikhonova,
  M.~V. Chekhova, and G.~Leuchs.
\newblock Bright squeezed-vacuum source with 1.1 spatial mode.
\newblock {\em Opt. Lett.}, 39:2403--2406, Apr 2014.

\bibitem{Yan12sqz}
Z.~Yan, X.~Jia, X.~Su, Z.~Duan, C.~Xie, and K.~Peng.
\newblock Cascaded entanglement enhancement.
\newblock {\em Phys. Rev. A}, 85:040305, Apr 2012.

\bibitem{Wu86sqz}
L.-A. Wu, H.~J. Kimble, J.~L. Hall, and H.~Wu.
\newblock Generation of squeezed states by parametric down conversion.
\newblock {\em Phys. Rev. Lett.}, 57:2520--2523, Nov 1986.

\bibitem{Suzuki06sqz}
S.~Suzuki, H.~Yonezawa, F.~Kannari, M.~Sasaki, and A.~Furusawa.
\newblock 7 db quadrature squeezing at 860 nm with periodically poled ktiopo4.
\newblock {\em Appl. Phys. Lett.}, 89:061116, Aug 2006.

\bibitem{Takeno07sqz}
Y.~Takeno, M.~Yukawa, H.~Yonezawa, and A.~Furusawa.
\newblock Observation of -9 db quadrature squeezing with improvement of phase
  stability in homodyne measurement.
\newblock {\em Opt. Expr.}, 15:4321--4327, Apr 2007.

\bibitem{Hetet07sqz}
G.~H{\'e}tet, O.~Gl{\"o}ckl, K.~A. Pilypas, C.~C. Harb, B.~C. Buchler, H.-A.
  Bachor, and P.~K. Lam.
\newblock Squeezed light for bandwidth-limited atom optics experiments at the
  rubidium d1 line.
\newblock {\em J. Phys. B: At. Mol. Opt. Phys.}, 40:221--226, Jan 2007.

\bibitem{Vahlbruch08sqz}
H.~Vahlbruch, M.~Mehmet, S.~Chelkowski, B.~Hage, A.~Franzen, N.~Lastzka,
  S.~Gossler, K.~Danzmann, and R.~Schnabel.
\newblock Observation of squeezed light with 10-db quantum-noise reduction.
\newblock {\em Phys. Rev. Lett.}, 100:033602, Jan 2008.

\bibitem{Eberle10sqz}
T.~Eberle, S.~Steinlechner, J.~Bauchrowitz, V.~H{\"a}ndchen, H.~Vahlbruch,
  M.~Mehmet, H.~M{\"u}ller-Ebhardt, and R.~Schnabel.
\newblock Quantum enhancement of the zero-area sagnac interferometer topology
  for gravitational wave detection.
\newblock {\em Phys. Rev. Lett.}, 104:251102, Jun 2010.

\bibitem{Heidmann87sqz}
A.~Heidmann, R.~J. Horowicz, S.~Reynaud, E.~Giacobino, C.~Fabre, and G.~Camy.
\newblock Observation of quantum noise reduction on twin laser beams.
\newblock {\em Phys. Rev. Lett.}, 59:2555--2557, Nov 1987.

\bibitem{Mertz91sqz}
J.~Mertz, T.~Debuisschert, A.~Heidmann, C.~Fabre, and E.~Giacobino.
\newblock Improvements in the observed intensity correlation of optical
  parametric oscillator twin beams.
\newblock {\em Opt. Lett.}, 16:1234--1236, Aug 1991.

\bibitem{Tapster88feedback}
P.~R. Tapster, J.~G. Rarity, and J.~S. Satchell.
\newblock Use of parametric down-conversion to generate sub-poissonian light.
\newblock {\em Phys. Rev. A}, 37:2963--2967, 1988.

\bibitem{Mertz91feedback}
J.~Mertz, A.~Heidmann, and C.~Fabre.
\newblock Generation of sub-poissonian light using active control with twin
  beams.
\newblock {\em Phys. Rev. A}, 44:3229--3238, Sep 1991.

\bibitem{Laurat03sqz}
J.~Laurat, T.~Coudreau, N.~Treps, A.~Ma{\^i}tre, and C.~Fabre.
\newblock Conditional preparation of a quantum state in the continuous variable
  regime: generation of a sub-poissonian state from twin beams.
\newblock {\em Phys. Rev. Lett.}, 91:213601, Nov 2003.

\bibitem{Hald99}
J.~Hald, J.~L. S{\o}rensen, C.~Schori, and E.~S. Polzik.
\newblock Spin squeezed atoms: A macroscopic entangled ensemble created by
  light.
\newblock {\em Phys. Rev. Lett.}, 83:1319--1322, Aug 1999.

\bibitem{Honda08sqz}
K.~Honda, D.~Akamatsu, M.~Arikawa, Y.~Yokoi, K.~Akiba, S.~Nagatsuka,
  T.~Tanimura, A.~Furusawa, and M.~Kozuma.
\newblock Storage and retrieval of a squeezed vacuum.
\newblock {\em Phys. Rev. Lett.}, 100:093601, Mar 2008.

\bibitem{Scholz09sm}
M.~Scholz, L.~Koch, R.~Ullmann, and O.~Benson.
\newblock Single-mode operation of a high-brightness narrow-band single-photon
  source.
\newblock {\em Appl. Phys. Lett.}, 94:201105, May 2009.

\bibitem{Wolfgramm11src}
F.~Wolfgramm, Y.~A. de~Icaza~Astiz, F.~A. Beduini, A.~Cer{\`e}, and M.~W.
  Mitchell.
\newblock Atom-resonant heralded single photons by interaction-free
  measurement.
\newblock {\em Phys. Rev. Lett.}, 106:053602, Feb 2011.

\bibitem{Matsko06review1}
A.~B. Matsko and V.~S. Ilchenko.
\newblock Optical resonators with whispering-gallery modes-part i: basics.
\newblock {\em J. Sel. T. Q. El.}, 12:3, Jan 2006.

\bibitem{Matsko06review2}
V.~S. Ilchenko and A.~B. Matsko.
\newblock Optical resonators with whispering-gallery modes-part ii:
  applications.
\newblock {\em J. Sel. T. Q. El.}, 12:15--32, Jan 2006.

\bibitem{Chiasera10WGMrev}
A.~Chiasera, Y.~Dumeige, P.~F{\'e}ron, M.~Ferrari, Y.~Jestin, G.~Nunzi~Conti,
  S.~Pelli, S.~Soria, and G.~C. Righini.
\newblock Spherical whispering-gallery-mode microresonators.
\newblock {\em Las. \& Phot. Rev.}, 4:457--482, Jan 2010.

\bibitem{Strekalov16rev}
D.~V. Strekalov, C.~Marquardt, A.~B. Matsko, H.~G.~L. Schwefel, and G.~Leuchs.
\newblock Nonlinear and quantum optics with whispering gallery resonators.
\newblock {\em J. Opt.}, 18:123002, Jan 2016.

\bibitem{Strekalov14SFG}
D.~V. Strekalov, A.~S. Kowligy, Y.-P. Huang, and P.~Kumar.
\newblock Optical sum-frequency generation in a whispering-gallery-mode
  resonator.
\newblock {\em New J. Phys.}, 16:053025, May 2014.

\bibitem{Savchenkov07THz}
A.~A. Savchenkov, A.~B. Matsko, M.~Mohageg, D.~V. Strekalov, and L.~Maleki.
\newblock Parametric oscillations in a whispering gallery resonator.
\newblock {\em Opt. Lett.}, 32:157--159, Jan 2007.

\bibitem{Fuerst10PDC}
J.~U. F{\"u}rst, D.~V. Strekalov, D.~Elser, A.~Aiello, U.~L. Andersen,
  C.~Marquardt, and G.~Leuchs.
\newblock Low-threshold optical parametric oscillations in a whispering gallery
  mode resonator.
\newblock {\em Phys. Rev. Lett.}, 105:263904, Dec 2010.

\bibitem{Beckmann11}
T.~Beckmann, H.~Linnenbank, H.~Steigerwald, B.~Sturman, D.~Haertle, K.~Buse,
  and I.~Breunig.
\newblock Highly tunable low-threshold optical parametric oscillation in
  radially poled whispering gallery resonators.
\newblock {\em Phys. Rev. Lett.}, 106:143903, Apr 2011.

\bibitem{Beckmann12coupling}
T.~Beckmann, K.~Buse, and I.~Breunig.
\newblock Optimizing pump threshold and conversion efficiency of whispering
  gallery optical parametric oscillators by controlled coupling.
\newblock {\em Opt. Lett.}, 37:5250--5252, Dec 2012.

\bibitem{Werner12BlueOPO}
C.~S. Werner, T.~Beckmann, K.~Buse, and I.~Breunig.
\newblock Blue-pumped whispering gallery optical parametric oscillator.
\newblock {\em Opt. Lett.}, 37:4224--4226, Oct 2012.

\bibitem{Werner15OPO}
C.~S. Werner, K.~Buse, and I.~Breunig.
\newblock Continuous-wave whispering-gallery optical parametric oscillator for
  high-resolution spectroscopy.
\newblock {\em Opt. Lett.}, 40:772--775, Mar 2015.

\bibitem{Fortsch15jopt}
M.~F{\"o}rtsch, T.~Gerrits, M.~J. Stevens, D.~Strekalov, G.~Schunk, J.~U.
  F{\"u}rst, U.~Vogl, F.~Sedlmeir, H.~G.~L. Schwefel, G.~Leuchs, S.~W. Nam, and
  C.~Marquardt.
\newblock Near-infrared single-photon spectroscopy of a whispering gallery mode
  resonator using energy-resolving transition edge sensors.
\newblock {\em J. Opt.}, 17:065501, Jan 2015.

\bibitem{Sizmann90SHGsqz}
A.~Sizmann, R.~J. Horowitz, G.~Wagner, and G.~Leuchs.
\newblock Observation of amplitude squeezing of the up-converted mode in second
  harmonic generation.
\newblock {\em Opt. Comm.}, 80:138--142, Dec 1990.

\bibitem{Kurz92SHsqz}
P.~Kurz, R.~Paschotta, K.~Fiedler, A.~Sizmann, G.~Leuchs, and J.~Mlynek.
\newblock Squeezing by second-harmonic generation in a monolithic resonator.
\newblock {\em Appl. Phys. B}, 55:216--225, 1992.

\bibitem{Drummond81SH}
P.~D. Drummond, K.~J. McNeil, and D.~F. Walls.
\newblock Non-equilibrium transitions in sub/second harmonic generation ii.
  quantum theory.
\newblock {\em Opt. Acta}, 28:211--225, 1981.

\bibitem{Pereira88SHGpumpSQZ}
S.~F. Pereira, M.~Xiao, H.~J. Kimble, and J.~L. Hall.
\newblock Generation of squeezed light by intracavity frequency doubling.
\newblock {\em Phys. Rev. A}, 38:4931, Nov 1988.

\bibitem{Hage10multipartite}
B.~Hage, A.~Samblowski, and R.~Schnabel.
\newblock Towards einstein-podolsky-rosen quantum channel multiplexing.
\newblock {\em Phys. Rev. A}, 81:062301, Jun 2010.

\bibitem{Pysher11ent_comb}
M.~Pysher, Y.~Miwa, R.~Shahrokhshahi, R.~Bloomer, and O.~Pfister.
\newblock Parallel generation of quadripartite cluster entanglement in the
  optical frequency comb.
\newblock {\em Phys. Rev. Lett.}, 107:030505, Jul 2011.

\bibitem{Brieussel16sqz}
A.~Brieussel, Y.~Shen, G.~Campbell, G.~Guccione, J.~Janousek, B.~Hage, B.~C.
  Buchler, N.~Treps, C.~Fabre, F.~Z. Fang, X.~Y. Li, T.~Symul, and P.~K. Lam.
\newblock Squeezed light from a diamond-turned monolithic cavity.
\newblock {\em Opt. Expr.}, 24:4042, Feb 2016.

\bibitem{GerryKnight}
C.~C. Gerry and P.~L. Knight.
\newblock {\em Introductory Quantum Optics}.
\newblock Cambridge {U}niversity {P}ress, 2005.

\bibitem{Chiao64Kerr}
R.~Y. Chiao, E.~Garmire, and C.~H. Townes.
\newblock Self-trapping of optical beams.
\newblock {\em Phys. Rev. Lett.}, 13:479--482, Oct 1964.

\bibitem{Kitagawa86sqz}
M.~Kitagawa and Y.~Yamamoto.
\newblock Number-phase minimum-uncertainty state with reduced number
  uncertainty in a kerr nonlinear interferometer.
\newblock {\em Phys. Rev. A}, 34:3974--3988, Nov 1986.

\bibitem{Shelby86sqz}
R.~M. Shelby, M.~D. Levenson, S.~H. Perlmutter, R.~G. DeVoe, and D.~F. Walls.
\newblock Broad-band parametric deamplification of quantum noise in an optical
  fiber.
\newblock {\em Phys. Rev. Lett.}, 57:691--694, Aug 1986.

\bibitem{Bergman91sqz}
K.~Bergman and H.~A. Haus.
\newblock Squeezing in fibers with optical pulses.
\newblock {\em Opt. Lett.}, 16:663--665, May 1991.

\bibitem{Rosenbluh91sqz}
M.~Rosenbluh and R.~M. Shelby.
\newblock Squeezed optical solitons.
\newblock {\em Phys. Rev. Lett.}, 66:153--156, Jan 1991.

\bibitem{Schmitt98sqz}
S.~Schmitt, J.~Ficker, M.~Wolff, F.~K{\"o}nig, A.~Sizmann, and G.~Leuchs.
\newblock Photon-number squeezed solitons from an asymmetric fiber-optic sagnac
  interferometer.
\newblock {\em Phys. Rev. Lett.}, 81:2446--2449, Sep 1998.

\bibitem{Krylov98sqz}
D.~Krylov and K.~Bergman.
\newblock Amplitude-squeezed solitons from an asymmetric fiber interferometer.
\newblock {\em Opt. Lett.}, 23:1390--1392, Sep 1998.

\bibitem{Friberg96sqz}
S.~R. Friberg, S.~Machida, M.~J. Werner, A.~Levanon, and T.~Mukai.
\newblock Observation of optical soliton photon-number squeezing.
\newblock {\em Phys. Rev. Lett.}, 77:3775--3778, Oct 1996.

\bibitem{Spalter98sqz}
S.~Sp{\"a}lter, M.~Burk, U.~Str{\"o}{\ss}ner, A.~Sizmann, and G.~Leuchs.
\newblock Propagation of quantum properties of subpicosecond solitons in a
  fiber.
\newblock {\em Opt. Expr.}, 2:77--83, Feb 1998.

\bibitem{Riek17sqz}
C.~Riek, P.~Sulzer, M.~Seeger, A.~S. Moskalenko, G.~Burkard, D.~V. Seletskiy,
  and A.~Leitenstorfer.
\newblock Subcycle quantum electrodynamics.
\newblock {\em Nature}, 541:376--379, Jan 2017.

\bibitem{Moskalenko15vac}
A.~S. Moskalenko, C.~Riek, D.~V. Seletskiy, G.~Burkard, and A.~Leitenstorfer.
\newblock Paraxial theory of direct electro-optic sampling of the quantum
  vacuum.
\newblock {\em Phys. Rev. Lett.}, 115:263601, Dec 2015.

\bibitem{Riek15vac}
C.~Riek, D.~V. Seletskiy, A.~S. Moskalenko, J.~F. Schmidt, P.~Krauspe,
  S.~Eckart, S.~Eggert, G.~Burkard, and A.~Leitenstorfer.
\newblock Direct sampling of electric-field vacuum fluctuations.
\newblock {\em Science}, 350:420--423, Oct 2015.

\bibitem{Levandovsky99sqz}
D.~Levandovsky, M.~Vasilyev, and P.~Kumar.
\newblock Amplitude squeezing of light by means of a phase-sensitive fiber
  parametric amplifier.
\newblock {\em Opt. Lett.}, 24:984--986, Jul 1999.

\bibitem{Heersink05sqz}
J.~Heersink, V.~Josse, G.~Leuchs, and U.~L. Andersen.
\newblock Efficient polarization squeezing in optical fibers.
\newblock {\em Opt. Lett.}, 30:1192--1194, May 2005.

\bibitem{Dong08sqz}
R.~Dong, J.~Heersink, J.~F. Corney, P.~D. Drummond, U.~L. Andersen, and
  G.~Leuchs.
\newblock Experimental evidence for raman-induced limits to efficient squeezing
  in optical fibers.
\newblock {\em Opt. Lett.}, 33:116--118, Jan 2008.

\bibitem{Margalit98Xsqz}
M.~Margalit, C.~X. Xu, E.~P. Ippen, and H.~A. Haus.
\newblock Cross phase modulation squeezing in optical fibers.
\newblock {\em Opt. Expr.}, 2:72--76, Feb 1998.

\bibitem{Hirosawa05sqz}
K.~Hirosawa, H.~Furumochi, A.~Tada, F.~Kannari, M.~Takeoka, and M.~Sasaki.
\newblock Photon number squeezing of ultrabroadband laser pulses generated by
  microstructure fibers.
\newblock {\em Phys. Rev. Lett.}, 94:203601, May 2005.

\bibitem{Milanovic10sqz}
J.~Milanovic, M.~Lassen, U.~L. Andersen, and G.~Leuchs.
\newblock A novel method for polarization squeezing with photonic crystal
  fibers.
\newblock {\em Opt. Expr.}, 18:1521--1527, Jan 2010.

\bibitem{Rarity05fiber}
J.~G. Rarity, J.~Fulconis, J.~Duligall, W.~J. Wadsworth, and P.~S.~J. Russell.
\newblock Photonic crystal fiber source of correlated photon pairs.
\newblock {\em Opt. Expr.}, 13:534--544, Jan 2005.

\bibitem{Fan07src}
J.~Fan and A.~Migdall.
\newblock A broadband high spectral brightness fiberbased two-photon source.
\newblock {\em Opt. Expr.}, 15:2915--2920, Feb 2007.

\bibitem{Nold10THG}
J.~Nold, P.~H{\"o}lzer, N.~Y. Joly, G.~K.~L. Wong, A.~Nazarkin, A.~Podlipensky,
  M.~Scharrer, and P.~S.~J. Russell.
\newblock Pressure-controlled phase matching to third harmonic in ar-filled
  hollow-core photonic crystal fiber.
\newblock {\em Opt. Lett.}, 35:2922--2924, Sep 2010.

\bibitem{Finger15sqz}
M.~A. Finger, T.~S. Iskhakov, N.~Y. Joly, M.~V. Chekhova, and P.~S.~J. Russell.
\newblock Raman-free, noble-gas-filled photonic-crystal fiber source for
  ultrafast, very bright twin-beam squeezed vacuum.
\newblock {\em Phys. Rev. Lett.}, 115:143602, Oct 2015.

\bibitem{Vogl15CLEO_Hg}
U.~Vogl, N.~Y. Joly, P.~S.~J. Russell, C.~Marquardt, and G.~Leuchs.
\newblock Squeezed light and self-induced transparency in mercury-filled hollow
  core photonic crystal fibers.
\newblock Jun 2015.

\bibitem{Bradley13Kagome}
T.~D. Bradley, Y.~Wang, M.~Alharbi, B.~Debord, C.~Fourcade-Dutin, B.~Beaudou,
  F.~Gerome, and F.~Benabid.
\newblock Optical properties of low loss (70 db/km) hypocycloid-core kagome
  hollow core photonic crystal fiber for rb and cs based optical applications.
\newblock {\em J. Lightwave Tech.}, 31:2752--2755, Jan 2013.

\bibitem{Chembo10comb}
Y.~K. Chembo, D.~V. Strekalov, and N.~Yu.
\newblock Spectrum and dynamics of optical frequency combs generated with
  monolithic whispering gallery mode resonators.
\newblock {\em Phys. Rev. Lett.}, 104:103902, Mar 2010.

\bibitem{Del'Haye07comb}
P.~Del'Haye, A.~Schliesser, O.~Arcizet, T.~Wilken, R.~Holzwarth, and T.~J.
  Kippenberg.
\newblock Optical frequency comb generation from a monolithic microresonator.
\newblock {\em Nature}, 450:1214--1217, Dec 2007.

\bibitem{Liang15OPO}
W.~Liang, A.~A. Savchenkov, Z.~Xie, J.~F. McMillan, J.~Burkhart, V.~S.
  Ilchenko, C.~W. Wong, A.~B. Matsko, and L.~Maleki.
\newblock Miniature multioctave light source based on a monolithic microcavity.
\newblock {\em Optica}, 2:40, Jan 2015.

\bibitem{Clemmen:2009dn}
S.~Clemmen, K.~P. Huy, W.~Bogaerts, R.~G. Baets, P.~Emplit, and S.~Massar.
\newblock {Continuous wave photon pair generation in silicon-on-insulator
  waveguides and ring resonators}.
\newblock {\em Opt. Expr.}, 17(19):16558--16570, September 2009.

\bibitem{Azzini:2012rc}
S.~Azzini, D.~Grassani, M.~J. Strain, M.~Sorel, L.~G. Helt, J.~E. Sipe,
  M.~Liscidini, M.~Galli, and D.~Bajoni.
\newblock {Ultra-low power generation of twin photons in a compact silicon ring
  resonator}.
\newblock {\em Opt. Expr.}, 20(21):23100--23107, October 2012.

\bibitem{Engin:2013qd}
E.~Engin, D.~Bonneau, C.~M. Natarajan, A.~S. Clark, M.~G. Tanner, R.~H.
  Hadfield, S.~N. Dorenbos, V.~Zwiller, K.~Ohira, N.~Suzuki, H.~Yoshida,
  N.~Iizuka, M.~Ezaki, J.~L. O'Brien, and M.~G. Thompson.
\newblock {Photon pair generation in a silicon micro-ring resonator with
  reverse bias enhancement}.
\newblock {\em Opt. Expr.}, 21(23):27826, November 2013.

\bibitem{Guo:2014db}
Y.~Guo, W.~Zhang, S.~Dong, Y.~Huang, and J.~Peng.
\newblock {Telecom-band degenerate-frequency photon pair generation in silicon
  microring cavities}.
\newblock {\em Opt. Lett.}, 39(8):2526--2529, April 2014.

\bibitem{Grassani:2015zl}
D.~Grassani, S.~Azzini, M.~Liscidini, M.~Galli, M.~J. Strain, M.~Sorel, J.~E.
  Sipe, and D.~Bajoni.
\newblock {Micrometer-scale integrated silicon source of time-energy entangled
  photons}.
\newblock {\em Optica}, 2(2):88--94, February 2015.

\bibitem{Wakabayashi:2015lq}
R.~Wakabayashi, M.~Fujiwara, K.-i. Yoshino, Y.~Nambu, M.~Sasaki, and T.~Aoki.
\newblock {Time-bin entangled photon pair generation from {Si} micro-ring
  resonator}.
\newblock {\em Opt. Expr.}, 23(2):1103, January 2015.

\bibitem{Suo:2015fv}
J.~Suo, S.~Dong, W.~Zhang, Y.~Huang, and J.~Peng.
\newblock {Generation of hyper-entanglement on polarization and energy-time
  based on a silicon micro-ring cavity}.
\newblock {\em Opt. Expr.}, 23(4):3985--3995, February 2015.

\bibitem{Dutt15on-chip_sqz}
A.~Dutt, K.~Luke, S.~Manipatruni, A.~L. Gaeta, P.~Nussenzveig, and M.~Lipson.
\newblock On-chip optical squeezing.
\newblock {\em Phys. Rev. Appl.}, 3:044005, Apr 2015.

\bibitem{Hoff15sqz}
U.~B. Hoff, B.~M. Nielsen, and U.~L. Andersen.
\newblock Integrated source of broadband quadrature squeezed light.
\newblock {\em Opt. Expr.}, 23:12013--12036, May 2015.

\bibitem{Purdy13sqz}
T.~P. Purdy, P.-L. Yu, R.~W. Peterson, N.~S. Kampel, and C.~A. Regal.
\newblock Strong optomechanical squeezing of light.
\newblock {\em Phys. Rev. X}, 3:031012, Jul 2013.

\bibitem{Painter13sqz}
A.~H. Safavi-Naeini, S.~Gr{\"o}blacher, J.~T. Hill, J.~Chan, M.~Aspelmeyer, and
  O.~Painter.
\newblock Squeezed light from a silicon micromechanical resonator.
\newblock {\em Nature}, 500:185--189, Aug 2013.

\bibitem{Teich85sub}
M.~C. Teich and B.~E.~A. Saleh.
\newblock Observation of sub-poisson franck-hertz light at 253.7 nm.
\newblock {\em JOSA B}, 2:275--282, Feb 1985.

\bibitem{Schottky37}
W.~Schottky and E.~Spehnke.
\newblock Raumladungsschw{\"a}chung des schroteffekts.
\newblock {\em Wiss. Ver{\"o}ff. Siemens-Werke}, 16:1--18, 1937.

\bibitem{Yamamoto87sqz}
Y.~Yamamoto and S.~Machida.
\newblock High-impedance suppression of pump fluctuation and amplitude
  squeezing.
\newblock {\em Phys. Rev. A}, 35:5114--5130, Jun 1987.

\bibitem{Machida87sqz}
S.~Machida, Y.~Yamamoto, and Y.~Itaya.
\newblock Observation of amplitude squeezing in a constant-current- driven
  semiconductor laser.
\newblock {\em Phys. Rev. Lett.}, 58:1000--1003, Mar 1987.

\bibitem{Machida88sqz}
S.~Machida and Y.~Yamamoto.
\newblock Ultrabroadband amplitude squeezing in a semiconductor laser.
\newblock {\em Phys. Rev. Lett.}, 60:792--794, Feb 1988.

\bibitem{Richardson91sqz}
W.~H. Richardson, S.~Machida, and Y.~Yamamoto.
\newblock Squeezed photon-number noise and sub-poissonian electrical partition
  noise in a semiconductor laser.
\newblock {\em Phys. Rev. Lett.}, 66:2867--2870, Jun 1991.

\bibitem{Marin95sqz}
F.~Marin, A.~Bramati, E.~Giacobino, T.-C. Zhang, J.~P. Poizat, J.-F. Roch, and
  P.~Grangier.
\newblock Squeezing and intermode correlations in laser diodes.
\newblock {\em Phys. Rev. Lett.}, 75:4606--4609, Dec 1995.

\bibitem{Maurin05laser}
I.~Maurin, I.~Protsenko, J.-P. Hermier, A.~Bramati, P.~Grangier, and
  E.~Giacobino.
\newblock Light intensity-voltage correlations and leakage-current excess noise
  in a single-mode semiconductor laser.
\newblock {\em Phys. Rev. A}, 72:033823, Sep 2005.

\bibitem{Wang93sqz}
H.~Wang, M.~J. Freeman, and D.~G. Steel.
\newblock Squeezed light from injection-locked quantum well lasers.
\newblock {\em Phys. Rev. Lett.}, 71:3951--3954, Dec 1993.

\bibitem{Freeman93sqz}
M.~J. Freeman, H.~Wang, D.~G. Steel, R.~Craig, and D.~R. Scifres.
\newblock Wavelength-tunable amplitude-squeezed light from a room-temperature
  quantum-well laser.
\newblock {\em Opt. Lett.}, 18:2141--2143, Dec 1993.

\bibitem{Wolfl02sqz}
F.~Wolfl, R.~G. Ispasoiu, J.~F. Ryan, and A.~M. Fox.
\newblock Photon-number squeezing in a free-running quantum-well laser
  operating at 980 nm.
\newblock {\em J. Opt. B: Quantum Semiclass. Opt.}, 4:129--133, Feb 2002.

\bibitem{Uemukai05sqz}
M.~Uemukai, S.~Nozu, and T.~Suhara.
\newblock High-efficiency ingaas qw distributed bragg reflector laser with
  curved grating for squeezed light generation.
\newblock {\em J. Sel. T. Q. El.}, 11:1143--1147, Jan 2005.

\bibitem{Yamamoto86sqz}
Y.~Yamamoto, N.~Imoto, and S.~Machida.
\newblock Amplitude squeezing in a semiconductor laser using quantum
  nondemolition measurement and negative feedback.
\newblock {\em Phys. Rev. A}, 33:3243--3261, May 1986.

\bibitem{Buchler99squash}
B.~C. Buchler, M.~B. Gray, D.~A. Shaddock, T.~C. Ralph, and D.~E. McClelland.
\newblock Suppression of classic and quantum radiation pressure noise by
  electro-optic feedback.
\newblock {\em Opt. Lett.}, 24:259--261, Feb 1999.

\bibitem{Shapiro87feedback}
J.~H. Shapiro, G.~Saplakoglu, S.-T. Ho, P.~Kumar, B.~E.~A. Saleh, and M.~C.
  Teich.
\newblock Theory of light detection in the presence of feedback.
\newblock {\em JOSA B}, 4:1604--1620, Oct 1987.

\bibitem{Mancini2000squash}
S.~Mancini, D.~Vitali, and P.~Tombesi.
\newblock Motional squashed states.
\newblock {\em J. Opt. B: Quantum Semiclass. Opt.}, 2:190--195, 2000.

\bibitem{Caldeira85damping}
A.~O. Caldeira and A.~J. Leggett.
\newblock Influence of damping on quantum interference: An exactly soluble
  model.
\newblock {\em Phys. Rev. A}, 31:1059--1066, Feb 1985.

\bibitem{Leuchs05dissipation}
G.~Leuchs and U.~Andersen.
\newblock The effect of dissipation on non-classical states of the radiation
  field.
\newblock {\em Las. Phys.}, 15:129--134, 2005.

\bibitem{Eberly80revival}
J.~H. Eberly, N.~B. Narozhny, and J.~J. Sanchez-Mondragon.
\newblock Periodic spontaneous collapse and revival in a simple quantum model.
\newblock {\em Phys. Rev. Lett.}, 44:1323--1326, May 1980.

\bibitem{Rempe87revival}
G.~Rempe, H.~Walther, and N.~Klein.
\newblock Observation of quantum collapse and revival in a one-atom maser.
\newblock {\em Phys. Rev. Lett.}, 58:353--356, Jan 1987.

\end{thebibliography}
\printindex

\end{document}